%% file: lepton-anom-mag-mom.tex
\documentclass[onecolumn,sort&compress,numbers]{els-mrw} % For numbered references Style 

\usepackage{amsmath,amssymb,amsfonts,amsthm,makeidx,graphicx,color}
\usepackage{hyperref}
\usepackage{wrapfig}
\usepackage{txfonts}
\usepackage{helvet}
\usepackage{mathbbol}

\input definitions.tex

\begin{document}

%%%%%%%%%%%%%%%%%%%%%%%%%%%%%%%%%%%%%%%%%%%%%%%%%%%%%%%%%%%%%%%%
%% the following items are mandatory: 
%% - title
%% - author names
%% - affiliation details
%% - abstract
%% - keywords

%% Precise, concise, and informative description of the focus of this work. Avoid abbreviations and formulae in the title
\chapter{Lepton anomalous magnetic moments: Theory}\label{chap1}

%% All author names and affiliations, and email address for corresponding author
\author[1]{Hartmut Wittig}%

\address[1]{\orgname{Johannes Gutenberg University Mainz},
  \orgdiv{Institute for Nuclear Physics and PRISMA$^++$ Cluster of
    Excellence}, \orgaddress{Johann Joachim 
    Becher-Weg 45, 55099 Mainz, Germany}}
%\address[2]{\orgname{GSI Helmholtz Centre for Heavy Ion Research},
%  \orgdiv{Helmholtz Institute Mainz}, \orgaddress{Staudinger Weg 18,
%    55128 Mainz, Germany}} 

%\articletag{Chapter Article tagline: update of previous edition, reprint.}

\maketitle

%%%%%%%%%%%%%%%%%%%%%%%%%%%%%%%%%%%%%%%%%%%%%%%%%%%%%%%%%%%%%%%%
%% the following item is mandatory: 
%% 100-150 word summary of the chapter
\begin{abstract}[Abstract]
The anomalous magnetic moment of a lepton encodes the fraction of the
lepton's interaction strength with an external magnetic field, which
is generated by quantum corrections. Lepton anomalous magnetic moments
are sensitive probes of fundamental interactions and play a pivotal
role in the quest for ``new physics'' that may be able to explain the
shortcomings of the Standard Model. This chapter introduces the basic
concepts and describes the calculation of the individual contributions
arising from electromagnetism, the strong and the weak interactions.
\end{abstract}

%% 5-10 words that embody the key topics in the chapter. What terms would someone put into a search engine if they were looking for a chapter like this?
\begin{keywords}
 	Lepton magnetic moments \sep Precision observables \sep
        Multi-loop calculations \sep Dispersion theory \sep Lattice
        Quantum Chromodynamics
\end{keywords}

%%%%%%%%%%%%%%%%%%%%%%%%%%%%%%%%%%%%%%%%%%%%%%%%%%%%%%%%%%%%%%%%
%% the following item is optonal: 
%% - Single figure visually illustrating the key topic/method/outcome described in the chapter
\begin{figure}[h]
  \centering
  \includegraphics[width=0.97\textwidth]{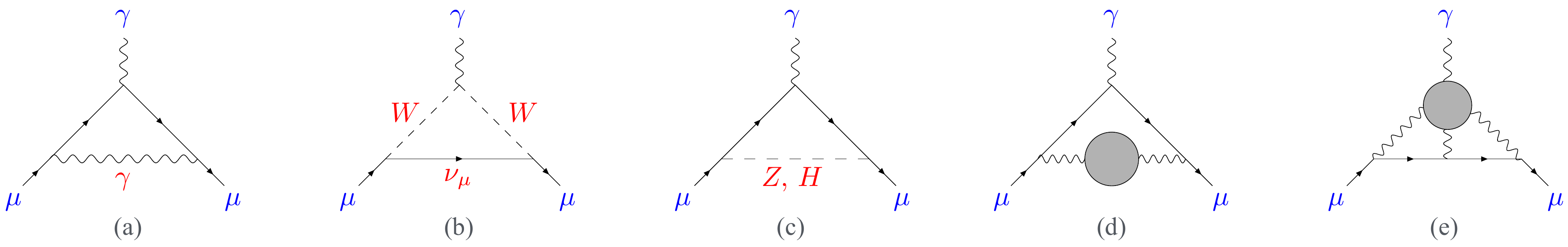}
  \caption{Diagrams representing the leading-order electromagnetic and
    weak contributions to the muon anomalous magnetic moment (diagrams
    (a)--(c)), as well as the hadronic vacuum polarisation (diagram
    (d)) and hadronic light-by-light scattering (diagram (e))
    contributions that arise from the strong interaction.}
  \label{fig:diagramsLO}
\end{figure}

%%%%%%%%%%%%%%%%%%%%%%%%%%%%%%%%%%%%%%%%%%%%%%%%%%%%%%%%%%%%%%%%
%% the following item is optional: 
%% - System of abbreviations/terms/symbols used in the specific field of study/community. List and define
%\begin{glossary}[Nomenclature]
%	\begin{tabular}{@{}lp{34pc}@{}}
%%        ChPT & Chiral Perturbation Theory \\
%%        hQCD & Holographic QCD \\
%        BSM  & Beyond-Standard Model \\
%        EW   & Electroweak \\
%		HVP  & Hadronic Vacuum Polarisation\\
%		HLbL & Hadronic Light-by-Light Scattering\\
%        SM   & Standard Model \\
%	\end{tabular}
%\end{glossary}

%%%%%%%%%%%%%%%%%%%%%%%%%%%%%%%%%%%%%%%%%%%%%%%%%%%%%%%%%%%%%%%%
%% the following item is mandatory: 
%% List of the key points and topics a reader can expect to learn from this chapter 

\section*{Key Points}

\begin{itemize}
  \item Anomalous magnetic moments arise from quantum corrections
    involving the electromagnetic, weak and strong interactions
  \item The anomalous magnetic moment of the muon is a sensitive probe
    of the Standard Model
  \item Discrepancies between the measured values of leptonic
    anomalous magnetic moments and the Standard Model prediction, if
    observed, provide a strong hint for physics beyond the Standard
    Model
  \item Electromagnetic and weak contributions account for 99.994\% of
    the value of the muon's anomalous magnetic moment
  \item Strong interaction contributions dominate the uncertainty of
    the Standard Model prediction for the muon anomalous magnetic
    moment
  \item The most important methodologies to determine the
    contributions from the strong interaction are dispersion integrals
    using experimentally measured hadronic cross sections and
    numerical calculations in lattice Quantum Chromodynamics
\end{itemize}

%%%%%%%%%%%%%%%%%%%%%%%%%%%%%%%%%%%%%%%%%%%%%%%%%%%%%%%%%%%%%%%%
%% the following items are mandatory: 
%% - Section: Introduction 
%% - further sections
%% - Section: Conclusion

\input intro
\input QED

\input HVP_DD
\input HVP_lat
\input HLbL
\input EW
\input conclusion

\begin{ack}[Acknowledgments]%
I it a pleasure to thank Gilberto Colangelo, Achim Denig, Simon
Kuberski, Dominik St\"ockinger and Thomas Teubner for comments on the
manuscript. I am grateful to Makiko Nio for discussions on QED
corrections. This work was partially supported by Deutsche
Forschungsgemeinschaft (German Research Foundation, DFG) through the
Collaborative Research Center 1660 ``Hadrons and Nuclei as Discovery
Tools'', and through the Cluster of Excellence ``Precision Physics,
Fundamental Interactions and Structure of Matter'' (PRISMA+ EXC
2118/1), funded within the German Excellence strategy (Project
No.\ 390831469).
\end{ack}

%%%%%%%%%%%%%%%%%%%%%%%%%%%%%%%%%%%%%%%%%%%%
%% Optional: A list of references to other relevant works/articles/websites which are not cited in the text but that would further enhance a readers understanding of this topic
%\seealso{article title article title}

%%%%%%%%%%%%%%%%%%%%%%%%%%%%%%%%%%%%%%%%%
%% Mandatory: Bibliography using bibtex 
\bibliographystyle{Numbered-Style} %% for Numbered Reference Style
\bibliography{jointbiblio}
%\bibliography{biblio,MCevent,Expts,pipi_Exp}

\end{document}

%% file: definitions.tex
\newcommand{\be}{\begin{equation}}
\newcommand{\ee}{\end{equation}}
\newcommand{\bea}{\begin{eqnarray}}
\newcommand{\eea}{\end{eqnarray}}
\newcommand{\eq}[1]{Eq.\,(\ref{#1})}

\newcommand{\rmO}{{\textrm{O}}}

\newcommand{\Nf}{N_{\rm{f}}}

\newcommand{\GeV}{{\rm{GeV}}}

\newcommand{\ahvp}{a_\mu^{\rm{hvp}}}
\newcommand{\alohvp}{a_\mu^{\rm{lo,\,hvp}}}
\newcommand{\ahlbl}{a_\mu^{\rm{hlbl}}}
\newcommand{\awin}{a_\mu^{\rm{win}}}
\newcommand{\aew}[1]{a_\mu^{{\rm{EW}}\,{#1}}}
\newcommand{\GF}{G_{\rm{F}}}
\newcommand{\rb}[1]{\raisebox{1.5ex}[-1.5ex]{#1}}

%% file: intro.tex
\section{Introduction} \label{sec:intro}

Lepton anomalous magnetic moments have played a crucial role in the
development of the Standard Model of particle physics since the early
days of Quantum Field Theory, pushing the capabilities of both theory
and experiment to their limits. Since the development of Quantum
Mechanics in the 1920, it was accepted that the magnetic moment
induced by the electron's intrinsic angular momentum -- the spin --
was twice as large as the classical theory would suggest. This led to
the {\it ad hoc} introduction of the $g$-factor into the spin-induced
magnetic moment and setting $g=2$. The fact that this value turned out
to be a prediction of the Dirac equation gave further credence to this
view. When more refined experimental techniques became available in
the late 1940s, it was realised that the measured magnetic moment of
the electron deviated from Dirac's prediction by a small amount of
O($10^{-3}$). The first experimental observation of an ``anomalous''
contribution by Foley and Kusch \cite{Foley:1948zz,Kusch:1948mvb}
coincided with the development of Quantum Field Theory, and it is fair
to say that the first calculation of the anomalous contribution at
leading order by Julian Schwinger \cite{Schwinger:1948iu} was a
triumph for the newly established theory of Quantum Electrodynamics
(QED). Since then, lepton anomalous magnetic moments have continued to
provide a benchmark for our understanding of the basic constituents of
matter and the interactions among them. As the experimental
sensitivity improved over the years, increasingly refined theoretical
calculations were required, pushing the limits of our ability to
perform multi-loop calculations in QED, perturbative calculations in
the electroweak theory and eventually also the precise determination
of the contributions from the strong interaction which largely defies
a perturbative treatment.

At the time of writing (2025), the community at large is focussed
specifically on the muon's anomalous magnetic moment, $a_\mu$. The
E989 experiment at Fermilab has now published the final results from a
new measurement campaign
\cite{Muong-2:2021ojo,Muong-2:2023cdq,Muong-2:2025xyk} which, together
with the result from its predecessor experiment E821 at BNL
\cite{Bennett:2006fi}, has pushed the precision of the direct
measurement of $a_\mu$ to an impressive 124 parts per billion
(ppb). Given that the required experimental sensitivity for the first
detection of the anomaly by Foley and Kusch was at the per-mil level,
the enormous increase in precision of both experimental measurement
and theoretical prediction stands as an impressive testimony to the
progress that has been achieved. The precision of the measurement of
the anomalous magnetic moment of the electron, $a_e$, is even higher
and stands at a staggering 0.11~ppb. Whether or not the Standard Model
predictions for $a_\mu$ and, to a lesser extent, $a_e$ agree with
experiment has wide-ranging consequences. The answer to this question
determines whether the Standard Model (SM) remains -- at least for the
time being and despite all its deficiencies -- our best description of
the fundamental constituents of matter, or whether it must be
augmented or replaced by a more complete theory.

It is clear that a concise overview such as this cannot do justice to
the enormous amount of material on lepton anomalous magnetic
moments. Therefore, this article is focussed on introducing the main
concepts and summarising the current state of the art. A much more
complete and, in fact, exhaustive reference on the subject is the book
by Jegerlehner \cite{Jegerlehner:2017gek} which also contains many
explicit calculations. Another good reference is the review by
Jegerlehner and Nyffeler \cite{Jegerlehner:2009ry}. A review that
specifically focusses on lattice QCD for calculating the hadronic
contributions can be found in \cite{Meyer:2018til}. Finally, I would
like to point the reader to the White Papers on the subject that
appeared in 2020 \cite{Aoyama:2020ynm} and 2025
\cite{Aliberti:2025beg} which summarise the current knowledge on the
SM prediction for the muon anomalous magnetic moment.

%\comment{This article is structured as follows\ldots\ldots}

\subsection{Overview and definitions \label{sec:overview}}

In this subsection we elaborate on the above general motivation by
collecting the basic facts and definitions involving lepton magnetic
moments. In classical physics, a point-like particle of mass $m$,
electric charge $q$ and angular momentum $\vec{L}$ possesses a
magnetic moment $\vec\mu$, according to
\be\label{eq:muclassical}
  \vec{\mu}=\frac{q}{2m}\,\vec{L}\,.
\ee
This relation also holds for a quantum mechanical particle, provided
that $\vec{L}$ is restricted to orbital angular momentum. The spin of
a particle also gives rise to a magnetic moment. Specifically, if we
turn our attention to the electron, by setting $q=-e$ (where $e>0$
denotes the elementary charge), the spin-induced part of its magnetic
moment, $\vec\mu_s$, is given by
\be\label{eq:mudefinition}
  \vec\mu_s=-g\,\frac{e}{2m}\,\vec{S},\quad
  \vec{S}=\frac{1}{2}\hbar\vec{\sigma}, \ee
where $\vec{S}$ is the spin operator defined in terms of the Pauli
matrices $\vec{\sigma}$. The key difference to \eq{eq:muclassical} is
the appearance of the $g$-factor which, historically, was assigned the
{\it ad hoc} value $g=2$, in order to explain the observed spacing
between energy levels associated with the anomalous Zeeman effect.

The magnetic moment $\vec\mu_s$ is an intrinsic property of an
electrically charged particle with spin, which defines the interaction
strength of the particle with an external magnetic field. In Quantum
Mechanics, the interaction of an electron of charge $-e$ with an
external magnetic field, encoded in terms of the scalar and vector
potentials $\Phi$ and $\vec{A}$, respectively, is described by the
Pauli equation,
\be
  i\hbar\frac{\partial}{\partial t} \psi(\vec{x},t)=
  \left\{\frac{1}{2m}\left[\vec{\sigma}\cdot\left(
    \vec{p}+e\vec{A}\,\right)\right]^2  -e\Phi \right\}\psi(\vec{x},t)\,.
\ee
The Pauli equation is derived from the non-relativistic limit of the
Dirac equation and can be cast into a Schr\"odinger-like form, i.e.
\be\label{eq:pauli}
  i\hbar\frac{\partial}{\partial t}\psi(\vec{x},t)
  =\left\{\frac{1}{2m}\left(\vec{p}+e\vec{A}\,\right)^2
  -e\Phi+\frac{e\hbar}{2m}\,\vec{\sigma}\cdot\vec{B}
  \right\}\psi(\vec{x},t)\,.
\ee
Since the magnetic interaction strength is given by
$-\vec\mu_s\cdot\vec{B}$, equations (\ref{eq:mudefinition}) and
(\ref{eq:pauli}) imply $g=2$, which is the Dirac value of the
$g$-factor. However, in the late 1940s the magnetic interaction
strength of an electron could be measured much more precisely than
before, which led to the discovery that the $g$-factor of the electron
was slightly larger than expected, i.e. $g=2\cdot(1.00119\pm0.00005)$
\cite{Kusch:1948mvb,Foley:1948zz}. This suggested that the Dirac value
$g=2$ had to be modified by quantum corrections arising from virtual
interactions between a charged point particle (i.e. the electron) and
the photon field. The definition of the $g$-factor was therefore
augmented by what we now call the anomalous magnetic moment $a$
according to
\be
  g=2(1+a)\quad\Leftrightarrow\quad a={\textstyle\frac{1}{2}}(g-2).
\ee
It was Julian Schwinger who, in 1948, calculated the leading-order
correction in perturbation theory \cite{Schwinger:1948iu} (represented
by diagram~(a) in Fig.\,\ref{fig:diagramsLO}), using the then newly
formulated theory of Quantum Electrodynamics (QED). The result,
denoted by $a^{(2)}$ is a simple expression in terms of the
fine-structure constant $\alpha\equiv e^2/4\pi$, i.e.
\be\label{eq:Schwinger}
  a^{(2)}=\frac{\alpha}{2\pi}=0.001\,161\,40\ldots\,.
\ee
This value agreed with the experimental value up to the attainable
precision at the time.\footnote{The superscript on $a^{(2)}$ denotes
the order in the electric charge~$e$,} However, we now know that
Schwinger's result does not tell us the full story, since it does not
agree with the latest experimental value which is \cite{Fan:2022eto}
\be
   a_e^{\rm exp}=0.001\,159\,652\,180\,59\,(13)\quad [0.11\,{\rm ppb}].
\ee
Moreover, Schwinger's original calculation implied that the QED
correction was universal, i.e. identical for every lepton flavour,
$\ell=e,\,\mu,\,\tau$. In fact, the latest measurements of the muon
anomalous magnetic moment by the E989 experiment
\cite{Muong-2:2021ojo,Muong-2:2023cdq,Muong-2:2025xyk} at Fermilab,
when combined with the earlier measurements at Brookhaven
\cite{Bennett:2006fi}, yield an experimental average of
\be
   a_\mu^{\rm exp}=0.001\,165\,920\,715\,(145)\quad [124\,{\rm ppb}].
\ee
The fact that $a_e^{\rm exp}$ differs from $\alpha/2\pi$ is easily
understood when one realises that the anomalous magnetic moment
receives contributions beyond the leading (one loop) order in
$\alpha$. In addition, one must take the effects of the weak and
strong interactions into account, whose contributions at leading order
are shown in diagrams~(b)--(d) in Fig.\,\ref{fig:diagramsLO}. All
these corrections affect the electron and muon (and $\tau$) in
different ways, which explains why the anomalous magnetic moments
$a_\ell$ of all three known lepton flavours $\ell=e,\,\mu,\,\tau$
differ significantly. These observations illustrate why there is such
a high interest in lepton anomalous magnetic moments: the fact that
they can be measured so precisely makes them ideal probes for subtle
effects that either originate in the SM itself or in its extensions
which may provide solutions for yet unexplained phenomena such as dark
matter or the matter-antimatter asymmetry in the universe. Pictorially
speaking, all fundamental interactions leave their fingerprints in the
values of leptonic anomalous magnetic moments. This suggests a
particular strategy to search for physics beyond the Standard Model
(BSM), which compares precision measurements with equally precise,
SM-based theoretical predictions. Any observation of a non-zero
deficit between experiment and SM-based theoretical prediction could
then be attributed to BSM physics, i.e.
\begin{equation}
    a_\ell^{\rm exp}-(a_\ell^{\rm QED}+a_\ell^{\rm weak}+a_\ell^{\rm
      strong}) = a_\ell^{\rm BSM}\,.
\end{equation}
The muon anomalous magnetic moment is a particularly promising
quantity in the quest for new physics, owing to the fact that the
relative BSM contribution scales like \cite{Berestetskii:1956tkr}
\be\label{eq:BSMsensitivity}
\frac{a_\ell^{\rm BSM}}{a_\ell}\propto \frac{m_\ell^2}{M_{\rm BSM}^2},
\ee
where $m_\ell$ is the lepton mass and $M_{\rm BSM}$ the BSM
scale. Hence, the sensitivity of $a_\mu$ is enhanced relative to $a_e$
by a factor $(m_\mu/m_e)^2\approx 4\cdot10^4$, i.e. by four orders of
magnitude.\footnote{So-called chirally enhanced BSM scenarios
\cite{Crivellin:2021rbq} in which the BSM contribution scales linearly
in the lepton mass are the exception to this rule.}  Moreover, $a_\mu$
can be measured with a precision of 124\,ppb
\cite{Bennett:2006fi,Muong-2:2021ojo,Muong-2:2023cdq,Muong-2:2025xyk}
which is much more precise than what can currently be achieved
experimentally for $a_\tau$, which would be even more sensitive to
$M_{\rm BSM}$. However, the short mean life of the $\tau$, which
amounts to $2.903(5)\cdot10^{-13}$\,s presents a considerable
challenge to experimentalists. The PDG presently quotes $-0.057<
a_\tau <0.024$ at 95\% confidence
level\,\cite{ParticleDataGroup:2024cfk}.
A comprehensive discussion of different BSM scenarios can be found in
a recent review \cite{Athron:2025ets}.
The argument behind \eq{eq:BSMsensitivity} applies equally to quantum
corrections from heavier particles within the SM. For instance, the
contribution from the weak interaction scales like
${m_\ell^2}/{M_{W,\,Z}^2}$.
%
%\be
%   \frac{a_\ell^{\rm weak}}{a_\ell}\propto \frac{m_\ell^2}{M_{W,\,Z}^2}\,.
%\ee
%
Therefore, lepton anomalous magnetic moments have been benchmark
quantities for the development of Quantum Field Theory: In the 1950s
and 60s, the experimental sensitivity had reached a level that made it
possible to probe higher-order (two- and three-loop) effects in QED.
Since then, the attention has shifted to corrections from the weak and
strong interactions and to exposing BSM effects.

%% file: QED.tex
\section{QED contribution to lepton anomalous magnetic moments}
\label{sec:QED}

Historically, QED supplied the first estimate for the anomalous
magnetic moment of the electron. QED also contributes by far the
highest fraction to the SM estimate. For instance, it accounts for
99.994\% of the value of $a_\mu$ while that figure is even higher in
the case of $a_e$. Here we discuss the QED contribution up to
$10^{\rm{th}}$ order in the electric charge, starting with the basic
concepts and sketching the calculation performed by Schwinger which
produced the estimate in \eq{eq:Schwinger}. From this point on, we
will use natural units with $\hbar=c=1$.

QED describes the interactions between the photon field $A^\mu(x)$ and
the charged leptons $\ell=e,\,\mu,\,\tau$. It is defined in terms of
the (classical) Lagrangian
\be
   {\cal{L}}_{\rm{QED}} =-\frac{1}{4}F_{\mu\nu}(x)F^{\mu\nu}(x)
   +\sum_{\ell=e,\,\mu,\,\tau} \left\{ \bar{\psi}_\ell(x)\left(
   i\gamma^{\,\mu}\partial_\mu-m_\ell \right)\psi_\ell(x)
   -e\bar{\psi}_\ell(x)\gamma^{\,\mu} A_\mu(x)\psi_\ell(x) \right\}\,,
\ee
where $F^{\mu\nu}=\partial^{\,\mu} A^\nu-\partial^{\,\nu} A^\mu$ is
the field strength tensor, and $\psi_\ell$ denotes the four-component
Dirac spinor describing a fermion of mass~$m$ and electric
charge~$e$. The coupling between the photon and the lepton fields,
described by the last term, is realised by introducing the covariant
derivative $D_\mu$ via
\be
   D_\mu:=\partial_\mu+ieA_\mu(x)\,,
\ee
which not only yields the more compact form of the Lagrangian, i.e.
\be\label{eq:QEDLagrangian}
   {\cal{L}}_{\rm{QED}} =-\frac{1}{4}F_{\mu\nu}(x)F^{\mu\nu}(x)
   +\sum_{\ell=e,\,\mu,\,\tau} \bar{\psi}_\ell(x)\left(
   i\gamma^{\,\mu} D_\mu-m_\ell \right)\psi_\ell(x)
\ee
but also hints at the fact that QED is an Abelian gauge theory.

The Lagrangian is the basis for the derivation of the Feynman rules
which, in the case of QED, associate algebraic expressions with the
lepton and photon propagators, as well as for the lepton-photon
interaction vertex (see Fig.\,\ref{fig:Feynman}). These are the
elementary building blocks of Feynman diagrams that can be drawn for a
given process, classified according to the factors of electric
charge~$e$ that appear at a given order.

The relation between the lepton's $g$-factor $g_\ell$ and its
anomalous magnetic moment, $a_\ell$, is given by
\be
   g_\ell=2(1+a_\ell)\equiv g_\ell^{(0)}(1+a_\ell)\,,
\ee
where $g_\ell^{(0)}\equiv2$ denotes the tree-level value of the
$g$-factor that is derived from the Dirac equation.\\

\subsection{Lowest-order QED prediction for \texorpdfstring{$a_\ell$}{a	extunderscore l}}

Calculating the anomaly in QED requires the quantisation of the theory
of leptons and photons defined by the classical Lagrangian of
\eq{eq:QEDLagrangian}. This can be achieved through the canonical
procedure that elevates the classical lepton and photon fields to
operators acting in a Hilbert (Fock) space and then imposes suitable
commutation rules. Alternatively, one can use the path integral
formalism. None of the details pertaining to either procedure will be
discussed here within the available space, and instead the reader is
referred to standard textbooks \cite{Itzykson:1980rh, Peskin:1995ev,
  Weinberg:1995mt}. Renormalisation is another central topic in
Quantum Field Theory that this article cannot do justice
to. Renormalisation is a general procedure to treat infinities that
are encountered when evaluating loop integrals. It is a vast topic
that is inextricably linked to Quantum Field Theory, and again we
refer to the standard textbooks cited above.

\begin{figure}
    \centering
    \includegraphics[width=0.6\linewidth]{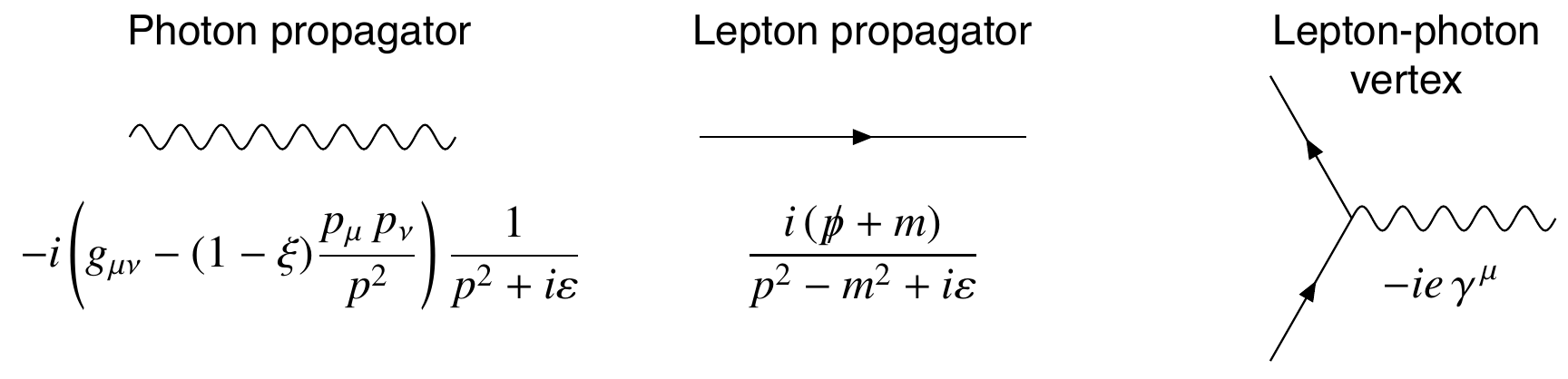}
    \caption{Feynman rules for Quantum Electrodynamics in momentum
      space. The lepton is characterised by its mass~$m$ and electric
      charge~$e$. Setting the gauge parameter $\xi$ to $\xi=1$
      corresponds to Feynman gauge, the choice $\xi=0$ is called
      Landau gauge.}
    \label{fig:Feynman}
\end{figure}

The starting point for the derivation of lepton anomalous magnetic
moments in the $S$-matrix element ${\cal{M}}$ that describes the
interaction between a lepton~$\ell$ with an external electromagnetic
current $j^{\,\mu}_{\rm em}(x)$, i.e.
\be
   {\cal{M}}(x)=\left\langle \ell(p')\left|j^{\,\mu}_{\rm em}(x)\right|
   \ell(p) \right\rangle\,,
\ee
where $p,\,p'$ denote the momenta of the incoming and outgoing
leptons, respectively. We are specifically interested in the limit of
vanishing momentum transfer, $(p'-p)^2\to0$. In momentum space, the
matrix element reads
\be
   \tilde{\cal{M}}(q)=\int d^4x\,e^{-iq{\cdot}x} \left\langle
   \ell(p')\left| j_{\rm em}^{\,\mu}(x) \right|\ell(p)
   \right\rangle\,, 
\ee
and noting that the lepton initial and final states are also
eigenstates of four-momentum, one finds
\be
   \tilde{\cal{M}}(q)=\int d^4x\,e^{-i(p-p'-q){\cdot}x} \left\langle
   \ell(p')\left| j_{\rm em}^{\,\mu}(0) \right|\ell(p)
   \right\rangle =(2\pi)^4\delta^{(4)}\Big(q-(p'-p)\Big)\left\langle
   \ell(p')\left| j_{\rm em}^{\,\mu}(0) \right|\ell(p)
   \right\rangle\,.
\ee
While the $\delta$-function guarantees momentum conservation, the
$T$-matrix element can be written as
\be\label{eq:Tmatrix}
   \left\langle \ell(p')\left| j_{\rm em}^{\,\mu}(0) \right|\ell(p)
   \right\rangle = (-ie)\, \bar{u}(p')\,\Gamma^{\,\mu}(p',p)\,u(p)\,,
\ee
where $u(p)$ and $\bar{u}(p')$ are Dirac spinors for the incoming and
outgoing lepton, and $\Gamma^\mu(p',p)$ is the vertex function. At
tree level it is given by $\Gamma^\mu(p',p)=\gamma^{\,\mu}$, in line
with the Feynman rule for the elementary lepton-photon vertex (see
Fig.\,\ref{fig:Feynman}). With the help of the Ward identity
$q_\mu\Gamma^\mu=0$ and by imposing Lorentz invariance, one derives
the general expression for the vertex function sandwiched between
Dirac spinors, which reads
\be\label{eq:FFdecom}
   \bar{u}(p') \Gamma^{\,\mu}(p',p) u(p) =
   \bar{u}(p') \left\{ F_1(q^2)\gamma^{\,\mu}
   +i\frac{\sigma^{\,\mu\nu}q_\nu}{2m_\ell} F_2(q^2) \right\} u(p)\,,
\ee
where we have also made use of the Gordon identity.\footnote{A
detailed derivation of the decomposition of the vertex function in
terms of the form factors $F_1$ and $F_2$ can be found in chapter~6 of
Peskin and Schroeder \cite{Peskin:1995ev}.}
The quantities $F_1$ and $F_2$ are commonly referred to as the Dirac
and Pauli form factors. We will now elaborate on their physical
interpretation. Choosing $\mu=0$ in \eq{eq:Tmatrix} one finds that
\eq{eq:FFdecom} turns into
\be
   \bar{u}(p') \Gamma^{\,0}(p',p) u(p) =
   \bar{u}(p') \left\{ F_1(q^2) +\frac{t}{4m_\ell^2} F_2(q^2) \right\}
   u(p)\,, 
\ee
where $t\equiv(p'-p)^2$, and hence once can define the electric form
factor $G_{\rm E}(q^2)$ as
\be
   G_{\rm E}(q^2)=F_1(q^2)+\frac{t}{4m_\ell^2} F_2(q^2)\,.
\ee
Furthermore, noting that integrating $j^{\,0}(x)$ over the spatial
volume yields the total electric charge, one derives that $F_1(0)=1$,
which means that $F_1(0)$ represents the electric charge in units of
$e$. Charge conservation then implies that $F_1(0)=G_{\rm E}(0)$
receives no further radiative corrections.
In a similar fashion one can show, by choosing $\mu=k=1,2,3$ that the
vertex function $\Gamma^{\,k}$ between spinors describing the incoming
and outgoing lepton is proportional to
\be
   G_{\rm M}(q^2)=F_1(q^2)+F_2(q^2)\,.
\ee
Specifically, for $t\equiv(p'-p)^2=0$, i.e.\ at vanishing momentum
transfer, $G_{\rm M}(0)$ yields the magnetic moment in units of the
Bohr magneton $e/(2m_\ell)$, i.e.
\be\label{eq:muFFs}
   \vec\mu_s = \frac{e}{2m_\ell}\left(F_1(0)+F_2(0)\right)\vec\sigma\,.
\ee
\begin{wrapfigure}{r}{5.6cm}
  \centering
  \includegraphics[width=0.18\textwidth]{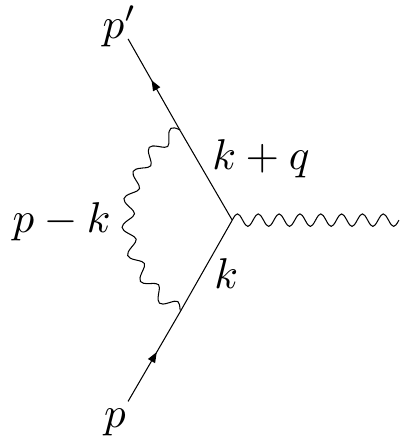}
  \caption{The vertex correction in QED at one loop.}
  \label{fig:oneloop}
\end{wrapfigure}
According to \eq{eq:mudefinition}, the spin-induced magnetic moment of
a particle with charge $e$ is given by
$\vec\mu_s=g_\ell(e/(2m_\ell))\,\vec{S}$. We can then solve
\eq{eq:muFFs} for the $g$-factor, which gives
\be
   g_\ell=2\left( F_1(0)+F_2(0) \right)\,,
\ee
and noting that $F_1(0)=1$ we find that the lepton's anomalous magnetic
moment is given by $F_2(0)$, i.e.
\be
   a_\ell\equiv{\textstyle\frac{1}{2}} (g_\ell-2) = F_2(0)\,.
\ee
At tree level, that is, without any radiative corrections from QED or,
indeed, any other Quantum Field Theory that is part of the SM or its
extensions, there is no contribution to the $g$-factors, and therefore
\be
   g_\ell = 2+\rmO(\alpha)\quad\Leftrightarrow\quad
   F_2(0)=\rmO(\alpha)\,. 
\ee
We will now proceed to sketch Schwinger's pioneering calculations
performed in 1948 \cite{Schwinger:1948iu}, which showed that the
leading QED correction is given by
\be
   F_2(0)=\frac{\alpha}{2\pi}\,.
\ee
The correction to the vertex function $\Gamma^\mu$ in QED at one loop
is determined from the diagram shown in Fig.\,\ref{fig:oneloop}
In order to compute it, we decompose the vertex function
$\Gamma^\mu(p',p)$ into a tree-level part $\gamma^{\,\mu}$ plus the
one-loop correction according to
\be
   \Gamma^\mu(p',p)=\gamma^{\,\mu}+\delta\Gamma^{\,\mu}(p',p)\,.
\ee
As we are interested in the scattering of the lepton $\ell$ off an
external potential $\tilde{A}_\mu^{\rm ext}$, we must evaluate --
after applying the Feynman rules -- the following momentum space
expression ($k\hspace{-4pt}/\equiv\gamma^{\,\mu} k_\mu$):
\begin{align}
   \bar{u}(p')\,\delta\Gamma^{\,\mu}(p',p)\,u(p)\tilde{A}_\mu^{\rm ext}
   &=\int\frac{d^4k}{(2\pi)^4}\,
   \frac{-ig_{\nu\lambda}}{(k-p)^2+i\epsilon} \bar{u}(p')
   (-ie\gamma^\nu) S_{\rm F}(k+q)\gamma^{\,\mu} S_{\rm F}(k)
   (-ie\gamma^\lambda)\, u(p)\tilde{A}_\mu^{\rm ext} \nonumber \\
   &= -\frac{ie^2}{(2\pi)^4} \int d^4k\, \bar{u}(p')
   \frac{\gamma^\nu(k\hspace{-4pt}/'+m_\ell)
     \gamma^{\,\mu}(k\hspace{-4pt}/+m_\ell)
     \gamma_\nu}{\left((k-p)^2+i\epsilon\right)\,
     \left({k'}^2-m_\ell^2+i\epsilon\right)\, \left({k}^2-m_\ell^2+i\epsilon\right)}
   \,u(p)\tilde{A}_\mu^{\rm ext} \nonumber \\
   &= \frac{2ie^2}{(2\pi)^4} \int d^4k\, \bar{u}(p')
   \frac{k\hspace{-4pt}/'\gamma^{\,\mu}k\hspace{-4pt}/
     +m_\ell^2\gamma_\mu -2m_\ell(k+k')^{\,\mu}}{\left((k-p)^2+i\epsilon\right)\,
     \left({k'}^2-m_\ell^2+i\epsilon\right)\, \left({k}^2-m_\ell^2+i\epsilon\right)}
   \,u(p)\tilde{A}_\mu^{\rm ext}\,, \label{eq:vertexcorr}
\end{align}
where we have abbreviated $(k+q)\equiv k'$, set the gauge parameter in
the photon propagator to~$\xi=1$, and inserted the expressions for the
fermion propagator $S_{\rm F}$ in momentum space (see
Fig.\,\ref{fig:Feynman}). Between the second and third line we have
simplified the expressions involving strings of $\gamma$-matrices
using the anticommutation relations.

The following steps are quite lengthy and will not be shown
explicitly. Detailed accounts are given in standard textbooks such as
\cite{Peskin:1995ev,Weinberg:1995mt,Scheck2013}. They involve the
introduction of Feynman parameters to rewrite and combine the
denominators, applying a Wick rotation and performing the integration
over momentum variables in terms of four-dimensional spherical
coordinates. Moreover, the momentum integrals exhibit both ultraviolet
and infrared divergencies which must be regulated. In the case of the
infrared divergence, this is accomplished by introducing a small
photon mass $m_\gamma$ which is eventually taken to zero in the final
result. Furthermore, the diagram for the vertex correction cannot be
considered in isolation but must be combined with the diagrams for the
lepton self energy. By comparing the expression that are derived from
\eq{eq:vertexcorr} with the form factor decomposition of
\eq{eq:FFdecom} one finally recovers Schwinger's result, i.e. the
contribution to $F_2(0)$ and hence to the anomalous magnetic moment at
order $e^2=4\pi\alpha$ in the electric charge, which reads
\be\label{eq:al1loop}
   a_\ell^{(2)}=\frac{\alpha}{2\pi} = 0.001\,161\,\ldots\,,
\ee
in excellent agreement with the measurement of the electron's
anomalous magnetic moment at the time, $a_e=0.001\,19(5)$, performed
by Kusch and Foley \cite{Foley:1948zz,Kusch:1948mvb}. At leading order
in $\alpha$, QED yields identical results for the anomalous magnetic
moments of all three charged leptons:
$a_e^{(2)}=a_\mu^{(2)}=a_\tau^{(2)}$. This no longer holds when higher
loop orders in QED are taken into account and also when including
contributions from the weak and strong interactions. It is interesting
to note that the $\rmO(\alpha)$ contribution deviates from the full SM
prediction for $a_e$ and $a_\mu$ by less than 0.4\%.

\subsection{Higher-order QED contributions}

We will now go beyond the one-loop calculation described in the
previous subsection and focus on higher-order
contributions. Specifying to the case of the muon for concreteness,
the general structure of its QED prediction is
\be\label{eq:amuQEDexp}
a_\mu^{\rm QED} = A_1 +A_{2}(m_\mu/m_e) +A_{2}(m_\mu/m_\tau)
+A_{3}(m_\mu/m_e,m_\mu/m_\tau)\,.
\ee
The terms $A_1,\,A_2$ and~$A_3$ admit a perturbative expansion in
powers of the fine-structure constant $\alpha$ according to
\be
   A_i= A_i^{(2)}\left(\frac{\alpha}{\pi}\right) 
       +A_i^{(4)}\left(\frac{\alpha}{\pi}\right)^2
       +A_i^{(6)}\left(\frac{\alpha}{\pi}\right)^3+\ldots
   \equiv\sum_{L=1}^{\infty}\,A_i^{(2L)}\left(\frac{\alpha}{\pi}\right)^L,
   \quad i=1,\,2,\,3\,.
\ee
$A_1$ describes the universal contribution that applies to all
leptons. In other words, $A_1$ represents the result that would be
obtained if there were only one species of leptons. The terms $A_2$ and
$A_3$ arise when a diagram contains a closed loop of a lepton whose
flavour $\ell'$ differs from that of the external lepton $\ell$,
i.e. $\ell'\neq\ell$.
To help distinguish the mass-dependent contributions, we will, from
now on, amend the terms $A_2$ and $A_3$ by a subscript that indicates
the flavour of the external lepton, i.e. we write $A_{2\ell}$,
$A_{3\ell}$ with $\ell=e,\,\mu,\,\tau$.
Since lepton anomalous magnetic moments are dimensionless quantities,
any non-universal flavour dependence can only enter via lepton mass
ratios, as already indicated in \eq{eq:amuQEDexp}. Thus, for the muon,
mass-dependent contributions arise whenever a $\tau$ lepton or an
electron contribute through a closed loop in a given diagram. The same
reasoning applies to the electron and $\tau$ anomalous magnetic moment
with the arguments of the corresponding $A_{2\ell}$'s in
\eq{eq:amuQEDexp} suitably adjusted, which implies that
$A_{2\ell}^{(2)}=0,\;\ell=e,\,\mu,\,\tau$. Mass-dependent
contributions involving two flavours require diagrams with at least
two closed lepton loops, which implies that such contributions start
only at three-loop order, i.e. $A_{3\ell}^{(2)}=A_{3\ell}^{(4)}=0$.

The relative size of the mass-dependent contribution $A_{2\ell}$
depends crucially on the mass ratios. For instance, for the muon,
$A_{2\mu}(m_\mu/m_e)$ dominates the non-universal part, since the
ratio $m_\mu/m_e$ gives rise to large logarithms
$\sim\ln(m_\mu/m_e)^2$. By contrast, $A_{2\mu}(m_\mu/m_\tau)$ is small
due to the decoupling of heavy particles in QED-like theories, which
produces a sub-leading contribution $\sim(m_\mu/m_\tau)^2$. For the
same reason, mass-dependent contributions to the electron anomalous
magnetic moment, $A_{2e}(m_e/m_\mu),\,A_{2e}(m_e/m_\tau)$ and
$A_{3e}(m_e/m_\mu, m_e/m_\tau)$ are all sub-leading.

\begin{figure}
    \centering
    \includegraphics[width=0.75\linewidth]{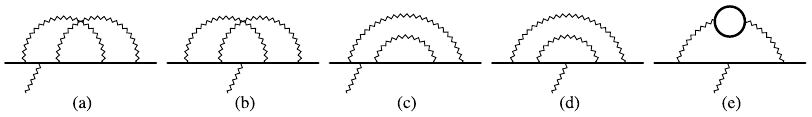}
    \caption{Fourth-order vertex diagrams. There are~7 diagrams in
      total (note that the time-reversed versions of~(a) and~(c) are
      not shown). Diagram~(e) gives rise to mass-dependent
      contributions whenever the mass of the lepton in the loop does
      not coincide with that of the extermal lepton. Photons are
      denoted by wavy lines, lepton loops by straight lines. Figure
      taken from~\cite{Aoyama:2012qma}.
    \label{fig:2loopdiags}}
    \bigskip
    \includegraphics[width=0.75\linewidth]{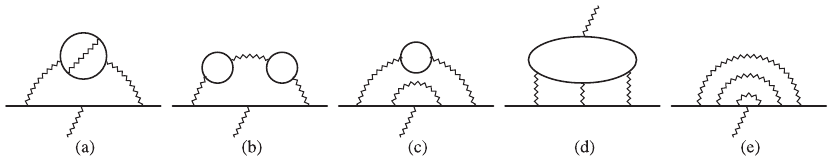}
    \caption{Sixth-order diagrams, divided into five gauge-invariant
      sets. The total number of vertex diagrams is 72. Set~(b) gives
      rise to doubly mass-dependent contributions collected in
      $A_{3\ell}$. Set~(d) corresponds to light-by-light
      scattering. Figure taken from~\cite{Aoyama:2012qma}.
    \label{fig:3loopdiags}}
    \bigskip
    \includegraphics[width=0.75\linewidth]{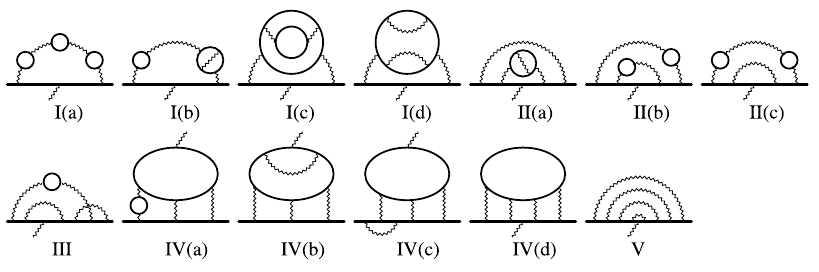}
    \caption{Eigth-order diagrams, divided into sets I--V and further
      subdivided into~13 gauge-invariant subsets. The total number of
      vertex diagrams is 891. Figure taken from~\cite{Aoyama:2012wj}.
    \label{fig:4loopdiags}}
    \bigskip
    \includegraphics[width=0.6\linewidth]{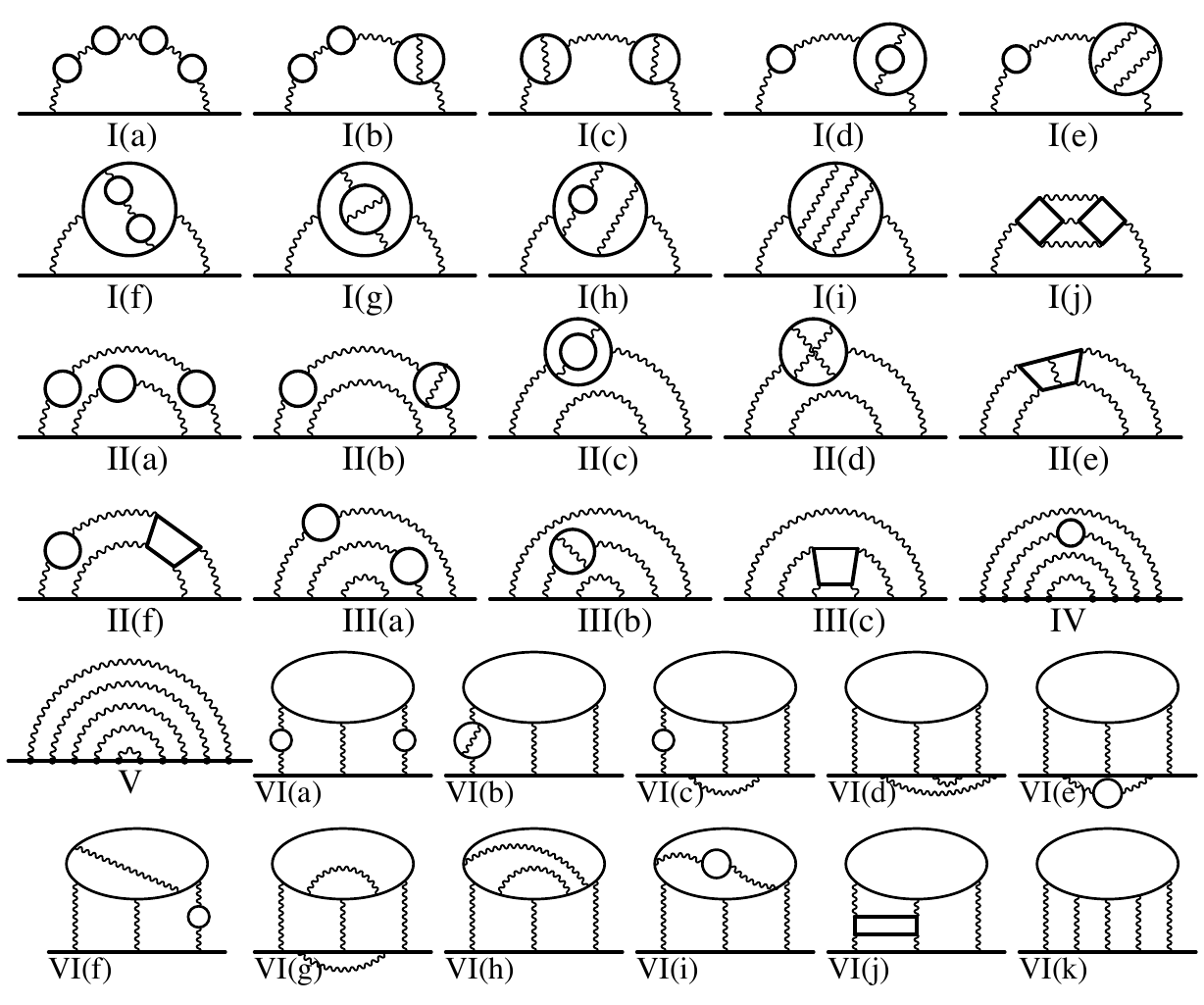}
    \caption{Sets I--VI of tenth-order diagrams, further subdivided
      into 32 gauge-invariant subsets. The total number of vertex
      diagrams is 12672. Figure taken from~\cite{Aoyama:2012wj}.
    \label{fig:5loopdiags}}
\end{figure}

We now provide a synopsis of the calculation of the terms
$A_{i\ell},\,\ell=e,\,\mu,\,\tau$ up to five-loop order ($10^{\rm th}$
order in the electric charge). The number of diagrams with distinct
topology at each order is listed in the second column in
Table\,\ref{tab:A1univ5loop}. All diagrams at a given order contribute
to the universal part $A_1$, including diagrams that contain lepton
loops whose flavour coincides with that of the external lepton. This
point is illustrated further in Fig.\,\ref{fig:2loopdiags} where the
diagrams arising at two-loop order are shown. Sub-classes of diagrams
that contain closed lepton loops must be evaluated separately whenever
the mass ratio of internal and external leptons differs from~1.

In the following we discuss the calculation of the universal
(mass-independent) coefficients $A_1^{(2L)}$ that contribute to
$a_e^{\rm QED},\,a_\mu^{\rm QED}$ and $a_\tau^{\rm QED}$ alike.
Moreover, the evaluation of $A_1$ does not involve any input
parameters that must be determined from experiment. The lowest order
involves the calculation of a single diagram sketched in the preceding
subsection and shown in Fig.\,\ref{fig:oneloop}. From \eq{eq:al1loop}
one reads off that
\be
   A_1^{(2)}=\frac{1}{2}\,.
\ee
At two loops, i.e. at $4^{\rm th}$ order in the electric charge, seven
vertex diagrams must be evaluated (see Fig.\,\ref{fig:2loopdiags}).
Six of them involve only photon lines and yield the expression
\cite{Jegerlehner:2017gek}
\be
 A_1^{(4)}(\hbox{photons})=-\frac{279}{144}+\frac{5\pi^2}{12} -\frac{\pi^2}{2}\ln2 +\frac{3}{4}\zeta(3)\,,
\ee
where $\zeta(n)$ denotes the Riemann zeta-function. Whenever the
lepton flavour $\ell'$ in diagram~(e) in Fig.~\ref{fig:2loopdiags}
coincides with that of the external lepton~$\ell$, i.e. for
$m_\ell/m_{\ell'}=1$, it contributes \cite{Jegerlehner:2017gek}
\be
 A_1^{(4)}(m_\ell/m_{\ell'}=1)=\frac{119}{36}-\frac{\pi^2}{3}\,,
\ee
so that the total universal contribution at order~4 is
\cite{Petermann:1957hs,SOMMERFIELD195826}
\be
 A_1=\frac{197}{144}+\left(\frac{1}{2}-3\ln2\right)\zeta(2)+\frac{3}{4}\zeta(3) = -0.328\,478\,965\,579\,193\,784\,582\ldots\,,
\ee
where we have substituted $\zeta(2)=\pi^2/6$.

At three loops ($6^{\rm th}$ order), the number of vertex diagrams
with different topology has grown to 72 (see
Fig.\,\ref{fig:3loopdiags} and Table~\ref{tab:A1univ5loop}). The
universal contribution $A_1^{(6)}$ has been calculated analytically by
Laporta and Remiddi \cite{Laporta:1996mq}, and its value agrees with
the numerical evaluation by Kinoshita \cite{Kinoshita:1995ym}. The
exact expression reads
\begin{align}
 A_1^{(6)} &= \frac{28259}{5184}+\frac{17101}{810}\pi^2
 -\frac{298}{9}\pi^2\ln2 +\frac{139}{18}\zeta(3) 
 +\frac{100}{3}\left\{a_4
 +\frac{1}{24}\ln^4{2} -\frac{1}{24}\pi^2\ln^2{2}\right\} 
 -\frac{239}{2160}\pi^4 +\frac{83}{72}\pi^2\zeta(3) 
 -\frac{215}{24}\zeta(5) \nonumber\\
 &= 1.181\,241\,456\,587\,200\ldots\,,
\end{align}
where $a_4\equiv{\rm Li}_4(1/2)$ is the polylogarithm
\be
{\rm Li}_4\left({\textstyle\frac{1}{2}}\right) =\sum_{n=1}^{\infty}\frac{1}{2^n\,n^4}=0.517\,479\,061\,673\,899\,386\ldots\,.
\ee

The four-loop ($8^{\rm th}$ order) contribution involves the
formidable task of evaluating 891 individual vertex diagrams (see Fig.\,\ref{fig:4loopdiags} and Table\,\ref{tab:A1univ5loop}).
Knowledge of the $8^{\rm th}$ order contribution, though small in
size, is mandatory to match the experimental precision. In 2017, the
universal contribution $A_1^{(8)}$ has been determined in
near-analytical form by Laporta \cite{Laporta:2017okg}, leaving only a
small number of integrals that must be evaluated numerically. As a
result, the mass-independent $8^{\rm th}$ order contribution is known
up to 1100 decimal places \cite{Laporta:2017okg}, the first few of
which are
\be
 A_1^{(8)} = -1.912\,245\,764\,926\,445\,574\ldots\,,
\ee
which agrees well with the numerical calculation of
Ref.~\cite{Aoyama:2014sxa}, which yields $A_1^{(8)}=-1.912 98 (84)$.
The full sequence of 1100 digits is listed in Table~1 of
Ref.~\cite{Laporta:2017okg}.

At five loops ($10^{\rm th}$ order), a staggering number of 12\,672
vertex diagrams must be evaluated. Figure~\ref{fig:5loopdiags}, taken
from Ref.~\cite{Aoyama:2012wj}, shows the 32 gauge-invariant subsets
that originate from classes~I-VI of topologies. Until now, only
numerical evaluations of the $10^{\rm th}$-order coefficients have
been performed. The first complete calculation of all universal and
mass-dependent contributions was reported by Aoyama, Hayakawa,
Kinoshita and Nio (AHKN) in~\cite{Aoyama:2012wj}, followed by several
updates \cite{Aoyama:2014sxa,Aoyama:2017uqe,Aoyama:2019ryr}.
Volkov~\cite{Volkov:2019phy,Volkov:2024yzc} calculated the
contribution to $A_1^{(10)}$ from Set~V independently and found that
the result differed by five standard deviations from the earlier work
by AHKN. After performing a diagram-by-diagram comparison, AHKN
identified 98 integrals that were responsible for the discrepancy due
to insufficient statistics in the Monte Carlo integration. The value
for the contribution from Set~V was updated
accordingly~\cite{Aoyama:2024aly}; They show good agreement with
Volkov's results. After combining the results for Set~V with the
remaining diagrams containing lepton loops, the two calculations are
consistent, with Volkov finding
$A_1^{(10)}=5.891(61)$~\cite{Volkov:2024yzc}, to be compared with
AHKN's $A_1^{(10)}=5.870(128)$~\cite{Aoyama:2024aly}. The value listed
in Table~\ref{tab:A1univ5loop} is based on a weighted average of the
two results, performed in~\cite{Aliberti:2025beg}.

\begin{table}[t]
    \centering
    \begin{tabular}{ccr@{.}l l}
    \hline\hline
    Coefficient $A_i^{(2L)}$ & \#Diagrams & \multicolumn{2}{l}{Value} & References \\
    \hline
    $A_1^{(2)}$  &     1 &    0&5 & \cite{Schwinger:1948iu} \\
    $A_1^{(4)}$  &     7 & $-0$&328\,478\,965\,579\,193\,\ldots & \cite{Petermann:1957hs,SOMMERFIELD195826} \\
    $A_1^{(6)}$  &    72 &    1&181\,241\,456\,587\,200\,\ldots & \cite{Laporta:1996mq,Kinoshita:1995ym} \\
    $A_1^{(8)}$  &   891 & $-1$&912\,245\,764\,926\,445\,\ldots & \cite{Laporta:2017okg,Aoyama:2014sxa} \\
    $A_1^{(10)}$ & 12672 &   5&887(55) & \cite{Aoyama:2017uqe,Volkov:2019phy,Volkov:2024yzc,Aoyama:2024aly,Aliberti:2025beg} \\
    \hline\hline
    \end{tabular}
    \caption{The coefficients describing the universal
      (mass-independent) QED contributions to lepton anomalous
      magnetic moments up to $10^{\rm th}$ (five-loop)
      order. Coefficients up to and including $8^{\rm th}$ order are
      exact.
    \label{tab:A1univ5loop}}
%\end{table}
%%
%\begin{table}[p]
%\centering
\bigskip
\begin{tabular}{l r@{.}l |l r@{.}l |l}
\hline\hline
Coefficient $A_{i\,e}^{(2L)}$ & \multicolumn{2}{l|}{\;\;Value(Error)} &
Coefficient $A_{i\,\mu}^{(2L)}$ & \multicolumn{2}{l|}{\quad Value(Error)} &
References\\
\hline
$A_{2\,e}^{(4)}(m_e/m_\mu)$      & 0&$519\,738\,676(24)\cdot10^{-6}$ & 
$A_{2\,\mu}^{(4)}(m_\mu/m_e)$    & 1&094\,258\,3093(76) &
\cite{Suura:1957zz,ELEND1966682,Li:1992xf} \\
$A_{2\,e}^{(4)}(m_e/m_\tau)$     & 0&$183\,790(25)\cdot10^{-8}$ & 
$A_{2\,\mu}^{(4)}(m_\mu/m_\tau)$ & 0&000\,078\,076(11) &
\cite{Suura:1957zz,ELEND1966682,Li:1992xf} \\
\hline
$A_{2\,e}^{(6)}(m_e/m_\mu)$      & $-0$&$737\,394\,164(24)\cdot10^{-5}$ & 
$A_{2\,\mu}^{(6)}(m_\mu/m_e)$    & 22&868\,379\,98(20) &
\cite{Samuel:1990qf,Li:1992xf,Laporta:1992pa,Laporta:1993ju} \\
$A_{2\,e}^{(6)}(m_e/m_\tau)$     & $-0$&$658\,273(79)\cdot10^{-7}$ & 
$A_{2\,\mu}^{(6)}(m_\mu/m_\tau)$ & 0&000\,360\,601(83) &
\cite{Samuel:1990qf,Li:1992xf,Laporta:1992pa,Laporta:1993ju} \\
$A_{3\,e}^{(6)}(m_e/m_\mu,m_e/m_\tau)$ & 0&$1909(1)\cdot10^{-12}$ &
$A_{3\,\mu}^{(6)}(m_\mu/m_e,m_\mu/m_\tau)$ & 0&000\,527\,738(75) &
\cite{Passera:2006gc,Ananthanarayan:2020acj}\\
\hline
$A_{2\,e}^{(8)}(m_e/m_\mu)$      & $0$&$916\,197\,070(37)\cdot10^{-3}$ & 
$A_{2\,\mu}^{(8)}(m_\mu/m_e)$    & 132&6852(60) &
\cite{Kinoshita:2002ns,Kinoshita:2004wi,Kurz:2013exa,Aoyama:2012wj}\\
$A_{2\,e}^{(8)}(m_e/m_\tau)$     & 0&$742\,92(12)\cdot10^{-5}$ & 
$A_{2\,\mu}^{(8)}(m_\mu/m_\tau)$ & 0&042\,4941(53) &
\cite{Kinoshita:2002ns,Kinoshita:2004wi,Kurz:2013exa,Aoyama:2012wj}\\
$A_{3\,e}^{(8)}(m_e/m_\mu,m_e/m_\tau)$ & 0&$746\,87(28)\cdot10^{-6}$ &
$A_{3\,\mu}^{(8)}(m_\mu/m_e,m_\mu/m_\tau)$ & 0&062\,722(10) & 
\cite{Kinoshita:2002ns,Kinoshita:2004wi,Kurz:2013exa,Aoyama:2012wj}\\
\hline
$A_{2\,e}^{(10)}(m_e/m_\mu)$      & $-0$&003\,82(39) & 
$A_{2\,\mu}^{(10)}(m_\mu/m_e)$    & 742&32(86) & \cite{Aoyama:2012wk}\\
$A_{2\,e}^{(10)}(m_e/m_\tau)$     & \multicolumn{2}{l|}{$\quad\rm{O}(10^{-5})$} & 
$A_{2\,\mu}^{(10)}(m_\mu/m_\tau)$ & $-0$&0656(45)  & \cite{Aoyama:2012wk} \\
$A_{3\,e}^{(10)}(m_e/m_\mu,m_e/m_\tau)$ & \multicolumn{2}{l|}{$\quad\rm{O}(10^{-5})$} &
$A_{3\,\mu}^{(10)}(m_\mu/m_e,m_\mu/m_\tau)$ & 2&011(10) & \cite{Aoyama:2012wk} \\
\hline\hline
\end{tabular}
\caption{Coefficients for the mass-dependent QED contributions up to
  $10^{\rm th}$ order. The two leftmost columns apply for the
  electron, the third and fourth column to the muon.
\label{tab:A23mass5loop}}
\bigskip
    \begin{tabular}{c r@{.}l r@{.}l | r@{.}l r@{.}l}
    \hline\hline
    $L$ & \multicolumn{2}{c}{$C_{L\,e}$}
        & \multicolumn{2}{c|}{$C_{Le}(\alpha/\pi)^{L}\cdot10^{12}$}
        & \multicolumn{2}{c}{$C_{L\,\mu}$}
        & \multicolumn{2}{c}{$C_{L\mu}(\alpha/\pi)^{L}\cdot10^{11}$} \\
    \hline
    1 &  0&5 & 1\,161\,409\,731&859(82) & 0&5 & 116\,140\,973&1859(82)\\
    2 &  $-0$&328\,478\,444\,002\,62(35) & $-1$\,772\,302&2456(3)
      &  0&765\,857\,4197(134) & 413\,217&6249(72) \\
    3 &  1&181\,234\,016\,818(8) & 14\,804&1104(0) 
      & 24&050\,509\,775(229) & 30\,141&9021(3) \\
    4 & $-1$&911\,321\,3909(12) & $-55$&641(0)
      & 130&8782(60) & 381&004(18) \\
    5 &  5&883(55) & 0&398(4) & 750&15(86) & 5&0726(58) \\
    \hline
      \multicolumn{3}{l}{Total 5-loop} & 1\,159\,652\,178&480(82)
      & \multicolumn{2}{c}{$\empty$} &    116\,584\,718&789(21) \\
    \hline\hline
    \end{tabular}
    \caption{The summed coefficients $C_{L\ell}$ at loop order~$L$ for
      the electron and the muon ($\ell=e,\,\mu$). After multiplication
      with the appropriate power of $(\alpha/\pi)$, we also provide
      the absolute contributions at each order. For the fine-structure
      constant $\alpha$, the value determined from atom interferometry
      on Rubidium~\cite{Morel:2020dww,Mohr:2024kco} has been
      chosen. The errors in $C_{Le}(\alpha/\pi)^L$ are dominated by
      the uncertainty in the fine-structure constant. By contrast, the
      errors on $C_{L\mu}(\alpha/\pi)^L$ arise from the uncertainties
      in the lepton mass ratios.
    \label{tab:QED5loop}}
\end{table}

Mass-dependent corrections require at least one closed lepton loop and
arise, therefore, at two-loop order and higher (see
Fig.~\ref{fig:2loopdiags}). Their evaluation also relies on the
knowledge of the mass ratio(s) $m_\ell/m_{\ell'}$ of the external
($\ell$) and virtual ($\ell'$) leptons. Specifically, there are two
types of asymptotic expansions, depending on whether
$m_\ell/m_{\ell'}<1$ or $m_\ell/m_{\ell'}>1$. All recent
determinations of mass-dependent contributions
%the coefficients $A_{2\ell}^{2L}$ and $A_{3\ell}^{2L}$ 
use as input the experimentally measured mass ratios listed as the
CODATA recommended values in~\cite{Mohr:2024kco}, i.e.
\be
 m_\mu/m_e=206.768\,2827(47)\,,\quad m_\mu/m_\tau=0.059\,4635(40)\,,
\ee
from which one derives the electron-to-$\tau$ mass ratio as
\be
 m_e/m_\tau = 0.287\,585(19)\cdot10^{-3}\,.
\ee
A comprehensive discussion of mass-dependent corrections up to
$10^{\rm th}$ order (five loops), including explicit expressions for
closed analytic forms, various asymptotic expansions, as well as plots
of classes of diagrams can be found in Chapter~4.1 of the book by
Jegerlehner~\cite{Jegerlehner:2017gek}. For more concise descriptions
we refer the reader to the White Papers by the Muon $g-2$ Theory
Initiative~\cite{Aoyama:2020ynm,Aliberti:2025beg}. Below we present a
summary of the main features.

The fourth-order (two-loop) mass-dependent correction was first
computed as an asymptotic expansion in the mass ratio $m_e/m_\mu$ in
Ref.~\cite{Suura:1957zz,Petermann:1957hs}. The analytic expression for
any mass ratio~$x$ was worked out in~\cite{ELEND1966682,Li:1992xf},
which was subsequently shown to assume the form~\cite{Passera:2006gc}
\be
 A_{2\ell}^{(4)}(1/x)=-\frac{25}{36}-\frac{\ln{x}}{3}+x^2(4+3\ln{x}) +\frac{x}{2}(1-5x^2)
 \left[\frac{\pi^2}{2}-\ln{x}\ln\left(\frac{1-x}{1+x}\right)-{\rm Li}_2(x)+{\rm Li}_2(-x)\right]
 +x^4 \left[\frac{\pi^2}{3}-2\ln{x}\ln\left(\frac{1}{x}-x\right) -{\rm Li}_2(x^2)\right]\,.
\ee
For $x>1$, one can use a series expansion, in order to avoid the cuts
that the logarithms develop in that regime. Furthermore, for numerical
evaluations it is more convenient to work with asymptotic expansions
that are presented in detail in Chapter 4.1 of
\cite{Jegerlehner:2017gek}. The corresponding expressions evaluate to
the values listed in rows~2 and~3 in Table~\ref{tab:A23mass5loop}.

While the mass-dependent contributions with one fermion loop at sixth
order, $A_2^{(6)}(1/x)$, are available in closed analytic form for any
value of the lepton mass ratio~$x$
\cite{Laporta:1992pa,Laporta:1993ju}, these expressions are not
practical to work with for numerical evaluations. Explicit expressions
for various mass ratios $x=m_e/m_\mu$ and $x=m_\tau/m_\mu$ are again
listed in Section~4.1 of Ref.~\cite{Jegerlehner:2017gek}. For the
muon, one finds a sizeable enhancement in
$A_{2\,\mu}^{(6)}(m_\mu/m_e)$, as signified by the large entry in the
fourth row of Table~\ref{tab:A23mass5loop}, which originates in the
leptonic light-by-light scattering diagrams shown as ``Set~(d)'' in
Fig.~\ref{fig:3loopdiags}. The contribution from
$A_{2\,\mu}^{(6)}(m_\mu/m_\tau)$, on the other hand, is strongly
suppressed due to the approximate decoupling of heavy fermions. At
$6^{\rm th}$ order, the mass-dependent contribution involving two
lepton loops enters for the first time (see diagram~(b) in
Fig.~\ref{fig:3loopdiags}). Numerical values for the sixth-order
coefficients $A_{2\ell}^{(6)}$ and $A_{3\ell}^{(6)}$ for
$\ell=e,\,\mu$ are listed in rows~4--6 in
Table~\ref{tab:A23mass5loop}, with the quoted errors arising from the
uncertainties in the respective lepton mass ratios.

At $8^{\rm th}$ order (four loops) one finds a large coefficient
$A_{2\mu}^{(8)}$ due to a double-logarithmic enhancement for
$m_\mu/m_e$. The various terms have been computed through numerical
integration~\cite{Aoyama:2012wk,Kinoshita:2005zr,Laporta:1993ds} and
in terms of asymptotic expansions~\cite{Kurz:2014wya, Kurz:2016bau}.
Furthermore, the $\tau$-lepton contributions to the eighth-order
coefficients have been checked independently in
Refs.~\cite{Aoyama:2012wk,Kurz:2013exa,Kataev:2012kn}. Numerical
values for the eighth-order coefficients $A_{2\ell}^{(8)}$ and
$A_{3\ell}^{(8)}$ are listed in rows~7--9 in
Table~\ref{tab:A23mass5loop}, with the quoted errors arising from the
numerical integration.

The relevant mass-dependent contributions arising at $10^{\rm th}$
order (five loops) have been computed
numerically~\cite{Aoyama:2012wj,Kinoshita:2005sm}, with asymptotic
expansions of some diagrams having been worked out by
Laporta~\cite{Laporta:1994md}. Further cross-checks of sub-classes
have been performed (see~\cite{Kataev:1992dg, Kataev:1994rw,
  Kataev:2006yh, Karshenboim:1993rt}). The values for the
coefficients $A_{2\ell}^{(10)}$ and $A_{3\ell}^{(10)}$ listed in
rows~10--12 in Table~\ref{tab:A23mass5loop} are taken
from~\cite{Aoyama:2019ryr,Aoyama:2020ynm}.

At twelfth order (six loops), the number of vertex diagrams increases
by another order of magnitude to about 120\,000. No actual
calculations have been performed yet. For the muon, one expects the
dominant contribution at $12^{\rm th}$ order to come from
light-by-light scattering diagrams, specifically from the
light-by-light scattering contribution to
$A_{2\mu}^{(12)}(m_\mu/m_e)$. Its size can be estimated by inserting
three vacuum polarisation loops into the $6^{\rm th}$-order
light-by-light scattering diagram, which yields
\be
  A_{2\mu}^{(12)}(m_\mu/m_e)\approx A_2^{(6)}(m_\mu/m_e) \left\{\frac{2}{3}\ln\frac{m_\mu}{m_e}-\frac{5}{9}\right\}^3 \approx 5400\,,
\ee
from which one infers that
$A_{2\mu}^{(12)}(m_\mu/m_e)(\alpha/\pi)^6\approx0.8\cdot10^{-12}$.
This is smaller than the experimental precision reached by the E989
experiment at Fermilab~\cite{Muong-2:2021ojo,Muong-2:2023cdq}, in
spite of the large estimated value for $A_{2\mu}^{(12)}$.

In Table~\ref{tab:QED5loop} we show the summed coefficients
\be\label{eq:CLell}
  C_{L\ell}\equiv A_1^{(2L)}+A_{2\ell}^{(2L)}+A_{3\ell}^{(2L)}\,,\quad L=1,\ldots,5\,,\quad\ell=e,\,\mu\,,
\ee
as well as the resulting QED contribution
\be
  a_\ell^{(2L)}=C_{L\ell}\,\left(\frac{\alpha}{\pi}\right)^L
\ee
at each loop order~$L$. The following observations can be made: For
the electron, the coefficients $C_{Le}$ receive their dominant
contributions from the universal part $A_1^{(2L)}$, as can be seen
from Tables~\ref{tab:A1univ5loop} and~\ref{tab:A23mass5loop}. As a
consequence, the coefficients $C_{Le}$ remain of order one and even
alternate between subsequent loop orders, signalling good convergence
of the perturbative series. By contrast, the coefficients $C_{L\mu}$
are dominated by the strongly enhanced mass-dependent contributions
$A_{2\mu}^{(2L)}(m_\mu/m_e)$ which grow rapidly for increasing
$L$. However, thanks to the smallness of $(\alpha/\pi)$, the
convergence of the expansion is not compromised.

For evaluating the QED corrections at each loop order (see
Table~\ref{tab:QED5loop}) one must pick a value for the fine-structure
constant $\alpha$ which enters as an input variable. Here we have
chosen the latest determination of $\alpha$ from atom interferometry
measurements on Rubidium (Rb) atoms, which
gives~\cite{Morel:2020dww,Mohr:2024kco}
\be
  \alpha^{-1}({\rm Rb})=137.035\,999\,2052(97)\,.
\ee
A similar measurement on Cesium (Cs) atoms
yields~\cite{Parker:2018sci, Mohr:2024kco}
\be
  \alpha^{-1}({\rm Cs})=137.035\,999\,045(27)\,,
\ee
which implies that there is a tension of 5.5 standard deviations
between the two measurements, i.e. $\alpha^{-1}({\rm
  Cs})-\alpha^{-1}({\rm Rb})=-0.160(29)\cdot10^{-6}$. A third method
to determine $\alpha$ is based on measuring $a_e$ and -- after
subtraction of the tiny electroweak and strong contributions --
equating the resulting experimental QED contribution to $a_e$ with the
five-loop perturbative expansion. The resulting value,
$\alpha^{-1}(a_e)=137.035\,999\,163(15)$, lies between the values
determined from Rb and Cs atom interferometry, i.e. $\alpha^{-1}({\rm
  Cs})-\alpha^{-1}(a_e)=-0.118(31)\cdot10^{-6}$ and $\alpha^{-1}({\rm
  Rb})-\alpha^{-1}(a_e)=+0.042(18)\cdot10^{-6}$. At the time of
writing (2025) this discrepancy has not been resolved. Discrepant
determinations of $\alpha$ translate into different estimates for the
QED contributions to $a_e$ and $a_\mu$, respectively. For instance,
for the muon one finds $a_\mu^{\rm QED}({\rm Cs})-a_\mu^{\rm QED}({\rm
  Rb})=0.137(36)\cdot10^{-11}$, which corresponds to a $3.8\sigma$
tension. However, given that the current experimental sensitivity for
$a_\mu$ is at the level of
$15\cdot10^{-11}$~\cite{Muong-2:2021ojo,Muong-2:2023cdq,
  Muong-2:2025xyk}, discrepant values of the fine-structure constant
are of no consequence. In the 2025 edition of the White Paper on the
muon $g-2$~\cite{Aliberti:2025beg}, the QED estimate is quoted as
\be
  a_\mu^{\rm QED} = (116\,584\,718.8\pm0.2)\cdot10^{-11}\,,
\ee
where the error accounts for the different determinations of $\alpha$
and includes uncertainties from the neglected $12^{\rm th}$-order
term.

While $a_e$ and $a_\mu$ have been measured with impressive precision,
the experimental knowledge of the anomalous magnetic moment of the
$\tau$ is rather poor. This is due to the short lifetime of the
$\tau$, which presents a considerable challenge to experimentalists to
come up with a measurement that is precise enough to allow for a
stringent test of the SM. Therefore, there has been relatively little
interest to push the calculation of $a_\tau^{\rm QED}$ beyond three
loops (sixth order). The sixth-order result for the $\tau$ anomalous
magnetic moment has been evaluated in 2007 by Eidelman and Passera
\cite{Eidelman:2007sb}. The mass-dependent coefficients are as
follows:
\begin{align}
    A_{2\tau}^{(4)}(m_\tau,m_e) &= \phantom{4}2.024\,284(55)\,,\quad
    A_{2\tau}^{(4)}(m_\tau,m_\mu)= 0.361\,652(38) \nonumber \\
    A_{2\tau}^{(6)}(m_\tau,m_e) &= 46.3921(15)\,,\qquad\,
    A_{2\tau}^{(6)}(m_\tau,m_\mu)=  7.010\,21(76)\,,\quad
    A_{3\tau}^{(6)}(m_\tau,m_e,m_\tau,m_\mu) =3.347\,97(41)\,.
    \label{eq:QEDtau_mass}
\end{align}
Summing up the universal and mass-dependent coefficients according to
\eq{eq:CLell} yields
\begin{equation}\label{eq:Ctau}
    C_{2\tau}=2.057\,457(93)\,,\quad C_{3\tau}=57.9315(27)\,.
\end{equation}
A comparison with the corresponding entries in
Tables\,\ref{tab:A23mass5loop} and~\ref{tab:QED5loop} shows the strong
growth of mass-dependent contributions to $a_\ell^{\rm QED}$ for
increasing lepton mass. The resulting estimate for the QED
contribution to $a_\tau$ quoted in \cite{Eidelman:2007sb} is
\begin{equation}\label{eq:atauQED}
    a_\tau^{\rm QED}=(117\,324\pm2)\cdot10^{-8}\,.
\end{equation}
As a word of caution we emphasise that the numerical values (and
errors) listed in eqs.\,(\ref{eq:QEDtau_mass})--(\ref{eq:atauQED}) are
based on the values of the lepton mass ratios and fine-structure
constant available when Ref.\,\cite{Eidelman:2007sb} was published
(2007) and have not been updated since. However, given that these
input parameters have not changed outside the errors quoted at the
time, one should not expect significant changes.

What would be required to extend the calculation of $a_\tau^{\rm QED}$
beyond sixth order?\footnote{I am grateful to Makiko Nio for
discussions on this point.}  For the muon, the mass-dependent
four-loop coefficients $A_{2\mu}^{(8)}$ were computed in
\cite{Kurz:2014wya, Kurz:2016bau} as an expansion in the mass ratio
$x=m_e/m_\mu$. The corresponding mass-dependent contributions
$A_{2\tau}(8)(m_\tau/m_e)$ and $A_{2\tau}(8)(m_\tau/m_\mu)$ can be
obtained simply by changing the numerical value of $x$ to the
appropriate mass ratios for the $\tau$. However, the contribution from
diagrams containing both electron and muon loops,
$A_{3\tau}(8)(m_\tau/m_e, m_\tau/m_\mu)$ cannot be treated in this
fashion and will require a dedicated calculation. While the expansion
technique has not been performed at the five-loop level, the necessary
diagrams can be computed numerically for the modified mass ratios in a
straightforward fashion. This will, however, require significant
numerical resources, and given the poor precision in the experimental
estimate for $a_\tau$, there is currently little incentive to make
this investment.

%% file: HVP_DD.tex
\section{Hadronic contributions}

The contributions from the strong interaction, although much smaller in
size than that from QED, have been the main focus of recent attention
in the case of the muon anomalous magnetic moment. At order $\alpha^2$
one encounters the leading-order hadronic vacuum polarisation (HVP)
contribution, which is represented by diagram~(d) in
Fig.~\ref{fig:diagramsLO}. The hadronic light-by-light scattering
(HLbL) contribution (diagram~(e) in Fig.~\ref{fig:diagramsLO} ) arises
at order $\alpha^3$. Together they contribute only about 60~ppm to the
SM prediction of $a_\mu$, which is much smaller than the two-loop QED
contribution that arises at order $\alpha^2$ (see
Table~\ref{tab:QED5loop}). However, HVP and HLbL contributions are
almost exclusively responsible for the total uncertainty assigned to
the SM prediction. By contrast, hadronic contributions play only a
minor role in the case of the electron anomalous magnetic moment,
$a_e$, as they account for only 1.5~ppb of the SM
prediction~\cite{Aliberti:2025beg}. The error assigned to the theory
estimate for $a_e$ is dominated by the uncertainty in the
determination of the fine-structure constant $\alpha$ which enters as
the key input quantity in the perturbative evaluation. For the $\tau$,
on the other hand, hadronic contributions are very prominent, with
recent evaluations \cite{Eidelman:2007sb,Giusti:2019hkz} suggesting
that the HVP contribution to $a_\tau$ is two orders of magnitude
larger than in the case of the muon.

Given the smallness of hadronic contributions in $a_e$ and the lack of
precision in experimental measurements of $a_\tau$, we will focus
primarily on $a_\mu$ in this section. By far the dominant contribution
from the strong interaction comes from the leading-order HVP
diagram~(d) in Fig.~\ref{fig:diagramsLO}. The grey ``blobs" indicate
the corrections due to quark loops in which quarks and antiquarks can
exchange any number of soft (i.e. low-energy) gluons which, in turn,
can also dissociate into virtual quark-antiquark pairs. Higher-order
HVP corrections, arising at order~$\alpha^3$, are shown in
Fig.~\ref{fig:HOHVP}. Although they are similar in magnitude compared
to HLbL, they have a much smaller impact on the value and overall
precision of the SM estimate.

Strong interaction effects are introduced via the hadronic
electromagnetic current:
\be\label{eq:JemMink}
   J^\mu(x)=\textstyle \frac{2}{3}(\bar{u}\gamma^\mu{u})(x)
                      -\frac{1}{3}(\bar{d}\gamma^\mu{d}))x)
                      -\frac{1}{3}(\bar{s}\gamma^\mu{s})(x)
                      +\frac{2}{3}(\bar{c}\gamma^\mu{c})(x)
   +\ldots\,,
\ee
i.e.\ the vector current in units of the proton charge summed over all
quark flavours and multiplied by their respective charge factors. The
contribution to the photon self-energy, which is the key part of
diagram~(d) in Fig.~\ref{fig:diagramsLO}, is encoded in the vacuum
polarisation amplitude $\Pi(q^2)$ defined as the Fourier-transformed
time-ordered product of two of these currents, i.e.
\be\label{eq:PmunuMink}
  (q^\mu q^\nu-q^2 g^{\mu\nu})\Pi(q^2) = ie^2\int d^4x\,
  {\rm e}^{iq{\cdot}x}\,\left\langle
  0\left|T\,J^\mu(x)J^\nu(0)\right|0\right\rangle\,.
\ee
The interactions between quarks and gluons are described by Quantum
Chromodynamics (QCD), a non-Abelian gauge theory based on the gauge
group SU(3)-colour. It is given in terms of the Lagrangian
\be\label{eq:QCDLagrangian}
  {\cal{L}}_{\rm QCD}=-\frac{1}{4}F_{\mu\nu}^a(x)F^{a\,\mu\nu}(x)+
  \sum_{f=u,d,s,\ldots} \bar{\psi}_f(x)\left(
   i\gamma^\mu D_\mu-m_f \right)\psi_f(x)\,,
\ee
where the sum runs over all quark flavours $f=u,d,s,\ldots$.
The covariant derivative $D_\mu$ is defined by
\be
  D_\mu:=\partial_\mu\mathbb{1}-ig\big({\textstyle\frac{1}{2}}\lambda^a\big)A_\mu^a(x)\,,
\ee
where $g$ is the gauge coupling, and the $3\times3$ matrices
$\lambda^a,\,a=1,\ldots,8$ denote the Gell-Mann matrices. They are the
generators of the gauge group SU(3) and satisfy the commutation
relations
\be
   \left[{\textstyle\frac{1}{2}}\lambda^a,\,{\textstyle\frac{1}{2}}\lambda^b \right]
   =if^{abc}{\textstyle\frac{1}{2}}\lambda^c\,,
\ee
where $f^{abc}$ denote the totally antisymmetric structure constants.
Thus, the covariant derivative generates the interaction between the
quark fields $\psi_f,\,\bar\psi_f$ and the gluon field $A_\mu^a(x)$.
The field strength tensor $F_{\mu\nu}^a$ is defined by
\be
  F_{\mu\nu}^a = \partial_\mu A_\nu^a-\partial_\nu A_\mu^a + if^{abc}A_\mu^b A_\nu^c\,.
\ee
The term proportional to $f^{abc}$ gives rise to gluon
self-interaction mediated by three-gluon and four-gluon vertices which
are both absent in the Abelian case. The combination $\alpha_s\equiv
g^2/4\pi$ is referred to as the strong coupling and represents the QCD
analogue of the fine-structure constant $\alpha=e^2/4\pi$.

As in any Quantum Field Theory, the parameters of QCD ``run" with the
energy scale, and the current reference value of $\alpha_s$ quoted in
the $\rm\overline{MS}$-scheme of dimensional regularisation is
\cite{ParticleDataGroup:2024cfk}
\be
   \alpha_s^{(5)}(M_Z^2)=0.1180(9)\,.
\ee
The strong coupling decreases logarithmically as the energy scale is
increased, which is the property of asymptotic freedom. By the same
token, $\alpha_s$ grows towards the low-energy regime such that
$\alpha_s\sim{\rm O}(1)$ at typical hadronic scales. This explains why
the contributions from the strong interaction to lepton anomalous
magnetic moments are not amenable to a perturbative
treatment. Specifically, the calculation of the hadronic contribution
to the photon self-energy $\Pi(q^2)$ in \eq{eq:PmunuMink} cannot be
performed in QCD perturbation theory. Nowadays one resorts to two
complementary approaches that allow for the precise determination of
both the HVP and the HLbL contributions with quantifiable
uncertainties: The traditional method -- the phenomenological
data-driven dispersive approach -- makes use of experimentally
measured hadronic cross section data for evaluating a dispersion
integral. The second method is based on numerical calculations
employing a discretised version of the QCD action (``Lattice QCD"). In
the remainder of this section we will discuss the general formalism
for both methodologies, followed by an overview of recent evaluations.

%\comment{Address the following topics: Dispersion relation, imaginary part, optical theorem, spectral function, $e^+e^-$ vs. $\tau$}

\subsection{Hadronic vacuum polarisation: The data-driven dispersive method}

The vacuum polarisation function $\Pi(q^2)$ defined in
\eq{eq:PmunuMink} contains the hadronic contribution to the photon
self-energy.
%\comment{$\Pi^{\mu\nu}$ has transversity built in -- needed?}
We now consider the timelike region of momentum transfers,
i.e. $s\equiv q^2>0$. 

\begin{wrapfigure}{r}{6cm}
  \centering
  \includegraphics[width=0.35\textwidth]{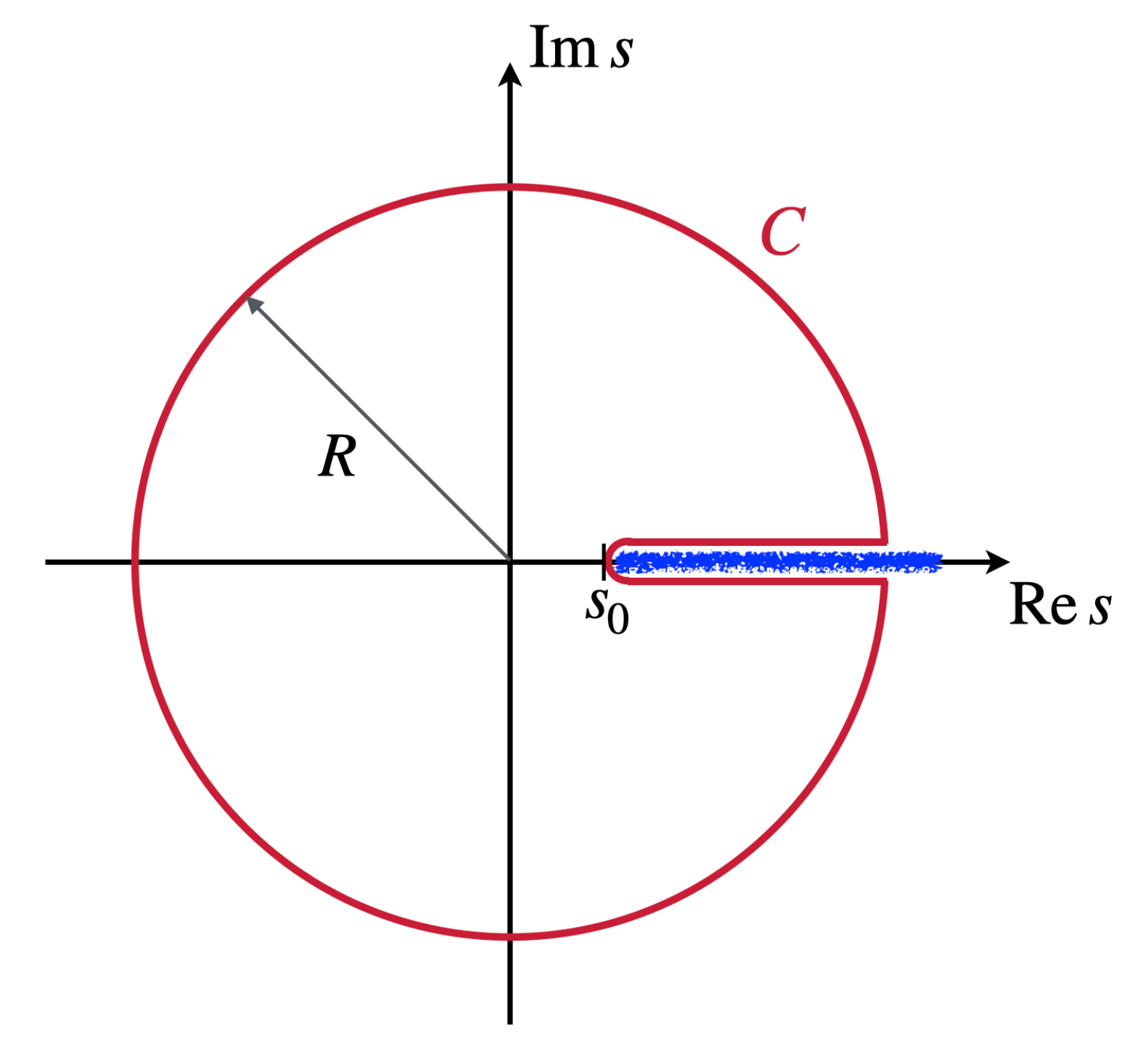}
  \caption{The contour $C$ of the integration path in \eq{eq:Cauchy}.
    The vacuum polarisation function is analytic except for a cut
    along the real axis represented by the thick blue line. The
    threshold value $s_0$ is identified with the squared pion mass.}
  \label{fig:Cauchy}
\end{wrapfigure}
Causality and unitarity imply that $\Pi(s)$ is analytic in the complex
$s$-plane, except for a cut along the positive real axis, starting at
the pion threshold, as shown in Fig.\,\ref{fig:Cauchy}. By means of
Cauchy's integral theorem, one can express the analytic vacuum
polarisation function in terms of a contour integral
\be\label{eq:Cauchy}
  \Pi(s)=\frac{1}{2{\pi}i}\oint_C ds'\,\frac{\Pi(s')}{s'-s}\,,
\ee
where $s'$ takes values along the curve~$C$. To convert the contour
integral into an integral along the cut, we introduce the
$i\epsilon$-prescription in the denominator, which ensures that the
imaginary part of $\Pi(s)$ changes sign when $s\to
s^\ast$.\footnote{This property is called the Schwarz reflection
principle.}  This implies
\be\label{eq:ImagPi}
  \lim_{\epsilon\to0}\left[ \Pi(s'+i\epsilon)-\Pi(s'-i\epsilon)\right]=2i\,{\rm Im}\,\Pi(s')\,.
\ee
Following the integration path along $C$ anticlockwise, noting that
the horizontal paths running parallel to the real axis give rise to an
integral over the left-hand side in \eq{eq:ImagPi}, and taking the
radius $R\to\infty$, we obtain
\be
  \Pi(s)=\frac{1}{\pi}\int_{s_0}^{\infty} ds'\,\frac{{\rm Im}\,\Pi(s')}{s'-s-i\epsilon} +I_C^{\infty}\,,
\ee
where $I_C^{\infty}$ denotes the contribution along the semi-circles
in the upper and lower planes.

To avoid any non-vanishing contributions from $I_C^{\infty}$ when the
vacuum polarisation function does not drop off sufficiently fast,
which will be the case when $\Pi(s)$ develops UV divergences, one must
perform a subtraction. This is accomplished by considering
$\Pi(s)-\Pi(0)$. The subtracted dispersion relation for the vacuum
polarisation function then reads
\be\label{eq:DRsub}
  \Pi(s)-\Pi(0)\equiv\Pi_{\rm ren}(s)=\frac{s}{\pi}\int_{s_0}^{\infty}ds'\,
  \frac{{\rm Im}\,\Pi(s')}{s'(s'-s-i\epsilon)}\,.
\ee
The next step involves expressing the imaginary part of the vacuum
polarisation function in terms of the spectral function which is given
by the total hadronic cross section $\sigma_{\rm
  tot}(e^+e^-\to\gamma^\ast\to\hbox{hadrons})$. The unitarity of the
$S$-matrix implies the optical theorem for the non-trivial part $T$ of
the $S$-matrix, i.e.
\be
  (T_{fi}-T_{if}^\ast)=i(2\pi)^4\sum_n\,\delta^{(4)}(P_f-P_n)\,T_{nf}^\ast T_{ni}\,,
\ee
where $T_{fi}$ is defined as the matrix element for the transition
$|i\,\rangle\to|f\,\rangle$, such that
$\langle{f}\,|T|{i}\,\rangle=(2\pi)^4\delta^{(4)}(P_f-P_i)\,T_{fi}$,
and the $\delta$-function ensures momentum conservation. Specifically,
in the forward scattering case, $|i\,\rangle\to|i\,\rangle$, the
relation becomes
\be
  {\rm Im}\,T_{ii}={\textstyle\frac{1}{2}}(2\pi)^4\sum_n\,\delta^{(4)}(P_i-P_n)\,\left|T_{ni}\right|^2\,,
\ee
where the sum on the right-hand side runs over all possible
intermediate states. When applied in the context of hadronic
contributions to the photon self-energy, the optical theorem implies
that the imaginary part of the vacuum polarisation function is
proportional to the total hadronic cross section:
\be\label{eq:Rratiodef}
  {\rm Im}\,\Pi(s)=\frac{e^2}{12\pi}\,R_{\rm had}(s),\quad 
  R_{\rm had}(s) = \sigma_{\rm tot}(e^+e^-\to\gamma^\ast\to\hbox{hadrons})\left/
  \,\left(\frac{4\pi\,\alpha(s)^2}{3s}\right)\right.\,.
\ee
The quantity $R_{\rm had}(s)$ is called the hadronic $R$-ratio and is
defined as the total hadronic cross section divided by the tree-level
cross section for $e^+e^-\to\gamma^\ast\to\mu^+\mu^-$ in the limit
$s\gg4m_\mu^2$. As a side remark, we note that the $R$-ratio is
identified with the spectral function that appears in the
K\"all\'en-Lehmann representation of the time-ordered product of two
electromagnetic currents \cite{Peskin:1995ev,Weinberg:1995mt}.

With these relations, the dispersion relation for the renormalised
vacuum polarisation function reads
\be
  \Pi_{\rm ren}(s)=\frac{\alpha\,s}{3\pi}\int_{s_0}^{\infty} ds'\,
  \frac{R_{\rm had}(s')}{s'(s'-s-i\epsilon)}\,.
\ee
We still have to clarify the value of the threshold $s_0$. The cut
along the real axis starts at the point where the centre-of-mass
energy is sufficiently large to sustain the production of a pion pair,
which implies $s_0=4m_{\pi^\pm}^2$. However, as we will see below,
when evaluating the dispersion integral using experimental data for
the cross section, the hadronic state may contain one or more photons,
which implies that hadron production starts at $\pi^0\gamma$, thereby
lowering the threshold to $s_0=m_{\pi^0}^2$.

In the low-energy regime, the $R$-ratio is dominated by the two-pion
channel, i.e.\ the contribution from
$e^+e^-\to\gamma^\ast\to\pi^+\pi^-$. In this situation, and below the
three-pion threshold, the $R$-ratio is given by the square of the
electromagnetic form factor of the pion:
\be\label{eq:pionFF}
  R_{\rm had}(s)=\frac{1}{4}\left(1-\frac{4m_\pi^2}{s}\right)^{3/2}\,
  \left|F_{\pi\pi}(s)\right|^2\,.
\ee
We will now proceed to sketch the derivation of the expression for the
dispersive evaluation of the leading-order HVP contribution $\ahvp$.
An in-depth discussion can be found in chapter~3.8
of~\cite{Jegerlehner:2017gek}.

\begin{figure}[tp]
\centering
\includegraphics[width=0.99\textwidth]{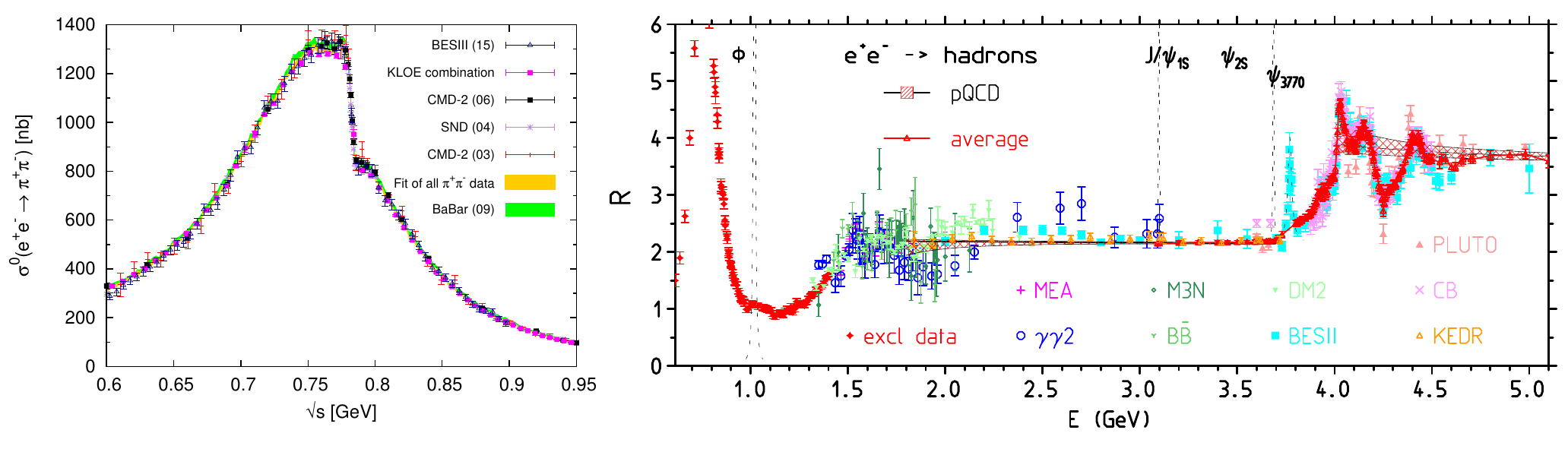}
\vspace{-0.3cm}
\caption{Left: Experimental data for the cross section
  $\sigma(e^+e^-\to\pi^+\pi^-)$ which dominates the dispersion
  integral (figure from \cite{Keshavarzi:2018mgv}). Right: the
  hadronic $R$-ratio for energies up to 5\,\GeV (figure from
  \cite{Jegerlehner:2017lbd}). \label{fig:Rratio}}
\end{figure}

We start by finding an expression for the internal photon line
including the hadronic ``blob'' in diagram~(d) of
Fig.\,\ref{fig:diagramsLO}. The corresponding photon propagator is
given by
\begin{align}
  \frac{-ig^{\mu\nu}}{q^2\left(1+\Pi_{\rm ren}(q^2)\right)}
  &=\frac{-ig^{\mu\nu}}{q^2}\,\left( 1-\Pi_{\rm ren}(q^2)
  +\left(\Pi_{\rm ren}(q^2)\right)^2 
  -\ldots\right) \nonumber\\
  &=\frac{-ig^{\mu\nu}}{q^2}+\frac{1}{\pi}\int_{s_0}^{\infty}
  \frac{ds}{s}\, 
  {\rm Im}\,\Pi(s)\,\frac{-ig^{\mu\nu}}{q^2-s}+\ldots
  \label{eq:DRphoton} 
\end{align}
where $q$ denotes the photon's momentum, and $\Pi_{\rm ren}(q^2)$ is
the hadronic contribution to the photon self-energy. In the second
line we have replaced $-\Pi_{\rm ren}(q^2)/q^2$ by the dispersion
relation in \eq{eq:DRsub}. The first term in the second line of
\eq{eq:DRphoton} is the free photon propagator which gives rise to the
one-loop QED correction computed by Schwinger. The second term encodes
the higher-order contribution arising from the strong interaction: It
consists of the propagator of a photon with mass~$m_\gamma=\sqrt{s}$,
i.e. $-ig^{\mu\nu}/(q^2-s)$, convoluted with the imaginary part of the
hadronic vacuum polarisation function. Computing the contribution
from the massive photon propagator alone proceeds by replacing
$-ig_{\nu\lambda}/((k-p)^2+i\epsilon)$ in \eq{eq:vertexcorr} by
$-ig_{\nu\lambda}/((k-p)^2-s+i\epsilon)$. The result, i.e. the $2^{\rm
  nd}$ order contribution for a massive photon with $m_\gamma^2=s$ is
given by \cite{Brodsky:1967sr}\footnote{Note that, by setting $s=0$
in~\eq{eq:amu-massive-photon}, one recovers Schwinger's result
$a_\mu^{(2)}(s=0)=\alpha/2\pi$.}
\be\label{eq:amu-massive-photon}
  a_\mu^{(2)}(s)=\frac{\alpha}{\pi}\,K_\mu^{(2)}(s)=
  \frac{\alpha}{\pi}\int_0^1 dz\,\frac{z^2(1-z)}{z^2+(s/m_\mu^2)(1-z)}\,.
\ee
To derive the dispersive expression for the leading-order HVP
contribution one must, in a second step, convolute $a_\mu^{(2)}(s)$
with the imaginary part of the vacuum polarisation function (see
\eq{eq:DRphoton}), which yields
\be\label{eq:LO-HVP-DR}
  \alohvp=\frac{\alpha}{\pi^2}\int_{s_0}^{\infty}\frac{ds}{s}\,
  {\rm Im}\,\Pi(s)\,K_\mu^{(2)}(s)=
  \frac{\alpha^2}{3\pi^2}\int_{s_0}^{\infty}\frac{ds}{s}\,
  R_{\rm had}(s)\,K_\mu^{(2)}(s)\,,
\ee
where we have substituted the $R$-ratio for ${\rm Im}\,\Pi(s)$ using
\eq{eq:Rratiodef}. For $s>4m_\mu^2$, the kernel function
$K_\mu^{(2)}(s)$ represented by the integral in
\eq{eq:amu-massive-photon} is conveniently expressed as
\be\label{eq:kernelfunc}
  K_\mu^{(2)}(s)=\frac{x^2}{2}(2-x^2) +\frac{(1+x^2)(1+x)^2}{x^2}
  \left(\ln(1+x)-x+\frac{x^2}{2} \right) +\frac{(1+x)}{(1-x)}x^2\ln{x},
  \qquad
  x=\frac{1-(1-4m_\mu^2/s)^{1/2}}{1+(1-4m_\mu^2/s)^{1/2}}\,.
%  x=\frac{1-\sqrt{1-4m_\mu^2/s}}{1+\sqrt{1-4m_\mu^2/s}}\,.
\ee
To discuss the properties of the convolution integral in
\eq{eq:LO-HVP-DR}, it is helpful to introduce a rescaled kernel
function $\hat{K}(s)$ defined by
\be
  \hat{K}(s)=\frac{3s}{m_\mu^2}\,K_\mu^{(2)}(s)\,
\ee
which has the value $\hat{K}(4m_\pi^2)=0.542\ldots$ at the pion
threshold and increases monotonically with $s$. The resulting integral
representation of the leading-order HVP contribution then reads
\be\label{eq:DDD}
  \alohvp=\left(\frac{\alpha m_\mu}{3\pi}\right)^2 \int_{s_0}^{\infty}
  ds\,\frac{R_{\rm had}(s)\,\hat{K}(s)}{s^2}\,.
\ee
The $R$-ratio in the integrand of \eq{eq:DDD} is convoluted with
$\hat{K}(s)/s^2$ where $\hat{K}(s)\sim1$ is a slowly varying function.
Therefore, the factor of $s^2$ in the denominator leads to a strong
enhancement of the $R$-ratio in the low-energy regime down to the pion
threshold. This precludes any attempt to evaluate the dispersion
integral using QCD perturbation theory. Instead, one supplies
experimental data for the total hadronic cross section
$e^+e^-\to\hbox{hadrons}$ and performs a numerical integration.
Experimental cross section data for the process $e^+e^-\to\pi^+\pi^-$
taken prior to the result from CMD-3 are shown in
Fig.\,\ref{fig:Rratio}. Indeed, one finds that the interval of
$\sqrt{s}=600-900$\,MeV contributes about 70\% to the dispersion
integral. This region is dominated by the two-pion channel and shows a
prominent peak due to the $\rho$-meson and a steep shoulder because of
$\rho-\omega$ mixing.
%This is also the region where the $R$-ratio is dominated by the pion
%form factor (see \eq{eq:pionFF}).
Data for the $R$-ratio in a wider interval are shown for illustration
in the right panel of Fig.\,\ref{fig:Rratio}.

Below $\sqrt{s}\simeq2$\,GeV, the $R$-ratio is determined by summing
the contributions from exclusive hadronic channels, i.e. $e^+e^-\to
\pi^+\pi^-$, $\pi^+\pi^-\pi^0$, $\pi^+\pi^-\pi^+\pi^-,\,\ldots$,
$K^+K^-, K_LK_S,\ldots$. Above $\sqrt{s}\simeq2$\,GeV where the
identification of exclusive channels becomes more difficult one
resorts to inclusive cross section data, and eventually, that is at
sufficiently high energy and away from flavour thresholds,
perturbative QCD can be applied. Narrow resonances in the charm and
bottom sector are included via suitable parameterisations of
experimental data.

While the evaluation of the dispersion integral is, in principle,
straightforward, there are several technical issues and subtleties
that we briefly address in the following paragraphs. A much more
in-depth discussion is presented in the first White Paper of the Muon
$g-2$ Theory Initiative \cite{Aoyama:2020ynm} and its successor
\cite{Aliberti:2025beg}.

By convention, the $R$-ratio is determined from the bare hadronic
cross section $\sigma_{\rm tot}^{(0)}(e^+e^-\to\hbox{hadrons})$,
meaning that all vacuum polarisation effects that enter the running of
the electromagnetic coupling, $\Delta\alpha$, must be subtracted in
order to avoid double counting (cf. \eq{eq:Rratiodef}).
%
%\be
%  R_{\rm had}(s) = \sigma_{\rm tot}^{(0)}(e^+e^-\to\gamma^\ast\to\hbox{hadrons})\left/
%  \,\left(\frac{4\pi\,\alpha^2}{3s}\right)\right.\,.
%\ee
%
While corrections due to QED vacuum polarisation can be computed in
perturbation theory, the corresponding hadronic contribution,
%$\Delta\alpha_{\rm had}(s)\equiv {\rm Re}\,\Pi_{\rm ren}(s)$, is also
%\comment{check conventions and exact relation to $\Pi_{\rm ren}(s)$}
$\Delta\alpha_{\rm had}(s)$, is also obtained from a dispersion
integral, i.e.
\be\label{eq:emrunning}
   \Delta\alpha_{\rm had}(s)=-\frac{\alpha\,s}{3\pi}\,{\rm P}\hspace{-3pt}\int_{s_0}^{\infty} ds'\,
   \frac{R_{\rm had}(s')}{s'(s'-s)}\,,
\ee
where ``P" denotes the principal value. In evaluating the subtraction
in this way, one relies on the same input data as in the determination
of $\alohvp$ itself. This is not a problem since a consistent
evaluation of both $\Delta\alpha_{\rm had}$ and $\alohvp$ can be
achieved via an iterative procedure.
Moreover, the hadronic cross section that enters the $R$-ratio is
considered inclusive of final-state radiation, implying that the final
state may contain one or more photons. This is the reason why the lower
bound on the dispersion integral is shifted from $4m_\pi^2$ to
$s_0=m_{\pi^0}^2$ since the threshold for hadron production sets in at
$\pi^0\gamma$.

There are two experimental approaches to measuring hadronic cross
sections up to a few GeV: The most obvious is to perform an energy
scan over the most relevant intervals in $\sqrt{s}$. The VEPP-2000
$e^+e^-$ machine \cite{Shatunov:2016VEPP,Shwartz:2016VEPP2} at the
Budker Institute in Novosibirsk has been constructed specifically for
this purpose, hosting the SND and CMD-2 / CMD-3
experiments~\cite{KHAZIN2008376}. Scans to measure hadronic cross
sections at energies $\sqrt{s}\gtrsim2$\,GeV have also been performed
by BESIII\,\cite{BESIII:2021wib} and at the KEDR detector.

At $e^+e^-$ colliders with a fixed centre-of-mass energy, one applies
the technique of Initial State Radiation (ISR), also dubbed
``Radiative Return". In the ISR approach, which has been applied at
BaBar, BESIII, CLEO and KLOE, the energy of the $e^+e^-$ collision is
reduced when a photon is radiated off the incoming electron or
positron. This necessitates the determination of the differential ISR
luminosity via
\be
  \frac{d\sigma_{\rm ISR}(\sqrt{s'})}{d\sqrt{s'}}=\frac{2\sqrt{s'}}{s}\,{\cal{W}}(s,E_\gamma,\theta_\gamma)\,\sigma(\sqrt{s'})\,,
\ee
where $\sqrt{s'}$ is the invariant mass of the virtual photon created
in the $e^+e^-$ collision, $\sqrt{s}$ is the fixed centre-of-mass
energy, and the quantity ${\cal{W}}(s,E_\gamma,\theta_\gamma)$ is the
so-called radiator function. It represents the probability for
radiating a photon with energy $E_\gamma$ and angle $\theta_\gamma$
and must be determined from a Monte Carlo event generator. One should
be aware that the use of Monte Carlo generators is not confined to
experiments employing the ISR technique, since soft initial and
final-state radiation occurs regardless of whether the collision
energy is tuned via the parameters of the accelerator or through
radiating a hard photon prior to the annihilation process. Moreover,
in order to meet the precision target, event generators must be able
to correctly describe processes of increasing complexity, including
hard final-state radiation which is relevant when measuring the
two-pion channel, or higher-order photon radiation emitted from
intermediate states.
Widely used event generators are {\sc Phokhara} \cite{Czyz:2004ua,
  Czyz:2004rj}, {\sc AfkQed}, {\sc Eva} \cite{Binner:1999bt,
  Czyz:2000wh, Campanario:2013uea} and {\sc BabaYaga}
\cite{CarloniCalame:2003yt, Balossini:2006wc, Budassi:2024whw}. The
``Radio Monte Carlow" effort has started a comprehensive review and
improvement of different Monte Carlo event generators, which is
published in Ref.\,\cite{Aliberti:2024fpq}.

An alternative to using $e^+e^-$ data in the evaluation of $\alohvp$
is provided by semi-leptonic $\tau$ decays, i.e. $\tau^-\to
H^-\nu_\tau$, where $H^-$ represents a charged hadronic state that is
related to the corresponding isovector final state~$H^0$ produced in
$e^+e^-\to H^0$ by an isospin rotation. This was first advocated in
\cite{Alemany:1997tn}, motivated by the fact that the $e^+e^-$ data
quality at the time was not sufficient (see also
\cite{Davier:2009ag,Jegerlehner:2011ti}).
%Originally, $\tau$ decays were suggested as a helpful alternative
%\cite{Alemany:1997tn} since the $e^+e^-$ data quality at the time was
%not sufficient.
More recently, $\tau$ decays have been reassessed following the
observation of significant tensions among different measurements of
the $e^+e^-\to\pi^+\pi^-$ cross section \cite{Achasov:2006vp,
  SND:2020nwa, CMD-2:2003gqi, Aulchenko:2006dxz, CMD-2:2006gxt,
  CMD-3:2023alj, CMD-3:2023rfe, KLOE:2008fmq, KLOE:2010qei,
  KLOE:2012anl, KLOE-2:2017fda, BaBar:2009wpw, BaBar:2012bdw,
  BESIII:2015equ}, which will be discussed in more detail below.

Data for the dominant decay mode $H^-=\pi^-\pi^0$ have been collected
at LEP \cite{OPAL:1998rrm,ALEPH:2005qgp,Davier:2013sfa}, CLEO
\cite{CLEO:1999dln} and Belle~\cite{Belle:2008xpe}. The main drawback
of the method is related to the fact that the isospin rotation has,
for now, not been determined with sufficient precision in a
model-independent way \cite{Aliberti:2025beg}.
To explain this in more detail, we discuss the ingredients in the
relevant dispersion integral based on the $\tau$ spectral function,
which takes the form
\be
  \left.\alohvp\right|_{\pi\pi}^{\tau} = \frac{\alpha^2}{12\pi^2}
  \int_{s_{\rm 0}}^{\infty}\frac{ds}{s}\,
  K_\mu^{(2)}(s)\left[\frac{1}{K_\Gamma(s)}\,
  \frac{d\Gamma_{\pi\pi(\gamma)}}{ds}\right]\,
  \frac{R_{\rm IB}(s)}{S_{\rm EW}^{\pi\pi}}\,,
\ee
where the kernel $K_\mu^{(2)}(s)$ is the same as in
\eq{eq:kernelfunc}, $s$ denotes the $\pi\pi$ invariant mass, and the
threshold is given by $s_{\rm 0}=4m_{\pi^\pm}^2$. The quantity in
square brackets consists of the photon-inclusive differential decay
rate as well as the kinematical factor $K_\Gamma(s)$ which is
proportional to the leptonic decay width $\Gamma_e(\tau\to
e\nu_\tau\bar\nu_e)$ and the CKM matrix element $|V_{ud}|$. The
isospin-breaking correction $R_{\rm IB}(s)$ is the crucial part in the
integrand. Without going into detail\footnote{A full account is given
in Section~2.3 of \cite{Aliberti:2025beg}.}, we note that $R_{\rm
  IB}(s)$ contains, among other terms, the long-distance QED
correction to $\tau^-\to\pi^-\pi^0\nu_\tau$ from both virtual and real
photon radiation \cite{Cirigliano:2001er, Cirigliano:2002pv,
  Flores-Baez:2006yiq, Miranda:2020wdg, Colangelo:2025iad,
  Colangelo:2025ivq}, final-state radiative corrections
\cite{Davier:2009ag}, as well as the ratio between electromagnetic
and weak pion form factors. It is the latter that produces the largest
uncertainty, as it is difficult to quantify in a model-independent
way.

\begin{wrapfigure}{r}{7cm}
  \centering
  \includegraphics[width=0.40\textwidth]{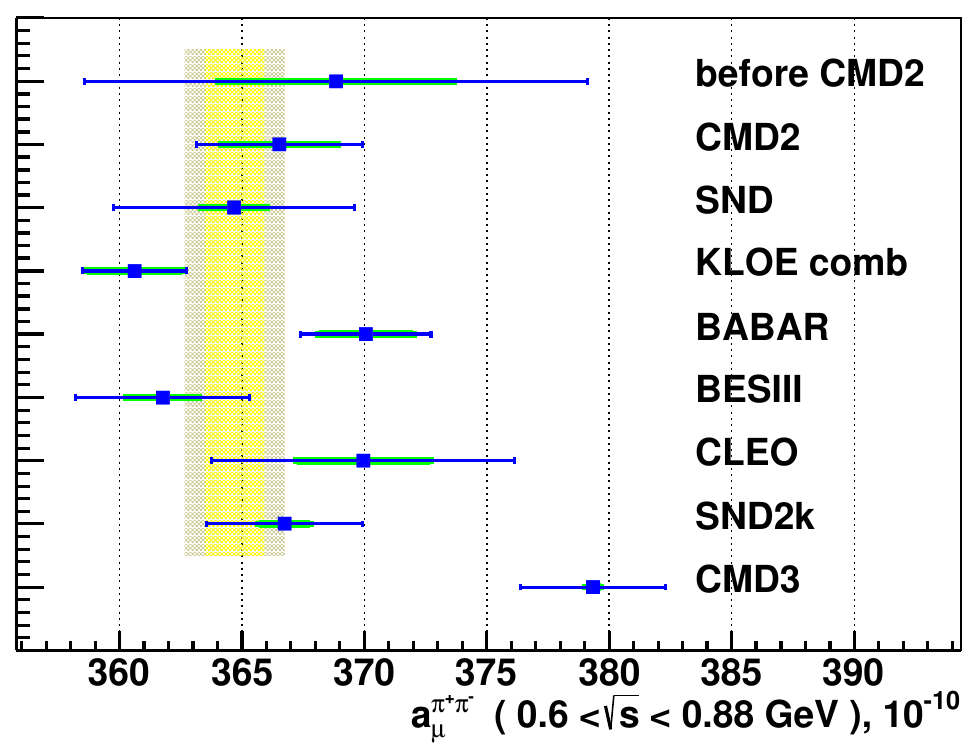}
  \caption{The contribution to $\alohvp$ extracted from the
    $e^+e^-\to\pi^+\pi^-(\gamma)$ cross section measured in the
    interval $\sqrt{s}=0.6-0.88$\,GeV (figure taken from
    \cite{CMD-3:2023alj}).}
  \label{fig:pipihvp}
\end{wrapfigure}
We now present the current status of determining the leading-order HVP
contribution via the dispersive approach. Evaluations of the
dispersion integral based on $e^+e^-\to\hbox{hadrons}$ have a long
history (see, e.g. \cite{Eidelman:1995ny, Davier:2010nc,
  Hagiwara:2011af, Davier:2017zfy, Jegerlehner:2017lbd,
  Jegerlehner:2018gjd, Keshavarzi:2018mgv, Colangelo:2018mtw,
  Ananthanarayan:2018nyx, Keshavarzi:2019abf, Davier:2019can,
  Hoferichter:2019mqg, Colangelo:2022prz, Davier:2023fpl,
  Stoffer:2023gba, Leplumey:2025kvv}). The approaches taken by different
groups vary in their choice of integration procedure, the region where perturbative QCD is applied, and the additional theoretical constraints
imposed.
At the release of the first White Paper in 2020 \cite{Aoyama:2020ynm},
tensions among different input data sets and different procedures to
determine $\alohvp$ could always be accommodated through a moderate
scaling of the error. The quoted result was based on merging the
results from Refs. \cite{Davier:2017zfy, Keshavarzi:2018mgv,
  Colangelo:2018mtw, Hoferichter:2019mqg, Davier:2019can,
  Keshavarzi:2019abf}, i.e.
\be\label{eq:LOHVP-DD}
   \left.\alohvp\right|_{\rm WP20} = 6931(40) \cdot 10^{-11}\quad [0.6\%]\,,
\ee
where the error accounts for experimental uncertainties, as well as
tensions in the input data for $e^+e^-$ hadronic cross sections and
among different analyses. The number in square brackets denotes the
relative precision, and the absolute uncertainty of
$\pm40\cdot10^{-11}$ dominated the error of the 2020 SM prediction,
$a_\mu^{\rm SM}$, which displayed a strong tension in excess of four
standard deviations with the direct measurement available in 2021 by
the Muon $g-2$ Collaboration~\cite{Muong-2:2021ojo}. However, with
the publication of the CMD-3 result \cite{CMD-3:2023alj,
  CMD-3:2023rfe} for the dominant $\pi^+\pi^-$ channel in 2023, the
tensions in the $e^+e^-$ hadronic cross section data have grown to a
level that signals a clear contradiction among the data. This is
illustrated in Fig.\,\ref{fig:pipihvp} where the two-pion contribution
to $\alohvp$ from the region $\sqrt{s}=600-880$\,MeV from different
experiments is shown: The previously existing tension of $2.8\sigma$
between KLOE \cite{KLOE:2008fmq, KLOE:2010qei, KLOE:2012anl,
  KLOE-2:2017fda} and BaBar \cite{BaBar:2009wpw, BaBar:2012bdw} is
small compared with the $5.1\sigma$ difference between KLOE and the
CMD-3 result (which also disagrees with its predecessor experiment,
CMD-2 \cite{Aulchenko:2006dxz, CMD-2:2006gxt}).

%%%Orig wrap
At face value, the CMD-3 data for the two-pion channel, when combined
with all other contributions, suggest that the SM prediction for
$a_\mu$ is, in fact, compatible with the direct measurement. As of now
no explanation to this puzzle of inconsistent measurements of the
$e^+e^-\to\pi^+\pi-$ cross section has been found. Some attempts at
clarifying the situation with the help of $\tau$ data have been made
\cite{Castro:2024prg,Davier:2025jiq}, but the fact that
isospin-breaking corrections relating the hadronic states in
$e^+e^-\to H^0$ and $\tau^-\to H^-\nu_\tau$ are currently not known in
a model-independent way requires further investigation.

The apparent contradictions in the $e^+e^-$ input data for the
dominant $\pi^+\pi^-$ channel were the main reason why the SM
prediction $a_\mu^{\rm SM}$ published in the 2025 update of the White
Paper \cite{Aliberti:2025beg} by the Muon $g-2$ Theory Initiative was
no longer based on the data-driven dispersive evaluation of $\alohvp$.
New and more precise experimental data combined with more refined
analyses are urgently required to resolve the situation. In addition,
the role of higher-order radiative corrections in Monte Carlo
generators that are necessary for the data-driven approach must be
better understood. We defer a more elaborate discussion of the
current situation to the end of section \ref{sec:LOHVPlat} where we
compare data-driven estimates with recent lattice QCD calculations.

At the end of this section, we briefly return to the relation between
$\alohvp$ and the hadronic contribution to the running of the
fine-structure constant, $\Delta\alpha_{\rm had}(q^2)$ which was
introduced in \eq{eq:emrunning}. This quantity appears in the relation
between the electromagnetic coupling $\alpha(q^2)$ at momentum
scale~$q^2$ and its value in the Thomson limit,
$\alpha\equiv\alpha(0)=1/137.035\,999\ldots$, i.e.
\be\label{eq:alpha_at_qsq}
   \alpha(q^2)=\frac{\alpha}{1-\Delta\alpha(q^2)}\,,\quad
   \Delta\alpha=\Delta\alpha_{\rm lep}+\Delta\alpha_{\rm had}
               +\Delta\alpha_{\rm top}\,,
\ee
where the subscripts ``lep'', ``had'' and ``top'' indicate the
contributions from leptons, hadrons and the top quark,
respectively. The value of $\alpha$ at the $Z$-pole, $\alpha(-M_Z^2)$,
is an important reference value. However, its determination from
electroweak data is not sufficiently precise to exploit the full
potential of future collider experiments. Likewise, efforts to
determine $\alpha(M_Z^2)$ from theory alone are limited by the same
type of hadronic contributions that affect lepton anomalous magnetic
moments. In our chosen convention, the hadronic running is given
directly in terms of the vacuum polarisation, $\Delta\alpha_{\rm
  had}(q^2)=-{\rm Re}\,\Pi_{\rm ren}(q^2)$ and, according to
\eq{eq:emrunning}, can be expressed in terms of a convolution integral
involving the $R$-ratio.
When space-like momenta are considered, $q^2=-Q^2$, the HVP
contribution $\alohvp$ can be evaluated from an alternative integral
representation involving $\Delta\alpha_{\rm had}$
\cite{Jegerlehner:2017gek}, i.e.
\be\label{eq:HVPalter}
   \alohvp=\frac{\alpha}{\pi}\int_0^1 dx\,(1-x)\Delta\alpha_{\rm had}(-Q^2)
          = \frac{\alpha^2}{6\pi^2}m_\mu^2 
            \int_0^1 dx\,x(2-x) \frac{D(Q^2)}{Q^2}\,,
   \quad Q^2\equiv Q^2(x)=\frac{x^2}{1-x}\,m_\mu^2\,,
\ee
where in the second equation we have introduced the Adler function
$D(Q^2)$, defined as the derivative of the hadronic shift in the
fine-structure constant \cite{Adler:1974gd, DeRujula:1976edq}
\be
   D(-s):=\frac{3\pi}{\alpha}\,s\,\frac{d}{ds}\Delta\alpha_{\rm had}(s)\,.
\ee
The Adler function can be represented in terms of a dispersion
integral
\be
   D(Q^2) = \frac{1}{Q^2}\int_{s_0}^{\infty}ds\,
            \frac{R_{\rm had}(s)}{(s+Q^2)^2}\,.
\ee
Moreover, $D(Q^2)$ can be computed in QCD perturbation theory
\cite{Eidelman:1998vc, Pich:2020gzz, Davier:2023hhn}, as well as in
lattice QCD \cite{Francis:2013fzp}.
Provided that $D(Q^2)$ is known reliably across the entire momentum
range, the Adler function approach allows one to express both
$\alohvp$ and $\Delta\alpha_{\rm had}$ in a space-like setting that
avoids integrating over resonances. In this way it is possible to
reduce the reliance on experimental data in the evaluation of
dispersion integrals.
The link between the $R$-ratio, $D(Q^2)$, the hadronic shift
$\Delta\alpha_{\rm had}$ and $\alohvp$ has been exploited not only to
determine $\Delta\alpha_{\rm had}(M_Z^2)$ (see, for instance,
\cite{Keshavarzi:2018mgv, Keshavarzi:2019abf, Davier:2019can,
  Jegerlehner:2019lxt, Ce:2022eix, Conigli:2025qvh}) but also to check
the consistency of the data-driven approach \cite{Davier:2023hhn}.
Interestingly, the integral representation in \eq{eq:HVPalter} forms
the basis for an experimental determination of $\alohvp$ planned by
the MUonE experiment, which is designed the measure the hadronic
running $\Delta\alpha_{\rm had}(-Q^2)$ in elastic muon-electron
scattering \cite{MUonE:2016hru}.

%% file: HVP_lat.tex
\subsection{Hadronic vacuum polarisation in lattice QCD} \label{sec:LOHVPlat}

The discussion at the end of the previous subsection shows that the
traditional data-driven dispersive formalism can run into problems
when the input data for hadronic cross sections show a significant
scatter that is not covered by the quoted uncertainties. This
highlights the need for an independent evaluation, based on an
alternative and systematically improvable theoretical approach which
is ultimately able to deliver the precision goal without relying
heavily on experimental input quantities and their intrinsic
uncertainties. Lattice QCD is a mature methodology that meets these
requirements. However, in order to have an impact on the debate
surrounding the muon $g-2$, lattice QCD must be able to determine
$\ahvp$ with an overall uncertainty well below one percent. As will
become apparent, this objective presents considerable but not
insurmountable challenges. Below we present a brief introduction to
the basic concepts of lattice QCD, leaving a more detailed discussion
in the context of the muon $g-2$ to Refs.~\cite{Meyer:2018til,
  Wittig:2020jtm}. For an in-depth introduction to lattice field
theory, one can consult several excellent textbooks on the
subject~\cite{Gattringer:2010zz, Rothe:2005nw, Smit:2002ug,
  Montvay:1994cy}.

Lattice QCD is a rigorous, non-perturbative treatment of the gauge
theory of quarks and gluons based on the regularised, Euclidean path
integral of QCD, which is defined as
\be\label{eq:QCDpath}
   Z=\int D[U]D[\bar\psi,\psi]\,
   e^{-S_{\rm G}[U]-S_{\rm F}[U,\bar\psi,\psi]}\,.
\ee
Here, $S_{\rm G}$ and $S_{\rm F}$ denote the Euclidean gluonic and
fermionic parts of the QCD action, and the integration is performed
over all gauge and fermionic fields. The above characterisation of
lattice QCD requires further explanation: ``Non-perturbative'' means
that approximations, such as expansions in powers of the coupling
constant, are absent in the determination of the
observable. ``Euclidean'' refers to the fact that lattice calculations
employ a Euclidean space-time metric such that spatial and temporal
components are treated on the same footing, implying that there is no
formal distinction between covariant and contravariant
four-vectors. Finally, the theory is regularised via the introduction
of a non-zero lattice spacing~$a$ between space-time points such that
$x_\mu=n_\mu a$. Considering, in addition, a finite space-time volume
of size $L^3\cdot T$, the QCD path integral $Z$ is mathematically well
defined and finite, for any suitable gauge-invariant discretisation of
the QCD action.

Gauge fields are represented in lattice QCD by elements of the gauge
group SU(3) which are called link variables, $U_\mu(x)\in\rm
SU(3)$~\cite{Wilson:1974sk}. The representation in terms of SU(3)
group elements avoids the gauge fixing procedure via the introduction
of the Faddeev-Popov determinant, which is necessary in the continuum
formulation of quantised gauge theories where the gauge potential
$A_\mu(x)$ is an element of the (non-compact) Lie algebra of SU(3).
The regularisation of the theory is realised by defining discretised
versions of $S_{\rm G}[U]$ and $S_{\rm{F}}[U,\bar\psi,\psi]$. There is
considerable freedom in choosing a particular discretisation, as long
as it reproduces the Euclidean action in the continuum when the
lattice spacing is formally taken to zero, $a\to0$, which implies
\be
 S_G[U]+S_{\rm F}[U,\bar\psi,\psi]
 \stackrel{a\to0}{\longrightarrow} \int d^4x\;\left\{
   -\frac{1}{2g_0^2}{\rm Tr}\,F_{\mu\nu}(x)^2 
   +\sum_{f=u,d,s,\ldots} \bar\psi_f(x)
    \left({\gamma_\mu D_\mu}+m_f\right)\psi_f(x) \right\}\,,
% S_G[U]+S_{\rm F}[U,\bar\psi,\psi]
% \,\stackrel{a\to0}{\longrightarrow}\, \int d^4x\;\left\{
%   -\frac{1}{4}{\rm Tr}\,F_{\mu\nu}(x)^2 
%   +\sum_{f=1}^{\Nf} \bar\psi_f(x)
%    \left({\,/\hspace{-5pt}D}^{\rm lat}[U]+m_f\right)\psi_f(x) \right\}\,,
\ee
where $F_{\mu\nu}$ denotes the field tensor in the continuum, $D_\mu$
is the covariant derivative, and the sum is taken over all active
quark flavours, $f=1,\ldots,\Nf$. With these preliminaries, the
expectation value $\langle\Omega\rangle$ of an observable $\Omega$ can
be expressed in terms of the lattice-regularised Euclidean path
integral as
\be
  \langle\Omega\rangle=\frac{1}{Z}\int
  \prod_{x\in\Lambda_{\rm E}}\prod_{\mu}dU_\mu(x)\,\Omega\,\prod_{f=1}^{\Nf}
  \det\left\{\,{/\hspace{-5pt}D}^{\rm lat}[U]+m_f\right\}\,e^{-S_{\rm G}[U]}\,,
\ee
where $\,{/\hspace{-5pt}D}^{\rm{lat}}[U]$ is the massless lattice
Dirac operator and $\Lambda_{\rm E}$ denotes the set of all lattice
sites. In this expression, the quark fields have been integrated out,
leaving a product of quark determinants, one for each quark
flavour. In a nutshell, the lattice formalism preserves gauge
invariance at the level of the regularised theory, defines observables
starting from the QCD action without any reference to perturbation
theory and allows for the stochastic evaluation of expectation values
by means of Monte Carlo integration.

The simplest choice for $S_{\rm G}[U]$ is the Wilson plaquette action
\cite{Wilson:1974sk}. Other widely used discretisations include the
L\"uscher-Weisz \cite{Luscher:1984xn} and Iwasaki
\cite{Iwasaki:2011np} actions that both show an accelerated rate of
convergence towards the continuum limit. There is a great variety of
discretisations of the fermionic part, which may be divided into three
main classes, i.e. staggered \cite{Kogut:1974ag, Susskind:1976jm},
Wilson \cite{Wilson:1974sk} and Ginsparg-Wilson \cite{Ginsparg:1981bj,
  Furman:1994ky, Neuberger:1997fp, Hasenfratz:1998ri, Luscher:1998pqa}
actions. These types differ in the way they deal with the so-called
``fermion doubling problem'', encapsulated in the Nielsen-Ninomiya
theorem \cite{Nielsen:1980rz, Nielsen:1981hk}, and by the extent to
which continuum symmetries are broken. Ginsparg-Wilson fermions,
realised either via the overlap operator \cite{Neuberger:1997fp,
  Neuberger:1998wv} or the domain wall formulation
\cite{Furman:1994ky}, describe a single fermion species and preserve
chiral symmetry at the expense of losing strict locality and the
necessity of generating gauge configurations on a five-dimensional
lattice. Wilson-type fermions, including their O($a$)-improved version
\cite{Sheikholeslami:1985ij} or the twisted-mass formalism
\cite{Frezzotti:2000nk} do not suffer from fermion doubling either,
but break chiral symmetry explicitly.
%are not rigorously protected against the appearance of small or
%negative eigenvalues of the Wilson-Dirac operator, as a result of
%explicit chiral symmetry breaking. 
Staggered fermions, while preserving a subgroup of chiral symmetry,
contain spurious fermionic degrees of freedom called ``tastes". The
partial degeneracy of the fermion spectrum in the staggered
formulation is usually removed by employing the ``rooting procedure''
which gives rise to additional lattice artefacts (the so-called
``taste breaking effects'') and non-localities.
%For completeness, we also mention that minimally doubled
%fermions~\cite{Karsten:1981gd, Wilczek:1987kw, Creutz:2007af,
%Borici:2007kz} describe a mass-degenerate doublet of quarks, which is
%useful in simulations with exact isospin, at the expense of breaking
%isotropy \cite{Capitani:2010nn}.
The extent to which one is confronted with broken continuum symmetries
or some degree of fermion doubling must be balanced against the
numerical effort in an actual lattice calculation. Staggered fermions
are the cheapest to simulate, overlap or domain wall fermions are the
most expensive. Different fermionic discretisations also differ by the
rate by which the continuum limit is reached.

After choosing a particular discretisation of the QCD action, the
calculation proceeds by generating a set of ensembles of gauge
configurations stochastically via a Markov process. The composition of
the gauge ensemble is determined by the statistical weight~$W_{\rm
  MC}$ of a particular configuration, which is given by the action
term in the path integral:
\be
   W_{\rm MC} \propto \prod_{f=1}^{\Nf}
  \det\left\{\,{/\hspace{-5pt}D}^{\rm lat}[U]+m_f\right\}\,e^{-S_{\rm G}[U]}\,,
\ee
which is why the procedure is also referred to as ``importance
sampling''.
For a given ensemble comprised of $N_{\rm cfg}$ gauge configurations,
the expectation value $\langle\Omega\rangle$ of an observable $\Omega$
is identified with the ensemble average $\overline\Omega$ in the limit
of infinite statistics, i.e.
\be
   \langle\Omega\rangle=\lim_{N_{\rm cfg}\to\infty}\overline\Omega\,.
\ee
As $N_{\rm cfg}$ is necessarily finite in an actual calculation, one
associates a statistical error $\sigma_\Omega$ with the result, which
is given as the square root of the variance
$\sigma_\Omega^2=\overline{\Omega^2}-{\overline\Omega}^2$.

The input parameters in any lattice calculation are the bare coupling
$g_0$ and the bare quark masses $m_f$. Their values are {\it a priori}
undetermined and must be fixed by matching a set of hadronic
observables to their physical values. Dimensionful quantities such as
hadron masses are obtained in units of the lattice spacing, and in
order to express them in physical units, one forms dimensionless
ratios in terms of a common reference quantity that sets the overall
scale and assigns a value in physical units to the lattice
spacing. For concreteness, let $am_{\rm PS}(m_1,m_2)$ denote the mass
of a generic pseudoscalar meson (in lattice units) composed from quark
flavours with masses $m_1$ and $m_2$. If, in addition, we consider the
mass of an octet baryon, $am_{\rm B}(m_1,m_1,m_2)$, we can make
contact with the physical situation by tuning the bare quark masses
$m_1$ and $m_2$ so that the ratio $am_{\rm PS}(m_1,m_2)/am_{\rm
  B}(m_1,m_1,m_2)$ assumes the physical pion-to-proton mass
ratio. This fixes $m_1$ and $m_2$ to the values corresponding to the
up- and down-quark masses. The procedure can then be extended in a
similar fashion to fix the values of the heavier quarks.
In this example, the overall scale is set by the proton mass. The
corresponding value of the lattice spacing in physical units is
obtained from
\be
   a^{-1}\,[{\rm MeV}]=
   \frac{m_{\rm p}\,[{\rm MeV}]}{am_{\rm B}(m_u,m_u,m_d)}\,.
\ee
The procedure of fixing the bare parameters of QCD through a set of
precisely known hadronic input quantities is called hadronic
renormalisation scheme. The particular set of input quantities (hadron
masses or decay constants) is mostly chosen for computational
convenience, provided that the error associated with the experimental
reference value is significantly smaller than the intrinsic
uncertainty of the lattice calculation. Once the theory is
``calibrated'' in this way, all other observables computed on the
lattice are genuine predictions of QCD.

In practice, lattice QCD calculations are mostly performed in the
isospin-symmetric limit in which the up- and down-quark masses are
degenerate and electromagnetic corrections are neglected. However, for
hadronic contributions to precision observables, such as lepton
magnetic moments, isospin-breaking corrections must be included to
reach a competitive level of precision.

The lattice approach to determining $\alohvp$ differs substantially
from the data-driven method. For once, lattice QCD does not compute
the $R$-ratio from first principles. Instead, the value of $\alohvp$
is accessible via a convolution integral over Euclidean momenta $Q^2$,
involving the subtracted vacuum polarisation amplitude,
$\hat\Pi(Q^2)$\cite{Lautrup:1969nc, Blum:2002ii}
\be\label{eq:HVPdefmom}
  \alohvp = \left(\frac{\alpha}{\pi}\right)^2 \int_0^\infty
  dQ^2\,f(Q^2)\,\hat\Pi(Q^2),\quad
  \hat\Pi(Q^2)=4\pi^2\left(\Pi_{\rm E}(Q^2)-\Pi_{\rm E}(0)\right)\,
\ee
The function $f(Q^2)$ in the integrand is known analytically, i.e.
\be\label{eq:HVPkernelmom}
   f(Q^2)=\frac{\hat{s}\,Z(\hat{s})^3}{m_\mu^2}\cdot
   \frac{1-\hat{s}\,Z(\hat{s})}{1+\hat{s}\,Z(\hat{s})^2},\quad
   Z(\hat{s})=-\frac{\hat{s}-\sqrt{\hat{s}^2+4\hat{s}}}{2\hat{s}},
   \quad \hat{s}\equiv Q^2/m_\mu^2\,,
%   f(Q^2)=\frac{m_\mu^2 Q^2 Z^3(1-Q^2 Z)}{1+m_\mu^2 Q^2 Z^2}, \quad
%   Z=-\frac{Q^2-\sqrt{Q^4+4m_\mu^2 Q^2}}{2m_\mu^2 Q^2}\,.
\ee
The amplitude $\Pi_{\rm E}(Q^2)$ in this Euclidean setting is obtained
from a lattice calculation of the polarisation tensor,
$\Pi_{\mu\nu}(Q)$, which, in turn, is related to the correlator of the
electromagnetic current $J_\mu(x)$, according to
\be\label{eq:PimunuEucl}
   \Pi_{\mu\nu}(Q)=i\int d^4x\,{\rm e}^{iQ\cdot x}\left\langle
   J_\mu(x)J_\nu(0) \right\rangle \equiv (Q_\mu
   Q_\nu-\delta_{\mu\nu}Q^2)\Pi_{\rm E}(Q^2)\,.
\ee
The last equality arises from current conservation and O(4)-invariance
which replaces Lorentz-invariance in the Euclidean formulation. The
electromagnetic current is defined as
\be\label{eq:JemEucl}
  J_\mu(x)={\textstyle\frac{2}{3}}(\bar{u}{\gamma_\mu}u)(x)
          -{\textstyle\frac{1}{3}}(\bar{d}{\gamma_\mu}d)(x)
          -{\textstyle\frac{1}{3}}(\bar{s}{\gamma_\mu}s)(x)
          +{\textstyle\frac{2}{3}}(\bar{c}{\gamma_\mu}c)(x) +\ldots.
\ee
Equations~(\ref{eq:PimunuEucl}) and~(\ref{eq:JemEucl}) are the
Euclidean analogues of the corresponding definitions in
Eqs.~(\ref{eq:PmunuMink}) and~(\ref{eq:JemMink}), with $Q^2=-q^2$ (but
omitting the explicit factors of the electric charge $e$ in
\eq{eq:PimunuEucl} in order to conform to the usual notation used in
the lattice QCD literature). The lattice formalism yields an inclusive
determination of $\alohvp$ and is not sensitive to individual hadronic
channels that play such a crucial role in the case of the data-driven
method. It is, however, possible to perform an exact decomposition
according to quark flavours or, alternatively, isospin channels.

\begin{figure}
    \centering
    \includegraphics[width=0.95\linewidth]{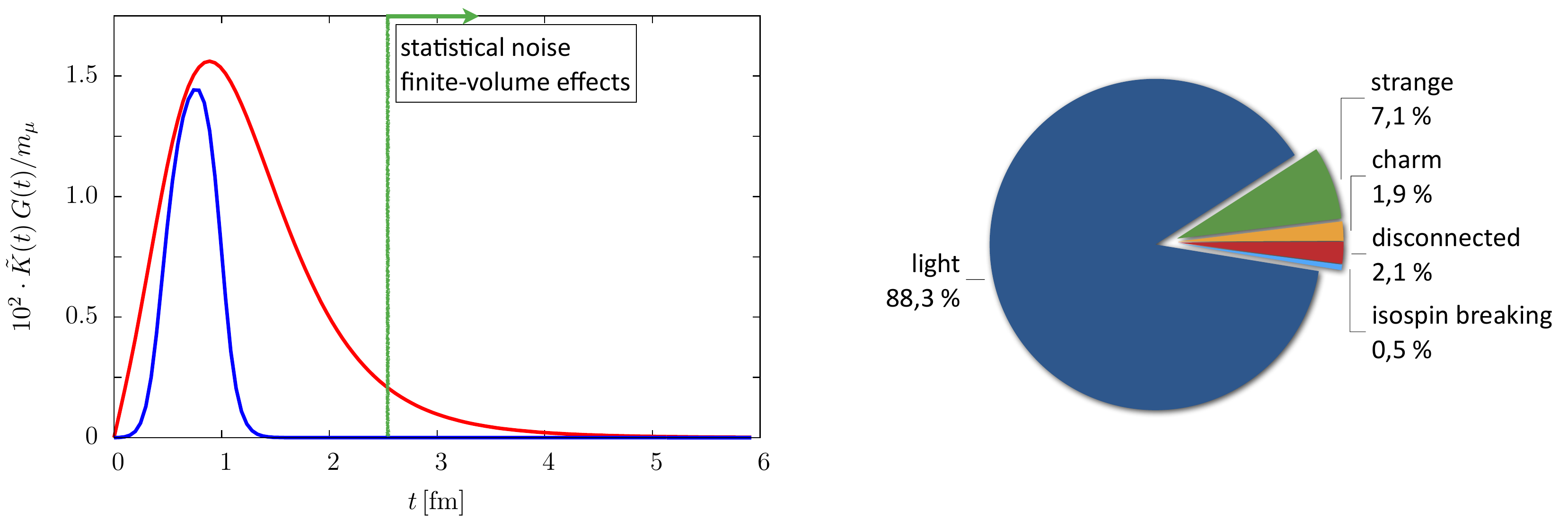}
    \caption{Left: The integrand of the time-momentum representation
      in \eq{eq:TMRdef} (red curve) and the corresponding integrand of
      the intermediate window observable (blue curve). Right: Pie
      chart depicting the fractions from individual quark flavours to
      $\alohvp$, as well as the contributions from quark-disconnected
      diagrams and isospin-breaking corrections, as evaluated in
      Ref.\,\cite{Djukanovic:2024cmq}.}
    \label{fig:HVPlat}
\end{figure}

The calculation of $\Pi_{\mu\nu}(Q)$ in lattice QCD is
straightforward. However, regarding the reliable determination of
systematic effects associated with the infrared regime (see below), it
is more convenient to use an integral representation over the
Euclidean time variable~$t$, known as the ``time-momentum
representation'' (TMR) \cite{Bernecker:2011gh}:
\be\label{eq:TMRdef}
   \alohvp = \left(\frac{\alpha}{\pi}\right)^2 \int_0^\infty
   dt\,\tilde{K}(t)\,G(t),\quad
   G(t)=-a^3\sum_{\vec{x}}\left\langle J_k(x)J_k(0)\right\rangle,
\ee
where $G(t)$ is the spatially summed correlator of the vector current,
which is a standard observable in lattice QCD. The kernel function
$\tilde{K}(t)$ is obtained from $f(Q^2)$ via\footnote{A detailed
discussion of $\tilde{K}(t)$, including a description how it can be
evaluated efficiently, can be found in Appendix~B of
\cite{DellaMorte:2017dyu}.}
\be
   \tilde{K}(t)=4\pi^2\int_0^\infty dQ^2\,f(Q^2)\left(
   t^2-\frac{4}{Q^2}\sin^2\left({\textstyle\frac{1}{2}}Qt\right)\right).
\ee
The evaluation of the TMR integral in \eq{eq:TMRdef} is easily
performed in lattice QCD and does not rely on experimental data,
except for simple input quantities such as hadron masses that fix the
hadronic renormalisation scheme and known with excellent experimental
precision.
However, the goal of determining $\alohvp$ with sub-percent precision
presents several challenges for lattice calculations that can be read
off from Fig.\,\ref{fig:HVPlat}. The left panel shows the integrand
$\tilde{K}(t)\,G(t)$ plotted against Euclidean time~$t$. Its tail
which starts around $t\gtrsim2.5\,\rm fm$ contributes about 3\% to the
value of $\alohvp$ but is subject to strong statistical fluctuations
due to the exponentially increasing statistical noise in the
correlator $G(t)$ as $t\to\infty$. The pie chart shown in the right
panel in Fig.\,\ref{fig:HVPlat} illustrates that there is a strong
hierarchy among the contributions from individual quark flavours: By
far the largest fraction is due to the so-called light-quark connected
contribution from up- and down-quark that accounts for almost 90\% of
the total value of $\alohvp$ and for which the problem of statistical
fluctuations is particularly acute.
The large-$t$ regime also carries the bulk of finite-volume effects
which the TMR integral must be accurately corrected for. At the other
end of the $t$-axis, i.e. for small Euclidean distances one encounters
strong discretisation effects (``lattice artefacts") which must be
removed via a careful extrapolation to the continuum limit. This
requires the ability to disentangle a complicated pattern of terms
beyond the leading corrections in the lattice spacing
\cite{Husung:2019ytz,Ce:2021xgd,Husung:2022kvi}. Finally, even though
they contribute only a small fraction to $\alohvp$, isospin-breaking
effects arising from unequal up- and down-quark masses and
electromagnetism must be accurately determined if lattice QCD is to
compete with the data-driven dispersive method.
Below we briefly describe the various computational strategies that
are adopted to address these challenges.

Since the light-quark connected contribution dominates the value of
$\alohvp$, it is crucial that it can be calculated with statistical
errors far below the percent level.
A widely used approach to identify the estimate of $\alohvp$ without
incurring unduly large statistical errors is the so-called ``bounding
method'' \cite{LehnerBounding2016, Borsanyi:2016lpl}, which monitors
the saturation of the TMR integrand by the contribution from the
lowest-lying $\pi\pi$-state that dominates $G(t)$ at large
distances. In addition, to allow for high statistical precision while
keeping the numerical cost at a manageable level, several noise
reduction strategies such as ``all-mode averaging'' \cite{Blum:2012uh}
and ``low-mode averaging'' (LMA) \cite{Giusti:2004yp, DeGrand:2004qw,
  Kuberski:2023zky, Moningi:2025dlf} are employed. LMA is based on the
exact treatment of the quark propagator in terms of a finite number of
low-lying eigenmodes of the lattice Dirac operator. The statistical
precision can be further improved by combining LMA with a dedicated
calculation of the energy levels in the isospin-1 channel
\cite{DellaMorte:2017khn, Gerardin:2019rua, Erben:2019nmx,
  Bruno:2019nzm, Paul:2023ksa, RBC:2024fic, Lahert:2024vvu,
  Djukanovic:2024cmq}. In this way the noise problem of the vector
correlator can be overcome completely.
Moreover, for many years, the statistical precision of lattice
calculations of $\alohvp$ was limited by so-called
``quark-disconnected'' diagrams that arise from propagators that
correspond to the quark's creation and annihilation at the same
point. Over the past decade, many techniques have been developed that
are able to control the high intrinsic statistical noise of these
diagrams \cite{McNeile:2006bz, Bali:2009hu, Stathopoulos:2013aci,
  Gulpers:2013uca, Francis:2014hoa, Gulpers:2015bba,
  DellaMorte:2017dyu, Giusti:2019kff}.

The second major challenge relates to the determination of
finite-volume effects. Even for a spatial box size of 6\,fm one
expects finite-volume corrections to the light-quark connected
contribution of about~3\%. A widely used method, originally proposed
by Meyer \cite{Meyer:2011um} and first applied in
\cite{Francis:2013fzp,DellaMorte:2017dyu} employs an effective field
theory description of the vector correlator in finite and infinite
volumes. The difference $\left[G(t,L)-G(t,\infty)\right]$ can be
computed from the respective spectral functions that are given -- both
in finite and infinite volumes -- in terms of the timelike pion form
factor.\footnote{The computation of the form factor in finite volume
involves a Lellouch-L\"uscher factor \cite{Lellouch:2000pv} which is
why the procedure is often referred to as the Meyer-Lellouch-L\"uscher
method.} Another analytic method is due to Hansen and Patella
\cite{Hansen:2019rbh,Hansen:2020whp} in which the finite-volume
correction is obtained from a sequence of exponentials multiplied by
the forward Compton amplitude of the pion. Finite-volume corrections
can also be computed in Chiral Perturbation Theory at NNLO
\cite{Aubin:2015rzx,Aubin:2020scy}. All three methods produce
compatible results, showing that finite-volume corrections can be
estimated reliably. Moreover, direct numerical determinations of
finite-volume effects obtained from the difference of results computed
on different volumes
\cite{Borsanyi:2020mff,RBC:2024fic,Djukanovic:2024cmq} and comparing
with the corresponding analytical calculations confirm that the latter
provide an accurate description.

Practically all recent lattice calculations of $\alohvp$ employ the
so-called ``RM123 approach'' \cite{deDivitiis:2011eh,
  deDivitiis:2013xla} to determine isospin-breaking corrections. This
method is based on an expansion about isosymmetric QCD in powers of
the electromagnetic coupling and the light quark mass difference.
When applied to the vector correlator $G(t)$ in the TMR integral, the
expansion produces a set of additional correlation functions featuring
the insertion of the scalar density operator (to account for strong
isospin breaking) or photon lines that arise from electromagnetic
corrections. Alternatively, one can simulate the QCD+QED path integral
directly \cite{Borsanyi:2013lga,Boyle:2017gzv}. The specific problems
due to treating the massless photon in a finite spatial box can be
addressed in a conceptually clean fashion by adoption C$^\star$
boundary conditions \cite{Lucini:2015hfa,Patella:2017fgk,RC:2025zoa}.

The fact that the short- and long-distance regions of the TMR
integrand shown in the left panel of Fig.\,\ref{fig:HVPlat} are the
main sources of uncertainty in lattice calculations has motivated the
introduction of so-called ``window observables''
\cite{RBC:2018dos}. The idea is to vary or, ideally, diminish the
influence of the specific effects that render the treatment of the
respective regions more difficult and costly.
Window observables are defined by convoluting the TMR kernel function
with smoothed step functions. Specifically, by defining
\be
\Theta(t,t',\Delta)={\textstyle\frac{1}{2}}
\Big( 1+\tanh\left[ (t-t')/\Delta \right] \Big)\,,
\ee
where $\Delta$ controls the width of the smoothed discontinuity, one
obtains the short-distance (SD), intermediate-distance (ID) and
long-distance (LD) window observables according to
\begin{align}
    (\alohvp)^{\rm SD}&=\left(\frac{\alpha}{\pi}\right)^2
  \int_0^{\infty} dt\,\tilde{K}(t)\,G(t)\,\Big[1-\Theta(t,t_0,\Delta)\Big]\,,\quad
  (\alohvp)^{\rm ID}\equiv\awin=\left(\frac{\alpha}{\pi}\right)^2
  \int_0^{\infty} dt\,\tilde{K}(t)\,G(t)\,\Big[\Theta(t,t_0,\Delta)-\Theta(t,t_1,\Delta)\Big]\,,
  \nonumber \\
  (\alohvp)^{\rm LD}&=\left(\frac{\alpha}{\pi}\right)^2
  \int_0^{\infty} dt\,\tilde{K}(t)\,G(t)\,\Theta(t,t_1,\Delta)\,.
  \label{eq:WINdef}
\end{align}
It is trivial to check that the three windows must add up to the full
leading-order HVP contribution $\alohvp$. The standard choice for the
parameters $t_0,t_1$ and $\Delta$ that has emerged, is $t_0=0.4$\,fm,
$t_1=1.0$\,fm and $\Delta=0.15$\,fm. Attention has mostly focused on
the intermediate window, $(\alohvp)^{\rm ID}\equiv\awin$, with the
corresponding integrand shown as the blue curve in the left panel of
Fig.\,\ref{fig:HVPlat}. The plot clearly shows that the regions of
large cutoff effects, high statistical noise and large finite-volume
corrections are strongly suppressed, making the intermediate-distance
window observable an ideal benchmark quantity: The precision of
$\awin$ is then only limited by statistics and has reached the per-mil
level \cite{Borsanyi:2020mff, Lehner:2020crt, Wang:2022lkq,
  Aubin:2022hgm, Ce:2022kxy, ExtendedTwistedMass:2022jpw,
  FermilabLatticeHPQCD:2023jof, RBC:2023pvn, Boccaletti:2024guq,
  MILC:2024ryz, FermilabLatticeHPQCD:2024ppc}. Moreover, the window
observable is easily translated into an integral that can be evaluated
using the experimentally measured $R$-ratio as input
\cite{Borsanyi:2020mff,Colangelo:2022vok, Boito:2022dry,
  Benton:2023dci, Benton:2024kwp}. This may serve to expose
significant differences between the lattice and data-driven dispersive
approaches.

\begin{figure}
    \centering
    \includegraphics[width=0.95\linewidth]{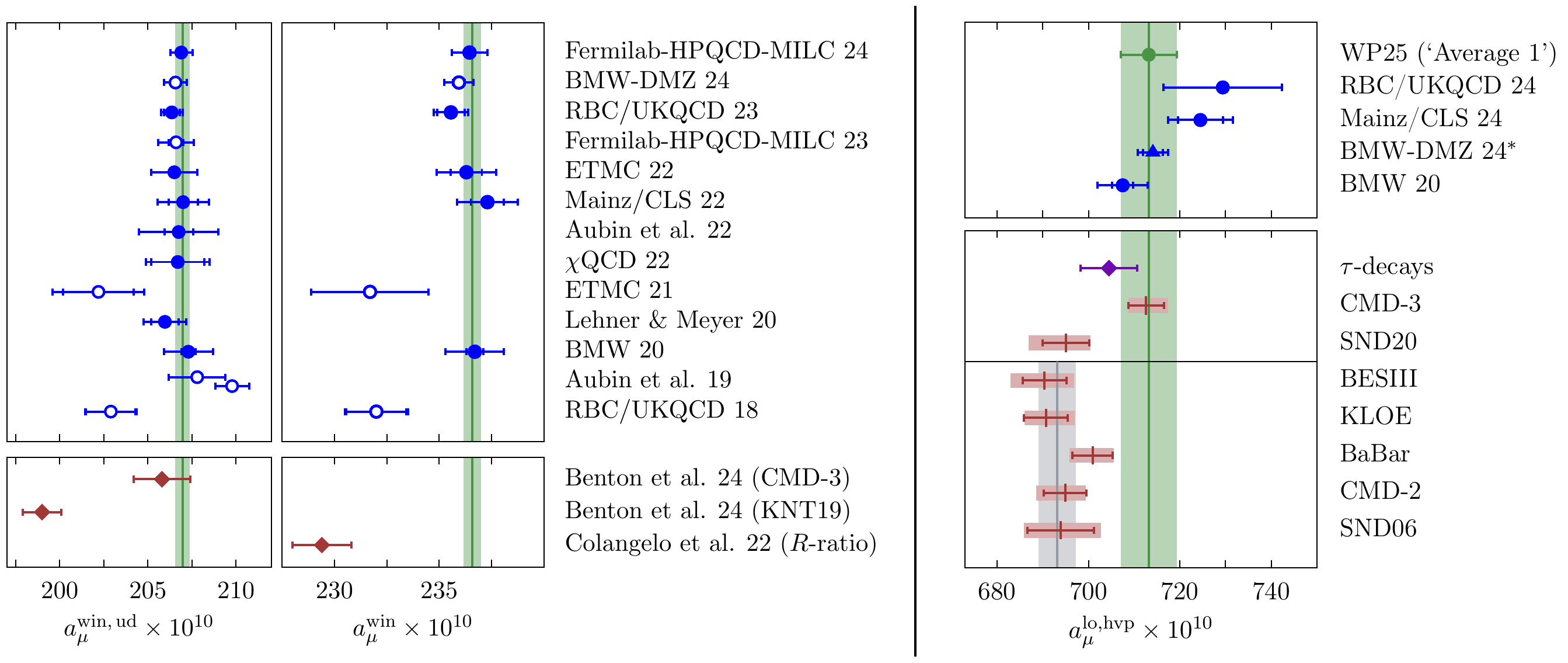}
    \caption{Left: Compilation of lattice results for the intermediate
      window observable \cite{RBC:2018dos, Aubin:2019usy,
        Borsanyi:2020mff, Lehner:2020crt, Giusti:2021dvd,
        Wang:2022lkq, Aubin:2022hgm, Ce:2022kxy,
        ExtendedTwistedMass:2022jpw, FermilabLatticeHPQCD:2023jof,
        RBC:2023pvn, Boccaletti:2024guq,
        FermilabLatticeHPQCD:2024ppc}, showing the dominant
      light-quark connected contribution $a_\mu^{\rm win,ud}$ on the
      far left and the result $\awin$ after adding the sub-leading
      contributions from the other quark flavours and isospin-breaking
      corrections. The green vertical bands represent weighted
      averages computed from the results shown as filled circles, as
      described in \cite{Aliberti:2025beg}. Open symbols represent
      values that either have been superseded by more recent
      calculations or results that have not been published yet. The
      corresponding results based on the data-driven dispersive method
      are shown as red diamonds in the lower part of the plots
      \cite{Colangelo:2020lcg,Benton:2024kwp}. Right: Comparison of
      recent lattice calculations \cite{Borsanyi:2020mff,
        Boccaletti:2024guq, Djukanovic:2024cmq, RBC:2024fic,
        RBC:2018dos} for $\alohvp$ and the data-driven method. The
      green vertical band is the consolidated lattice average from the
      2025 White Paper (WP25 ``Average\,1'') \cite{Aliberti:2025beg}
      shown in \eq{eq:LOHVP-lat}, which is based on a combination of
      many independent lattice results for the window observables. The
      result labelled BMW-DMZ\,24 \cite{Boccaletti:2024guq} is marked
      by an asterisk since it combines a lattice calculation with the
      data-driven dispersive approach in the long-distance regime,
      $t>2.8$\,fm, of the TMR integral. Results from the data-driven
      dispersive method are shown in the lower part. The outer bands
      on each data point represent the variation among analyses
      performed by different groups \cite{Colangelo:2018mtw,
        Colangelo:2022prz, Stoffer:2023gba, Leplumey:2025kvv,
        Keshavarzi:2019abf, Keshavarzi:2018mgv, Davier:2017zfy,
        Davier:2019can, Davier:2023fpl}, as summarised in
      \cite{Aliberti:2025beg}. The single point shown as a diamond is
      the estimate from hadronic $\tau$-decays as listed in WP25
      \cite{Aliberti:2025beg}. The grey band denotes the average from
      the 2020 White Paper \cite{Aoyama:2020ynm} (see
      \eq{eq:LOHVP-DD}), which was based on the results for
      $e^+e^-\to\pi^+\pi^-$ below the horizontal line.}
    \label{fig:window}
\end{figure}

A collection of lattice results for the intermediate window observable
is shown in Fig.\,\ref{fig:window}.
%The contribution from (mass-degenerate) $u,\,d$ quarks, which
%accounts for the bulk of the value of $\awin$, is shown in the left
%panel.
From the compilation in the leftmost panel one concludes that lattice
QCD produces consistent results for the dominant light-quark connected
contribution $a_\mu^{\rm win,ud}$ for a wide range of different
discretisations and with sub-percent precision. The only exceptions
are the calculations labelled RBC/UKQCD\,18~\cite{RBC:2018dos} and
ETMC\,21~\cite{Giusti:2021dvd} which have since been superseded by
RBC/UKQCD~23~\cite{RBC:2023pvn} and ETMC\,22
\cite{ExtendedTwistedMass:2022jpw}. After adding the contributions
from strange and charm quarks, as well as quark-disconnected
contributions and isospin-breaking corrections, one obtains the
results for $\awin$ plotted on the right of the left panel. One
observes excellent agreement among the latest lattice calculations
from each group, which have been performed for a wide range of
different discretisations, demonstrating that the lattice estimates
are very robust. The green vertical bands in both plots denote
weighted averages of the results represented by the filled circles. A
detailed description of the averaging procedure can be found in
section~3.4 of the 2025 White Paper \cite{Aliberti:2025beg}. For the
light-quark connected contribution $a_\mu^{\rm win,ud}$ one finds a
dramatic discrepancy of $6.8\sigma$ with the corresponding results
from the data-driven method based on the available $e^+e^-$ hadronic
cross section data published prior to the CMD-3 result (labelled
``Benton et al. (KNT19)'' in Fig.\,\ref{fig:window}) \cite{Benton:2024kwp}. By
contrast, if one replaces the data for the two-pion channel by the
CMD-3 result, the result agrees well with the lattice average, as
demonstrated by the point labelled ``Benton et
al. (CMD-3)''. Similarly, there is a tension of almost $5\sigma$
between the lattice average for $\awin$ and the data-driven estimate
by Colangelo et al. \cite{Stoffer:2023gba} that was performed before
the release of the CMD-3 result.

As was pointed out in \cite{Colangelo:2020lcg, Ce:2022kxy,
  ExtendedTwistedMassCollaborationETMC:2022sta}, the discrepancy
between lattice and data-driven evaluations can be traced to the
spectral function $R_{\rm had}(s)$ in the energy region
$\sqrt{s}=600-900$\,MeV which contains the contribution from the pion
form factor. This is also evidenced by the fact that the CMD-3 result
for the two-pion channel is significantly higher. The above discussion
shows that the intermediate window observable has fulfilled its role
as a benchmark quantity: It has established consistency among
different lattice calculations, while exposing strong tensions with
the data-driven method when the latter is based on $e^+e^-$ hadronic
cross section data for the dominant $\pi^+\pi^-$ channel taken prior
to CMD-3.

To assess the consequences for the SM estimate for $a_\mu$, it is
necessary to go beyond individual window observables and compare
estimates for the full leading-order HVP contribution
$\alohvp$. Lattice calculations of $\alohvp$ have been published
in Refs. \cite{Borsanyi:2017zdw, RBC:2018dos, Giusti:2018mdh,
  Giusti:2019hkz, Shintani:2019wai, FermilabLattice:2019ugu,
  FermilabLattice:2019dbx, Gerardin:2019rua, Aubin:2019usy,
  Borsanyi:2020mff, Lehner:2020crt, Aubin:2022hgm, Boccaletti:2024guq,
  RBC:2024fic, Djukanovic:2024cmq, ExtendedTwistedMass:2024nyi,
  FermilabLatticeHPQCD:2024ppc}. Out of these, there are three
calculations with at least percent-level precision
\cite{Borsanyi:2020mff, Boccaletti:2024guq, Djukanovic:2024cmq}, one
of which is a ``hybrid'' result \cite{Boccaletti:2024guq} that
combines a lattice calculation for a window observable up to 2.8\,fm
with the data-driven method in the long-distance tail of the TMR
integrand (see \eq{eq:TMRdef} and the left panel of
Fig.\,\ref{fig:HVPlat}). In addition there are many lattice results
for the short-distance \cite{Giusti:2021dvd, Wang:2022lkq,
  ExtendedTwistedMass:2022jpw, RBC:2023pvn, Kuberski:2024bcj,
  Boccaletti:2024guq, Spiegel:2024dec, MILC:2024ryz,
  ExtendedTwistedMass:2024nyi}, intermediate-distance
\cite{RBC:2018dos, Lehner:2020crt, Borsanyi:2020mff, Aubin:2022hgm,
  Wang:2022lkq, Ce:2022kxy, ExtendedTwistedMass:2022jpw, RBC:2023pvn,
  FermilabLatticeHPQCD:2023jof, Boccaletti:2024guq,
  ExtendedTwistedMass:2024nyi, FermilabLatticeHPQCD:2024ppc} and
long-distance \cite{RBC:2024fic, Djukanovic:2024cmq,
  ExtendedTwistedMass:2024nyi, FermilabLatticeHPQCD:2024ppc} window
observables that can be combined and summed up to form a global
average. The latter strategy has been adopted in the 2025 edition of
the White Paper (WP25), and the resulting average is represented as
the green vertical band in the rightmost panel of
Fig.\,\ref{fig:window}. It agrees well with the most recent
calculations of $\alohvp$ that are shown in the upper part of the
plot. Full details on the averaging procedure and the results that
were used as input can be found in section~3.6 of WP25
\cite{Aliberti:2025beg}. The 2025 White Paper estimate for $\alohvp$
is quoted as
\be\label{eq:LOHVP-lat}
   \left.\alohvp\right|_{\rm WP25} = 7132(61) \cdot 10^{-11}\quad [0.86\%]\,.
\ee
This is significantly higher and marginally less precise than the
previous White Paper estimate of \eq{eq:LOHVP-DD}, which is
represented as the grey vertical band in the lower part of the
right-hand side of Fig,\,\ref{fig:window}. The figure also shows the
results from the data-driven dispersive method using the
$e^+e^-\to\pi^+\pi^-$ cross section from different experiments as
input, as well as a recent re-evaluation of the spectral function from
hadronic $\tau$ decays \cite{Aliberti:2025beg}. The tensions between
lattice calculations and the data-driven dispersive method would
disappear only if the latter were evaluated using CMD-3 data
alone. The shift in the SM estimate for $\alohvp$ is the consequence
of the unresolved situation regarding the scatter in the $e^+e^-$ data
and the fact that the results from eight different lattice QCD
collaborations paint a very consistent picture.
It is this very shift that brings the SM estimate for $a_\mu$ into
agreement with the experimental measurement, at the current level of
precision \cite{Aliberti:2025beg}.

\subsection{Hadronic vacuum polarisation at higher orders and for other lepton flavours}

\begin{figure}[t]
    \centering
    \includegraphics[width=0.7\linewidth]{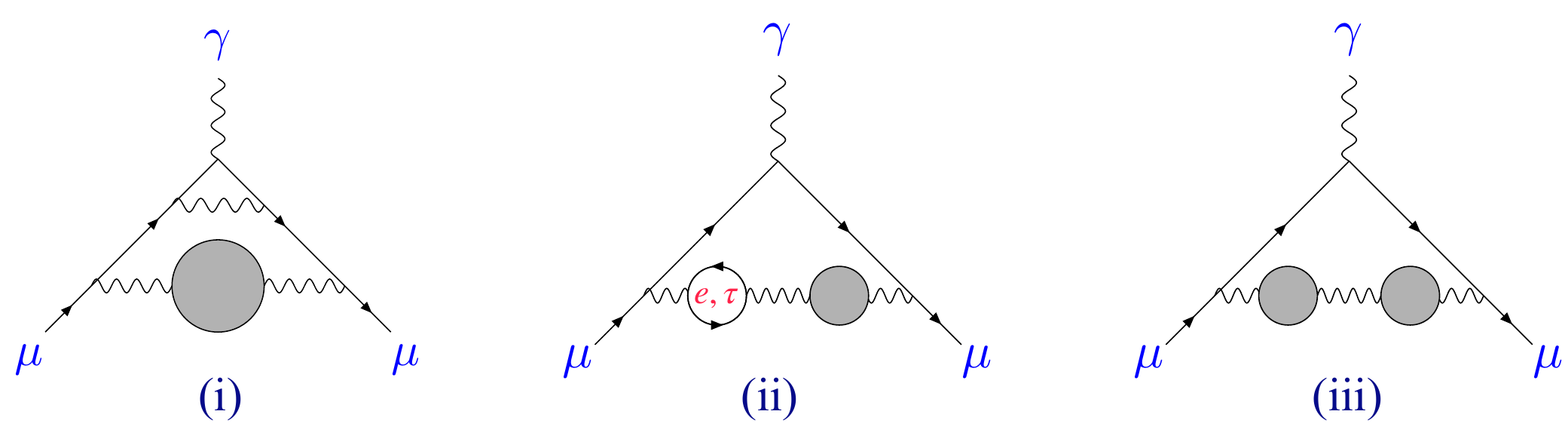}
    \caption{Diagrams representing the hadronic vacuum polarisation
      contribution to the muon anomalous magnetic moment at
      next-to-leading order. Diagram~(i) contains an additional photon
      line; Diagram~(ii) contains lepton loops whose flavour differs
      from that of the external lepton (here: the muon); Diagram~(iii)
      contains two insertions of the hadronic vacuum polarisation.}
    \label{fig:HOHVP}
\end{figure}

The enormous experimental sensitivity achieved in the direct
measurement of the muon anomalous magnetic moment implies that the HVP
contribution to the SM prediction of $a_\mu$ must be determined beyond
the leading order. Figure\,\ref{fig:HOHVP} shows the diagrams that
arise at next-to-leading order, which are suppressed by an additional
power of $\alpha$ relative to $\alohvp$. Below we sketch the
evaluation of these contributions, which is essentially based on the
same methodology as in the case of $\alohvp$, i.e. the data-driven
dispersive method and lattice QCD.

The necessary kernel functions for the data-driven method have been
worked out in \cite{Krause:1996rf,Barbieri:1974nc}, and a first
attempt was made already in Ref.\,\cite{Calmet:1976kd}. At the time of
the 2020 White Paper, data-driven evaluations had been published in
Refs.\,\cite{Hagiwara:2011af, Kurz:2014wya, Jegerlehner:2017gek,
  Keshavarzi:2019abf}. The estimate quoted in 2020
\cite{Aoyama:2020ynm}, i.e. $a_\mu^{\rm
  nlo,\,hvp}=-98.3(7)\cdot10^{-11}$ took into account a slightly
inflated error, reflecting tensions in the hadronic cross section data
and different analysis procedures at the time. The result from a first
lattice calculation was available \cite{Chakraborty:2018iyb} which,
however, lacked the necessary precision to be
competitive. Furthermore, the most important kernel functions arising
at next-to-next-to-leading order have been derived in
\cite{Kurz:2014wya}, and the resulting data-driven estimate of
$a_\mu^{\rm nnlo,\,hvp}=12.4(1)\cdot10^{-11}$ has been included in
both the 2020 and 2025 White Papers.
The impact of the CMD-3 result on the data-driven dispersive
evaluation of $a_\mu^{\rm nlo,\,hvp}$ has been studied in
\cite{DiLuzio:2024sps}, by replacing the data in the $\pi^+\pi^-$
channel by the CMD-3 dataset, leading to a higher estimate than what
was quoted in \cite{Keshavarzi:2019abf, Aoyama:2020ynm}. Therefore, in
the White Paper update of 2025, the results from
\cite{Keshavarzi:2019abf} and \cite{DiLuzio:2024sps} were adopted,
resulting in a quoted value that is the mean of the two results with
an error that covers the difference between the two estimates,
i.e. $a_\mu^{\rm nlo,\,hvp}= -99.6(1.3)\cdot10^{-11}$.

While the $R$-ratio is known for timelike kinematics, the relevant
kernel functions have also been worked out in the space-like regime
\cite{Nesterenko:2021byp,Balzani:2021del}. This opens the possibility
to determine the NLO HVP contribution by providing as input the
Euclidean vacuum polarisation amplitude $\Pi(Q^2)$ (see
\eq{eq:PimunuEucl}), which is accessible in lattice QCD. Moreover, in
order to conform with the standard practice of employing the
time-momentum representation in lattice calculations, the kernel
functions for the NLO contributions have been determined in
\cite{Balzani:2024gmu}. In this way, the evaluation of the
contributions represented by the diagrams in Fig.\,\ref{fig:HOHVP} is
greatly facilitated, as it requires the same spatially summed vector
correlator $G(t)$ as in the case of $\alohvp$. A first lattice evaluation
based on the TMR with a complete error budget has been published
in\,\cite{Beltran:2026ofp}.

\begin{table}
    \centering
    \begin{tabular}{l| r@{.}l c c | r@{.}l c c | r@{.}l c c }
    \hline\hline
    & \multicolumn{4}{c|}{electron / $10^{-12}$} & \multicolumn{4}{c|}{muon / $10^{-11}$}
    & \multicolumn{4}{c|}{tau / $10^{-8}$} \\
    \hline
    HVP,\,LO    &  1&89(3) & (D) & \cite{Keshavarzi:2019abf,DiLuzio:2024sps} 
     &  \multicolumn{2}{c}{\hspace{-11pt}7132(61)} & (L) & \cite{Aliberti:2025beg} & 340&2(2.1) & (D) & \cite{Keshavarzi:2019abf,DiLuzio:2024sps} \\
    HVP,\,NLO   &  $-0$&2263(35) & (D) & \cite{Keshavarzi:2019abf,DiLuzio:2024sps} 
     &  $-99$&6(1.3) & (D) & \cite{Aliberti:2025beg} & 7&85(4) & (D) & \cite{Keshavarzi:2019abf} \\
    HVP,\,NNLO  &  0&02799(17) & (D) & \cite{Jegerlehner:2017zsb}
     & 12&4(1)  & (D) & \cite{Aliberti:2025beg} & \multicolumn{4}{c}{./.} \\
    HLbL,\,LO   & 0&0351(23) & (D) & \cite{Hoferichter:2025fea} 
     & 112&6(9.6) & (L+D) & \cite{Aliberti:2025beg} & 3&77(29) & (D) &
     \cite{Hoferichter:2025fea}\\
    HLbL,\,NLO  & \multicolumn{4}{c|}{./.} & 2&8(6) & (L+D) & \cite{Colangelo:2014qya,Aliberti:2025beg} & \multicolumn{4}{c}{./.} \\
%    HVP,\,LO    &  1&89(3) & (D) & \cite{Keshavarzi:2019abf,DiLuzio:2024sps} 
%     &  713&2(6.1) & (L) & \cite{Aliberti:2025beg} & 340&2(2.1) & (D) & \cite{DiLuzio:2024sps} \\
%    HVP,\,NLO   &  $-0$&2263(35) & (D) & \cite{Keshavarzi:2019abf,DiLuzio:2024sps} 
%     &  $-9$&96(13) & (D) & \cite{Aliberti:2025beg} & 7&85(4) & (D) & \cite{DiLuzio:2024sps} \\
%    HVP,\,NNLO  &  0&02799(17) & (D) & \cite{Jegerlehner:2017zsb}
%     & 1&24(1)  & (D) & \cite{Aliberti:2025beg} & \multicolumn{4}{c}{./.} \\
%    HLbL        & 0&0351(23) & (D) & \cite{Hoferichter:2025fea} 
%     & 11&26(96) & (L+D) & \cite{Aliberti:2025beg} & 3&77(29) & (D) &
%     \cite{Hoferichter:2025fea}\\
    \hline\hline
    \end{tabular}
    \caption{Overview of hadronic corrections to lepton anomalous
      magnetic moments. The letters ``D'' and ``L'' indicate whether
      the estimate is based on the data-driven and lattice approaches,
      respectively. Notice the strong hierarchy among the size of the
      contributions among the three lepton flavours. The numbers are
      based on the most recent publications and/or summaries which can
      be consulted for a full list of references.}
    \label{tab:HVPsumm}
\end{table}

Although most of the recent efforts have focussed on the muon, the
contributions from the hadronic vacuum polarisation have also been
worked out for the electron and $\tau$. The overview presented in
Table \ref{tab:HVPsumm} shows the strong hierarchy of the size of the
HVP contribution among the three lepton flavours: for the electron,
hadronic contributions are tiny, yet large enough so that they must be
included in the SM prediction owing to the enormous experimental
precision that can be achieved. For the $\tau$-lepton, hadronic vacuum
polarisation effects are larger by two orders of magnitude relative to
the muon. However, there is as yet no direct measurement that is
precise enough to provide a test of the SM prediction for $a_\tau$.
The entries in the table are mostly based on the data-driven
dispersive method, although there exist a handful of lattice
calculations for the leading-order HVP contributions to the electron
and $\tau$ anomalous magnetic moments \cite{Borsanyi:2017zdw,
  Giusti:2019hkz, Giusti:2020efo}. When relying on the spectral
function from $e^+e^-$ data, the scatter among different measurements
of the dominant $\pi^+\pi^-$ channel has been taken into account in
the most recent evaluations \cite{DiLuzio:2024sps} by scaling the
error accordingly.
Earlier data-driven evaluations of the HVP contributions to the
electron and $\tau$ anomalous magnetic moments can be found in
Refs. \cite{Krause:1996rf, Eidelman:2007sb, Jegerlehner:2017zsb,
  Keshavarzi:2019abf}.

%% file: HLbL.tex
\subsection{Hadronic light-by-light scattering}

Another important type of hadronic contribution to lepton anomalous
magnetic moments is due to hadronic light-by-light scattering (HLbL),
represented by Diagram~(e) in Fig.\,\ref{fig:diagramsLO}. Even though
it is suppressed by another power of $\alpha$ relative to that of the
leading-order HVP, it accounts for the second largest uncertainty in
the current SM prediction of $a_\mu$. In the case of the electron this
contribution is mostly insignificant, since the HLbL contribution to
$a_e$ is an order of magnitude smaller than the total error which is
dominated by the uncertainty in the input variable
$\alpha$~\cite{Hoferichter:2025fea}. Hadronic contributions to
$a_\tau$ are quite sizeable in general~\cite{Keshavarzi:2019abf,
  Hoferichter:2025fea} (see Table\,\ref{tab:HVPsumm}), but given the
poor sensitivity of the experimental measurement, the fairly accurate
determination of $a_\tau^{\rm hlbl}$ cannot be utilised yet to perform
a stringent test of the SM. This puts the spotlight on the HLbL
contribution to the muon anomalous magnetic moment, $\ahlbl$, which is
mostly discussed in this subsection.

For the muon, the relative suppression of HLbL compared to the
leading-order HVP implies that a determination of $\ahlbl$ to within
10\% is sufficient to allow for a meaningful comparison with the
current experimental value of $a_\mu$. Still, given the complexity in
evaluating the HLbL diagram, it was long thought to be impossible that
$\ahlbl$ could be determined even at this level of precision.

\begin{figure}
    \centering
    \includegraphics[width=0.98\linewidth]{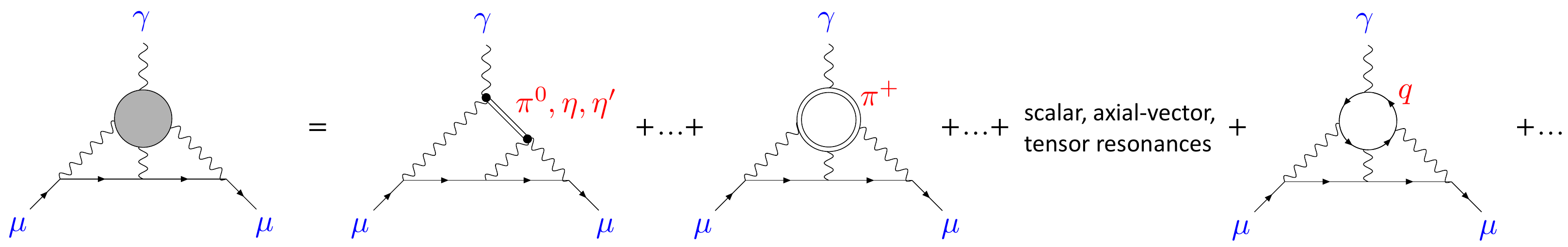}
    \caption{The HLbL diagram expressed as a sum of individual
      intermediate hadronic states and quark loops. The right-hand
      side describes the contributions from single pseudoscalar meson
      exchange, pseudoscalar box contributions, exchange of scalar,
      axial-vector and tensor resonances, as well as quark loops
      (Figure adapted from \cite{Nyffeler:2017ohp}).}
    \label{fig:HLbLdecomp}
\end{figure}

Significant progress was made in the 1990s after the realisation that
the HLbL diagram could be decomposed into a set of individual hadronic
contributions, each of which is described by a particular meson
exchange process as indicated in Fig.\,\ref{fig:HLbLdecomp}. These
contributions can be expressed and evaluated in terms of the
transition form factors (TFFs) for the individual sub-processes.
The dominant contribution is due to the exchange of single
pseudoscalar mesons, followed by heavier scalar mesons (such as the
$f_0(980)$ and $a_0(980)$), as well as axial-vector and tensor mesons
at increasingly higher energies. Furthermore, there are contributions
from charged pion and kaon loops. All of these are intrinsically
non-perturbative. At high enough loop momenta, the HLbL diagram is
described by quark loops that can be evaluated in QCD perturbation
theory.

Up to 2014, the HLbL contribution $\ahlbl$ was determined along
these lines using hadronic models supplemented by chiral symmetry and
large-$N_c$ arguments \cite{deRafael:1993za, Hoferichter:2018dmo}, as
well as perturbative QCD. These efforts culminated in the so-called
``Glasgow consensus'' of 2009 which relied on the available
calculations at the time \cite{Bijnens:1995cc, Bijnens:1995xf,
  Bijnens:2001cq, Hayakawa:1995ps, Hayakawa:1996ki, Hayakawa:1997rq,
  Knecht:2001qf, Melnikov:2003xd, Bijnens:2007pz,
  Jegerlehner:2009ry,Prades:2009tw} and which entered most of the SM
estimates published prior to the first White Paper in 2020. However,
the fact that the contributions shown on the right-hand side of
Fig.\,\ref{fig:HLbLdecomp} cannot be separated unambiguously in model
calculations leads to a double-counting problem and proves to be a
major limitation regarding the achievable precision.
A much more systematic and rigorous approach was outlined and
implemented in a series of papers in 2014 and thereafter
\cite{Colangelo:2014dfa, Colangelo:2014pva, Pauk:2014rfa,
  Pauk:2014rta, Colangelo:2015ama, Hagelstein:2017obr}. This not only
solved the double-counting issue encountered in model calculations but
also opened the way for a data-driven dispersive treatment, similar to
that applied in the case of HVP. However, unlike in the case of the
HVP contribution, it is impossible to treat all intermediate hadronic
states in terms of a single dispersion integral, because the
dispersion relation takes a different form for each intermediate
state.

The particular strength of the dispersive treatment of exclusive
intermediate states lies in the description of the regime of low
photon virtualities. Indeed, as will be described in the paragraphs
below, the most important contributions which account for about two
thirds of the total $\ahlbl$ can be quantified with a precision of
5\%.
The high-energy behaviour of the HLbL contribution must be treated
separately and gives rise to the so-called short-distance constraints
whose relevance was first pointed out by Melnikov and
Vainshtein~\cite{Melnikov:2003xd}.

\begin{wrapfigure}{r}{5cm}
  \centering
  \includegraphics[width=0.18\textwidth]{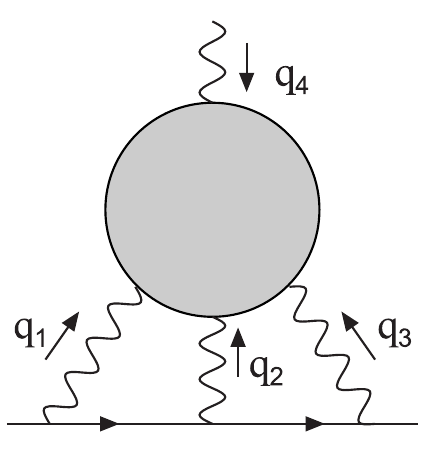}
  \caption{The HLbL diagram showing the momentum assignment to the
    internal $(q_1, q_2, q_3)$ and external $(q_4)$ photon lines.
    Figure taken from \cite{Bijnens:2020xnl}.}
  \label{fig:HLbLvirtualities}
\end{wrapfigure}
Understanding the transition between the low-energy region (where the
dispersive approach provides an accurate and quantitative description)
and the perturbative regime is challenging.
It is for this reason that additional information from analytic and
phenomenological methods such as holographic QCD
\cite{Maldacena:1997re, Aharony:1999ti, Leutgeb:2019gbz,
  Cappiello:2019hwh, Leutgeb:2021mpu}, rational approximants
\cite{Chisholm:1973, HughesJones:1976, Peris:2006ds, Masjuan:2007ay,
  Masjuan:2009wy, Masjuan:2012wy, Escribano:2013kba,
  Escribano:2015yup, Masjuan:2017tvw, Estrada:2024cfy} and functional
methods based on Dyson-Schwinger and/or Bethe-Salpeter equations
\cite{Eichmann:2016yit, Eichmann:2019tjk, Raya:2019dnh,
  Eichmann:2019bqf, Eichmann:2021zuv, Aguilar:2024dlv} are considered
for providing further guidance.
Moreover, even though these approaches rely on approximations to some
extent, they provide important cross-checks for those dominant
channels that are very well controlled via the dispersive method, as
well as complementary information on sub-leading contributions for
which the dispersive formalism has not yet reached the desired level
of precision.

The HLbL contribution is also amenable to lattice QCD
calculations. First ideas to compute it on the lattice by combining
QCD and QED were proposed \cite{Hayakawa:2005eq} but proved much
harder to implement with the desired precision than in the case of the
HVP contribution \cite{Blum:2014oka}. These early attempts have now
been superseded by elaborate formalisms in position space
\cite{Blum:2015gfa, Green:2015mva, Asmussen:2022oql}, similar in
spirit to the TMR for the HVP but a lot more complicated in
practice. A second line of activity has been the calculation of the
two-photon-pseudoscalar transition form factor \cite{Gerardin:2016cqj,
  Gerardin:2019vio, ExtendedTwistedMass:2023hin, Gerardin:2023naa,
  Lin:2024khg} that can be compared directly to the corresponding
determination via the dispersive method \cite{Hoferichter:2018kwz,
  Hoferichter:2018dmo, Holz:2024diw, Holz:2024lom}.

The literature on HLbL is vast, owing to the extremely varied set of
formalisms that are applied and to the complexity of the
evaluation. In this subsection we will merely sketch the various
approaches and give a status report on current evaluations.

To describe the derivation of the data-driven dispersive formalism we
largely follow Refs. \cite{Colangelo:2017fiz, Aliberti:2025beg} which
can be consulted for further details. The starting point is the HLbL
tensor which is defined in terms of the time-ordered product of four
electromagnetic currents $J^\mu$, i.e.
\begin{equation}\label{eq:HLbLtensor}
  \Pi^{\mu\nu\lambda\sigma}(q_1, q_2, q_3) = -i \int d^4x_1 d^4x_2 d^4x_3
  \,e^{-i(q_1{\cdot}x_1+q_2{\cdot}x_2+q_3{\cdot}x_3)}\,
  \left\langle 0\left| 
  T\left\{J^\mu(x_1)J^\nu(x_2)J^\lambda(x_3)J^\sigma(0)\right\}
  \right|0\right\rangle\,,
\end{equation}
where $q_1, q_2, q_3$ denote the four-momenta of the internal photons
as depicted in Fig.\,\ref{fig:HLbLvirtualities}. The HLbL tensor is
the analogue of the HVP tensor in \eq{eq:PmunuMink}. Since the
dispersive formalism is focussed on the contribution from the light
$(u, d, s)$ quark flavours to the HLbL diagram, the electromagnetic
current in this definition is restricted to the three-flavour theory,
i.e.
\begin{equation}
   J^\mu=\textstyle \frac{2}{3}\bar{u}\gamma^\mu{u}
                -\frac{1}{3}\bar{d}\gamma^\mu{d}
                -\frac{1}{3}\bar{s}\gamma^\mu{s}\,.
\end{equation}
Contracting $\Pi^{\mu\nu\lambda\sigma}$ with polarisation vectors
yields the hadronic contribution to the helicity amplitudes for
off-shell photon-photon scattering $\gamma^\ast(q_1, \mu)
\gamma^\ast(q_2, \nu)\to\gamma^\ast(-q_3, \lambda)
\gamma^\ast(q_4,\sigma)$.

In the next step one decomposes the HLbL tensor into a generating set
of gauge-invariant Lorentz structures, with scalar coefficient
functions that are free of kinematic singularities.\footnote{In the
simpler case of the HVP contribution, there is only one such
coefficient function, $\Pi(q^2)$, shown in \eq{eq:PmunuMink}.}
Following the work of Bardeen, Tung and Tarrach \cite{Bardeen:1968ebo,
  Tarrach:1975tu}, one identifies 54 such structures
\cite{Colangelo:2015ama}, only seven of which are independent while
the others are related through crossing. The $\Pi_i$'s can be shown to
be free from kinematic singularities, which implies that the limit of
vanishing external momentum, $q_4\to0$ can be taken.
It is then possible to derive an expression for $\ahlbl$ in terms of
the coefficient functions $\Pi_i$ integrated over the loop momenta
$q_1$ and $q_2$ ($q_3$ being fixed by momentum conservation). Noting
that the $\Pi_i$'s depend on $q_1^2$, $q_2^2$ and $q_1{\cdot}q_2$, one
can perform five integrals of the eight-dimensional integration
analytically after Wick-rotating to Euclidean momenta.
The redundancy among the $\Pi_i$'s can be reduced by applying
projection operator techniques~\cite{Aldins:1970id, Barbieri:1974nc,
  Jegerlehner:2017gek} to obtain a minimal set of 12 independent
kernel functions, $\bar\Pi_1,\bar\Pi_2,\ldots,\bar\Pi_{12}$ that are
linear combinations of the original coefficient
functions~\cite{Colangelo:2015ama, Colangelo:2017fiz, Bijnens:2024jgh,
  Hoferichter:2024fsj}. As detailed in \cite{Colangelo:2017fiz}, the
master formula for $\ahlbl$ is then obtained as
\begin{equation}
    \ahlbl= \frac{2\alpha^3}{3\pi^2}\int_{0}^{\infty} dQ_1
    \int_{0}^{\infty} dQ_2 \int_{-1}^{1} d\tau\,
    \sqrt{1-\tau^2}\,Q_1^2 Q_2^2\,\sum_{i=1}^{12} T_i(Q_1, Q_2, \tau)\,
    \bar\Pi_i(Q_1, Q_2, \tau),\quad Q_1\equiv|Q_1|,\quad Q_2\equiv|Q_2|\,,
\end{equation}
with Euclidean photon virtualities $Q_i^2=-q_i^2,\,i=1, 2, 3$ that
satisfy
\begin{equation}
    s=q_3^2=-Q_3^2=-Q_1^2-2Q_1Q_2\tau-Q_2^2,\quad
    t=q_2^2=-Q_2^2,\quad u=q_1^2=-Q_1^2\,.%\quad q_4^2=0.
\end{equation}
%
%The HLbL contribution is obtained in the limit of vanishing external momentum $q_4$. 
The formalism proceeds by deriving dispersion relations for each of
the twelve scalar functions expressed in terms of the respective
kinematic variables, by means of analyticity.
In the implementation of the dispersive framework, the contributions
from one- and two-meson intermediate states have been explicitly
accounted for, as will be discussed below.

By far the dominant contribution (about $50-60$\%) to $\ahlbl$ is due
to the pion-pole contribution which is represented by the first
diagram on the right-hand side of Fig.\,\ref{fig:HLbLdecomp} (along
with the sub-leading contributions from the $\eta$ and $\eta'$). The
pion pole is determined through the transition form factor
$F_{\pi^0\gamma^\ast\gamma^\ast}(q_1^2, q_2^2)$ which describes the
decay of a neutral pion into two virtual photons,
$\pi^0(q_1+q_2)\to\gamma^\ast(q_1, \mu)\gamma^\ast(q_2, \nu)$ and
which is defined by
\begin{equation}
    i\int d^4x\,e^{iq_1{\cdot}x}\,\left\langle 0\left|
    T\{J_\mu(x)J_\nu(0)\}\right|\pi^0(q_1+q_2)\right\rangle =
    \epsilon_{\mu\nu\lambda\sigma}\,q_1^\lambda q_2^\sigma\,
    F_{\pi^0\gamma^\ast\gamma^\ast}(q_1^2, q_2^2)\,.
\end{equation}
For real photons and in the limit of massless quarks, the pion TFF
$F_{\pi^0\gamma^\ast\gamma^\ast}$ is related to the Adler-Bell-Jackiw
anomaly \cite{Adler:1969gk, Bell:1969ts, Bardeen:1969md}
\begin{equation}
    F_{\pi^0\gamma^\ast\gamma^\ast}(0,0) = \frac{1}{4\pi^2\,F}\,,
\end{equation}
where $F$ is the pion decay constant in the chiral limit. The
high-energy behaviour in both the singly and doubly virtual case is
obtained from a light-cone expansion of the two electromagnetic
currents~\cite{Lepage:1979zb,Lepage:1980fj} and from the operator
product expansion \cite{Nesterenko:1982dn, Novikov:1983jt,
  Gorsky:1987mu}, respectively, which yields
\begin{equation}
    \lim_{Q^2\to\infty}\,Q^2F_{\pi^0\gamma^\ast\gamma^\ast}(-Q^2,0) = 2F\,\qquad
    \lim_{Q^2\to\infty}\,Q^2F_{\pi^0\gamma^\ast\gamma^\ast}(-Q^2,-Q^2) = {\textstyle\frac{2}{3}}F\,.
\end{equation}
The high-energy behaviour in the singly virtual case is called the
Brodsky-Lepage limit.

The contribution of $F_{\pi^0\gamma^\ast\gamma^\ast}$ to $\ahlbl$ is
obtained by integrating products of doubly and singly virtual TFFs
convoluted with appropriate weight functions $w_1,\,w_2$ specified
in~\cite{Jegerlehner:2009ry, Nyffeler:2016gnb, Masjuan:2017tvw,
  Hoferichter:2018kwz}
\begin{equation}\label{eq:pionpole}
    a_\mu^{\pi^0\textrm{-pole}}= \left\{\frac{\alpha}{\pi}\right)^3
    \int dQ_1\,dQ_2\,d\tau\,\left\{ w_1(Q_1, Q_2, \tau)\,
    F_{\pi^0\gamma^\ast\gamma^\ast}(-Q_1^2, -Q_3^2)\,
    F_{\pi^0\gamma^\ast\gamma^\ast}(-Q_2^2, 0) +w_2(Q_1, Q_2, \tau)\,
    F_{\pi^0\gamma^\ast\gamma^\ast}(-Q_1^2, -Q_2^2)\,
    F_{\pi^0\gamma^\ast\gamma^\ast}(-Q_3^2, 0)
    \right\}\,.
\end{equation}
The generalisation to the sub-leading contributions mediated by the
$\eta$ and $\eta'$ mesons is straightforward. It is interesting to
note that, while \eq{eq:pionpole} was derived prior to the dispersive
formalism, the latter reproduces the same expression when restricting
the intermediate hadronic states to the pion contribution.

Owing to the lack of accurate experimental data at the time, the first
calculations of the contributions arising from the pseudoscalar
$(\pi^0, \eta, \eta')$ pole were performed using phenomenological
descriptions for the TFFs based on vector meson dominance,
supplemented by constraints from the large-$N_c$
limit~\cite{Bijnens:1995xf, Hayakawa:1995ps, Hayakawa:1996ki,
  Hayakawa:1997rq, Bijnens:1995cc, Bijnens:2001cq, Knecht:2001qf}.
The dispersive formalism represented a major step forward for a
reliable determination of not only the pseudoscalar pole contribution
but rather the entire HLbL. Here we only sketch the main ingredients
of the determination, referring the reader to the respective sections
in the White Papers on the muon $g-2$ \cite{Aoyama:2020ynm,
  Aliberti:2025beg}, which provide a much more detailed account.

The task of determining the pion pole contribution via the dispersive
method entails the construction of the $\pi^0$ TFF
$F_{\pi^0\gamma^\ast\gamma^\ast}$ from its dominant singularities
\cite{Hoferichter:2014vra, Hoferichter:2018dmo, Hoferichter:2018kwz}.
The shape of the weight functions $w_{1, 2}$ that appear
in~\eq{eq:pionpole} and which are plotted in Fig.~58 of
\cite{Aoyama:2020ynm} suggests that the dominant contribution to the
integral arises from the region $Q_i<1$\,GeV. In this regime the
discontinuities can be reconstructed from experimental data for
$e^+e^-\to 2\pi,\,3\pi$ which serve as input for constraining the
electromagnetic pion form factor and the partial wave amplitude for
$\gamma^\ast\pi\to\pi\pi$, both of which enter the spectral
representation of the TFF. Skipping further details and referring the
reader to the detailed accounts presented in section~4.4
of~\cite{Aoyama:2020ynm} and section~5.5 of~\cite{Aliberti:2025beg},
we note that model-independent predictions for the singly and doubly
virtual TTFs are obtained in this way which, in turn, serve to
determine $a_\mu^{\pi^0\textrm{-pole}}$ via \eq{eq:pionpole}
\cite{Hoferichter:2014vra, Hoferichter:2018dmo, Hoferichter:2018kwz,
  Hoferichter:2021lct}. The formalism has been extended to the
sub-leading pseudoscalar pole contributions from the $\eta$ and
$\eta'$, respectively \cite{Hanhart:2013vba, Kubis:2015sga,
  Holz:2015tcg, Holz:2022hwz, Holz:2024diw, Holz:2024lom}.

The contribution from pseudoscalar poles can also be calculated in
lattice QCD. The corresponding formalism has originally been pioneered
in \cite{Gerardin:2016cqj} and subsequently applied to determine the
contribution from the pion pole \cite{Gerardin:2016cqj,
  Gerardin:2019vio, ExtendedTwistedMass:2022ofm, Gerardin:2023naa,
  Lin:2024khg} as well as the $\eta$ and $\eta'$. Lattice calculations
of pseudoscalar TFFs employ the same methods that are routinely used
in lattice QCD to compute form factors for a variety of problems,
ranging from semi-leptonic meson decays to nucleon structure. Here we
refrain from describing any further details and instead refer the
reader to the the FLAG report
\cite{FlavourLatticeAveragingGroupFLAG:2021npn} for a more general
discussion. The specific implementation in the case of computing the
pion TFF is described in the original paper \cite{Gerardin:2016cqj}.

\begin{figure}
    \centering
    \includegraphics[width=0.99\linewidth]{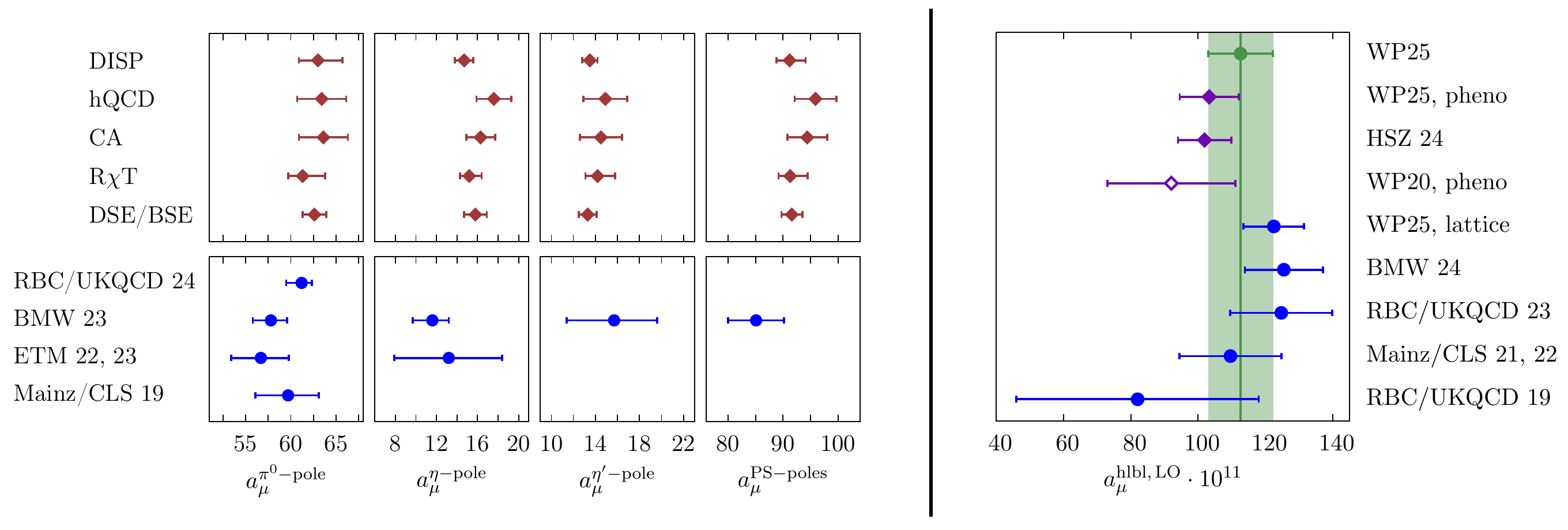}
    \caption{Left: Comparison of results for the contributions from
      pseudoscalar poles to the HLbL contribution in units of
      $10^{-11}$ from a variety of methods, including the dispersive
      approach (DISP) \cite{Hoferichter:2018kwz, Holz:2024diw,
        Hoferichter:2018dmo, Holz:2024lom}, holographic QCD (hQCD)
      \cite{Leutgeb:2022lqw}, Canterbury approximants (CA)
      \cite{Masjuan:2017tvw}, Rational Chiral Theory (R$\chi$T)
      \cite{Estrada:2024cfy} and functional methods (DSE/BSE)
      \cite{Eichmann:2019tjk}. Lattice results computed by RBC/UKQCD
      \cite{Lin:2024khg}, BMW \cite{Gerardin:2023naa}, ETM
      \cite{ExtendedTwistedMass:2022ofm} and Mainz/CLS
      \cite{Gerardin:2019vio} are shown in the bottom row of panels
      (figure adapted from \cite{Aliberti:2025beg}). Right: Summary of
      evaluations of $\ahlbl$, including the data-driven method
      (purple diamonds) \cite{Aoyama:2020ynm, Hoferichter:2024bae,
        Hoferichter:2024vbu, Aliberti:2025beg} and lattice QCD (blue
      circles) \cite{Blum:2019ugy, Chao:2021tvp, Chao:2022xzg,
        Blum:2023vlm, Fodor:2024jyn}. The open diamond shows the old
      data-driven estimate of WP20 which has been superseded. The
      green circle and vertical band represent the WP25 average of
      \eq{eq:HLbL_WP25} \cite{Aliberti:2025beg} (figure adapted from
      \cite{Aliberti:2025beg}).}
    \label{fig:PSpoles_HLbLsummary}
\end{figure}

The left panel of Fig.\,\ref{fig:PSpoles_HLbLsummary} compares the
evaluation of the pseudoscalar pole contributions using the dispersive
approach \cite{Hoferichter:2018dmo, Hoferichter:2018kwz, Holz:2024diw,
  Holz:2024lom} to a variety of other methods, including holographic
QCD (hQCD) \cite{Leutgeb:2022lqw}, Canterbury approximants (CA)
\cite{Masjuan:2017tvw}, rational chiral theory (R$\chi$T)
\cite{Estrada:2024cfy}, Dyson-Schwinger / Bethe-Salpeter equations
(DSE/BSE) \cite{Eichmann:2019tjk} and lattice QCD
\cite{Gerardin:2019vio, ExtendedTwistedMass:2022ofm, Gerardin:2023naa,
  Lin:2024khg}. The overall agreement -- both regarding the actual
value and the error estimate -- among different methods is striking,
suggesting that the influence of model assumptions and approximations
in other analytic theoretical approaches is weak. Although lattice QCD
has a tendency to produce slightly smaller estimates for the pion pole
contribution, the tension with the dispersive result is quite
small. After adding the contributions from the $\pi^0$, the $\eta$ and
$\eta'$, one finds in the data-driven dispersive formalism
\cite{Aliberti:2025beg}
\begin{equation}\label{eq:PSpoleDDD}
    \left.a_\mu^{\textrm{PS-pole}}\right|_{\textrm{disp}}
    =\Big(91.2^{+2.9}_{-2.4}\Big)\cdot10^{-11}\,.
\end{equation}

Before we discuss the derivation of the result for the full HLbL
contribution, we briefly sketch other theoretical approaches that are
routinely applied in analytic and phenomenological determination of
$\ahlbl$:

\begin{itemize}
    \item {\it Holographic QCD:} A hadronic model of QCD in the large
      $N_c$-limit can be constructed along the lines of the AdS/CFT
      correspondence \cite{Maldacena:1997re, Aharony:1999ti},
      i.e.\ the conjectured mapping between a conformally invariant
      supersymmetric Yang-Mills theory (CFT) in the limit of infinite
      't~Hooft coupling and a five-dimensional classical theory of
      gravity on anti-de Sitter (AdS) space.
    %The fifth dimension corresponds to the energy scale of the
    %holographically dual Quantum Field Theory that exists on the
    %conformal boundary of AdS space.
      Holographic QCD (hQCD) is too crude a model to meet the
      precision requirements for a competitive determination for the
      HVP contribution. However, it is capable of providing useful and
      complementary information on HLbL at the 10\%
      level. Furthermore, hQCD naturally satisfies short-distance
      constraints \cite{Leutgeb:2019gbz, Cappiello:2019hwh,
        Leutgeb:2021mpu}, as well as the Brodsky-Lepage asymptotic
      behaviour of the pseudoscalar TFFs. The strength of the approach
      lies in the ability to provide results for the axial-vector and
      tensor sectors that are still poorly constrained when applying
      the dispersive framework.
    \item {\it Rational approximants:} Certain classes of rational
      functions such as Pad\'e approximants can be rigorously applied
      to describe form factors and spectral functions. In particular,
      under certain conditions one can prove theorems that establish
      the convergence of a sequence of rational approximants to a
      model-independent description \cite{Peris:2006ds,
        Masjuan:2007ay, Masjuan:2009wy}. Furthermore, limiting cases
      arising from the high-energy behaviour can also be incorporated
      into the framework. While Padé approximants have been
      successfully applied to the $\pi^0$ TFF in the singly virtual
      case \cite{Masjuan:2012wy, Escribano:2013kba, Escribano:2015yup,
        Masjuan:2017tvw}, the doubly virtual case requires a
      generalisation of the formalism to Canterbury approximants (CA)
      \cite{Chisholm:1973, HughesJones:1976, Sanchez-Puertas:2016auw}.
      Until recently, Canterbury approximants have been the only
      source for the $\eta$ and $\eta'$ pole contributions, by means
      of a computation of the respective TFFs \cite{Estrada:2024cfy}.
      Resonance Chiral Theory (R$\chi$T), first considered in
      \cite{Ecker:1989yg} (see also \cite{Ruiz-Femenia:2003jdx,
        Cirigliano:2006hb, Mateu:2007tr, Guo:2009hi}) is another
      theoretical framework that extends chiral effective theory by
      supplementing the Lagrangian of Chiral Perturbation Theory
      \cite{Gasser:1983yg, Gasser:1984gg} by explicit degrees of
      freedom describing light vector mesons. This has been applied in
      the context of HLbL in \cite{Estrada:2024cfy}.
    \item {\it Functional methods:} Coupled Dyson-Schwinger and
      Bethe-Salpeter equations (DSEs, BSEs) have been used to tackle a
      variety of problems in hadron physics. In principle, DSEs
      provide exact relations among the $n$-point correlation
      functions of QCD. However, in practical applications the
      truncation of the DS series is unavoidable, which incurs an
      intrinsic error that can be difficult to quantify. In the
      context of the HLbL contribution, two approaches have been
      followed: The ``direct'' method seeks to express the entire HLbL
      amplitude in terms of quark propagators and quark-photon
      vertices, each of which are computable via DSEs and BSEs
      \cite{Eichmann:2016yit}. In the ``indirect'' approach the
      relevant TFFs for specific sub-contributions to the HLbL
      contribution are computed in the DSE / BSE formalism and then
      used as input in the dispersive approach. In this way, results
      for the pseudoscalar pole and pseudoscalar box contributions
      have been computed \cite{Eichmann:2019tjk, Raya:2019dnh,
        Eichmann:2019bqf}, as well as contributions due to
      axial-vector and scalar exchange \cite{Eichmann:2024glq}. Errors
      are typically estimated by considering variations of the
      parameters that enter the evaluation.
\end{itemize}

In order to quantify the sum of all hadronic intermediate states that
contribute to HLbL, one must add the contributions from pion and kaon
boxes as well as $S$-wave rescattering to the result of
\eq{eq:PSpoleDDD}, all of which carry a negative sign. According to
the decomposition shown in Fig.\,\ref{fig:HLbLdecomp}, the result must
still be matched to the high-energy behaviour of the HLbL tensor
\cite{Melnikov:2003xd}. The latter gives rise to the short-distance
constraints (SDCs) which serve to determine the HLbL contribution in
kinematical regimes where two or three photon virtualities that are
integrated over become large. SDCs are derived by applying the
operator product expansion to the electromagnetic quark currents in
\eq{eq:HLbLtensor}. For the kinematical situations mentioned above
this has been done in \cite{Melnikov:2003xd, Bijnens:2019ghy,
  Bijnens:2020xnl, Bijnens:2021jqo, Bijnens:2022itw, Bijnens:2024jgh}.
The appropriate matching to the contributions of hadronic intermediate
states derived in the dispersive framework can be organised
efficiently in terms of a scale $Q_0$ that separates the low- and
high-energy regimes \cite{Hoferichter:2024bae, Hoferichter:2024vbu}.
Summing up all contributions and sources of uncertainty, one obtains
the current estimate of the total HLbL contribution in the data-driven
dispersive approach \cite{Aliberti:2025beg}, i.e.\
\begin{equation}\label{eq:HLbLdisp}
    \left.\ahlbl\right|_{\textrm{disp}}=\Big(103.3^{+8.8}_{-8.6}\Big)
    \cdot 10^{-11}\,.
\end{equation}
A full discussion of the evaluation of the different ingredients,
including a comparison between analytic approaches that form the basis
of this result are found in section~5.10 of\,\cite{Aliberti:2025beg}.
The estimate in \eq{eq:HLbLdisp} is not the result shown in
Table~\ref{tab:HVPsumm}. As indicated in the table, the latter is
obtained via the combination of the data-driven dispersive method and
lattice QCD calculations of $\ahlbl$ which we now turn our attention
to.

The general framework for lattice QCD calculations of $\ahlbl$ is
based on the same underlying formalism and techniques that have been
introduced in section~\ref{sec:LOHVPlat}. Two complementary lines of
activity regarding the HLbL contribution are pursued: The first is
aimed at performing a direct determination of $\ahlbl$ by calculating
the diagram on the left-hand side in Fig.\,\ref{fig:HLbLdecomp}. The
second effort is devoted to computing the TFFs for the dominant
pseudoscalar pole contribution, i.e. the first diagram on the
right-hand side of Fig.\,\ref{fig:HLbLdecomp}.

The HLbL diagram consists of a quark loop embedded in a structure
containing photon and muon lines. The quark loop corresponds to a
four-point correlation function of the electromagnetic quark current,
and its calculation is relatively straightforward in lattice
QCD. However, in the absence of an analytic expression for the effects
of the surrounding photon and muon lines, the calculation of the HLbL
diagram presents considerable difficulties. This situation is quite
unlike the one encountered in the case of the HVP where the relevant
two-point correlation function can be readily convoluted and
integrated with a kernel function that encodes the rest of the HVP
diagram (see \eq{eq:TMRdef}).

\begin{wrapfigure}{r}{5cm}
  \centering
  \includegraphics[width=0.19\textwidth]{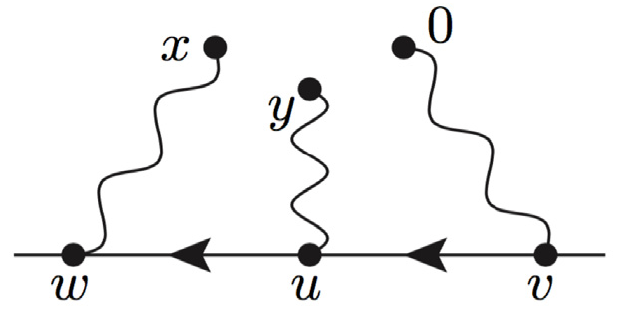}
  \caption{The QED part of the HLbL diagram. The muon has
    momentum~$p$, and after integrating over the photon vertices at
    $u, v$ and $w$ one obtains he position-space QED kernel
    ${\cal{L}}_{[\rho,\sigma];\mu\nu\lambda}(p, x, y)$. Figure taken
    from \cite{Asmussen:2018oip}.}
  \label{fig:HLbLkernel}
\end{wrapfigure}
Therefore, the first proposal to compute $\ahlbl$ directly on the
lattice \cite{Hayakawa:2005eq} was based on the concept of including
the muon line into the lattice calculation and computing the matrix
element of the electromagnetic current between muon initial and final
states in QCD+QED. Expectation values in QCD+QED can be defined by
including the QED dynamics in the Markov chain Monte Carlo, which is
formally accomplished by augmenting the QCD path integral of
\eq{eq:QCDpath} by an integration over the photon field and including
the QED action in the exponent. Without going into details, we note
that the key idea in Refs. \cite{Hayakawa:2005eq, Blum:2014oka} was to
recover the HLbL diagram by means of a non-perturbative subtraction of
correlation functions that differ by terms of O($\alpha^4$). The main
drawback of this strategy is the high level of statistical noise
incurred by the subtraction, as was observed in the concrete
application of the method published in \cite{Blum:2014oka}. As a
consequence, this approach was abandoned in favour of a stochastic
treatment of the QED contribution, which proceeds by explicit
insertions of photon propagators in position space and sampling over
all photon vertex positions \cite{Blum:2015gfa, Blum:2016lnc}. This
produced a first lattice QCD estimate of the HLbL contribution by the
RBC/UKQCD collaboration~\cite{Blum:2019ugy} with a relative total
error of 45\% and hence about twice as large as the data-driven
dispersive estimate at the time.

While this new method does not suffer from the large level of
intrinsic statistical noise, it is still affected by a conceptual
issue related to the explicit treatment of the photon field on the
lattice. Since the photon is a massless unconfined particle, its
inclusion in a lattice calculation may lead to strong finite-volume
effects that fall off as powers of the inverse volume instead of being
exponentially suppressed~\cite{Duncan:1996xy}. Several methods have
been proposed to deal with this problem~\cite{Gockeler:1989wj,
  Polley:1990tf, Duncan:1996xy, Hayakawa:2008an, Endres:2015gda,
  Lucini:2015hfa} (see \cite{Patella:2017fgk} for a review), mostly by
introducing {\it ad hoc} prescriptions for dealing with the zero modes
of the photon field \cite{Duncan:1996xy, Hayakawa:2008an}. The most
widely used method is the so-called QCD$_{\textrm{L}}$ prescription
\cite{Hayakawa:2008an} in which the spatial zero modes of the photon
field are set to zero.\footnote{The majority of lattice calculations
of electromagnetic corrections to the HVP contribution discussed in
section~\ref{sec:LOHVPlat} are also based on the QCD$_{\textrm{L}}$
prescription.}

Any conceptual issues related to QCD$_{\textrm{L}}$ can be avoided by
an alternative method developed by the Mainz group
\cite{Green:2015mva, Asmussen:2016lse, Asmussen:2017bup,
  Asmussen:2018lcw, Asmussen:2019act, Asmussen:2022oql}, which treats
the muon and photon lines semi-analytically in infinite volume. The
aim is to provide an analytic kernel function in position space that
can be convoluted with the four-point correlation function computed in
QCD and integrated over. Specifically, the expression for the Pauli
form factor at zero momentum transfer, $F_2(0)$, which is identified
with the HLbL contribution, is derived as~\cite{Green:2015mva,
  Meyer:2018til}
\be
  F_2(0) = \frac{m_\mu\,e^6}{3} \int
  d^4x\,d^4y\,{\cal{L}}_{[\rho,\sigma];\mu\nu\lambda}(p, x, y)\,
  i\widehat\Pi_{\rho;\mu\nu\lambda\sigma}(x,y),\quad
  i\widehat\Pi_{\rho;\mu\nu\lambda\sigma}(x,y) = -\int d^4z\,
  z_\rho\left\langle J_\mu(x)J_\nu(y)J_\sigma(z)J_\lambda(0)
  \right\rangle \,.
\ee
The QED kernel function ${\cal{L}}_{[\rho,\sigma];\mu\nu\lambda}(p, x,
y)$ is represented by the diagram on the left-hand side of
Fig.\,\ref{fig:HLbLkernel} after integration over the vertices $u, v$
and $w$. The corresponding algebraic expression is obtained from the
Feynman rules and reads
\be
  {\cal{L}}_{[\rho,\sigma];\mu\nu\lambda}(p, x, y) =
  \frac{1}{16m_\mu^2} \int d^4u\,d^4v\,d^4w\,G(w-x)G(u-y)G(v)\,
  e^{-ip\cdot(w-v)}\,{\rm Tr}\left\{[\gamma_\rho, \gamma_\sigma]
  (-i/{\hspace{-5pt}p}+m_\mu)\gamma_\mu S(w-u)\gamma_\nu
  S(u-v)\gamma_\lambda (-i/{\hspace{-5pt}p}+m_\mu) \right\}\,,
\ee
where $G(x)=1/(4\pi^2x^2)$ is the massless scalar propagator, and the
muon propagator $S(x)$ is given by
\be
  S(x)=\int\frac{d^4p}{(2\pi)^4}\,\frac{-i/{\hspace{-5pt}p}+m_\mu}{p^2+m_\mu^2}
  \,e^{ip{\cdot}x}\,.
\ee
The kernel function ${\cal{L}}_{[\rho,\sigma];\mu\nu\lambda}(p, x, y)$
still depends on the muon momentum $p=im_\mu\hat\epsilon$. Since
$F_2(0)$ is a Lorentz scalar, one may replace
${\cal{L}}_{[\rho,\sigma];\mu\nu\lambda}(p, x, y)$ by its average over
the direction of the muon's momentum, i.e.
\be
  {\cal{L}}_{[\rho,\sigma];\mu\nu\lambda}(p, x, y) \to
  \bar{\cal{L}}_{[\rho,\sigma];\mu\nu\lambda}(x, y) :=
  \frac{1}{2\pi^2}\int d\Omega_{\hat\epsilon}\,
  {\cal{L}}_{[\rho,\sigma];\mu\nu\lambda}(p=im_\mu\hat\epsilon, x, y)\,,
\ee
where the integration is performed over the solid angle
$\Omega_{\hat\epsilon}$. This yields the master formula for $\ahlbl$
in terms of a convolution integral in position space:
\be
  \ahlbl = \frac{m_\mu\,e^6}{3} \int
  d^4x\,d^4y\,\bar{\cal{L}}_{[\rho,\sigma];\mu\nu\lambda}(x, y)\,
  i\widehat\Pi_{\rho;\mu\nu\lambda\sigma}(x,y),\quad
  i\widehat\Pi_{\rho;\mu\nu\lambda\sigma}(x,y) = -\int d^4z\,
  z_\rho\left\langle J_\mu(x)J_\nu(y)J_\sigma(z)J_\lambda(0)
  \right\rangle \,.
\ee
The key advantage of this formulation is that the QED kernel function
$\bar{\cal{L}}_{[\rho,\sigma];\mu\nu\lambda}(x, y)$ is
infrared-finite, thereby avoiding the conceptual issue related to {\it
  ad hoc} prescriptions such as QED$_{\textrm{L}}$. Furthermore, owing
to current conservation, one can add total derivatives of the vector
current to the kernel which, upon partial integration, only produce
boundary terms so that the value of the integral remains
unchanged~\cite{Asmussen:2019act, Blum:2017cer}. This property can be
exploited to optimise the spatial profile of the kernel function in
actual lattice calculations, so that the influence of the short- and
long-distance regimes is reduced, similarly to the window observables
used for computing the HVP contribution in lattice QCD.
The semi-analytic calculation of the kernel is described in
\cite{Asmussen:2022oql}, and the computer code for the numerical
implementation has been made public in the same reference. A variant
of the method which projects the kernel function to the muon's rest
frame was developed by the RBC/UKQCD collaboration \cite{Blum:2017cer}
and applied in \cite{Blum:2023vlm}.

\begin{figure}
    \centering
    \includegraphics[width=0.97\linewidth]{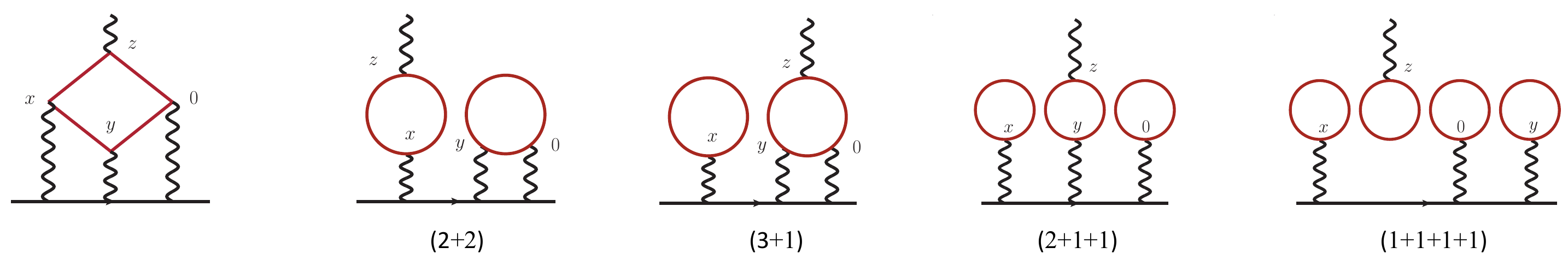}
    \caption{Collection of quark-connected and -disconnected diagrams
      that must be evaluated in lattice calculations of $\ahlbl$. Only
      the fully connected diagram and the $(2+2)$ disconnected diagram
      make significant contributions (figure adapted from
      \cite{Chao:2020kwq}).}
    \label{fig:HLbLDiags}
\end{figure}

The calculation of the four-point correlation function of the
electromagnetic quark current requires the evaluation of the
quark-connected and quark-disconnected diagrams shown in
Fig.\,\ref{fig:HLbLDiags}. At the time of writing (2025), independent
direct calculations of the complete HLbL contribution have been
performed and published by three groups, i.e. RBC/UKQCD
\cite{Blum:2019ugy, Blum:2023vlm}, Mainz/CLS \cite{Chao:2021tvp,
  Chao:2022xzg} and BMW \cite{Fodor:2024jyn}, using a variety of
different discretisations of the QCD action and formalisms. The first
calculation published by RBC/UKQCD \cite{Blum:2019ugy} was based on
explicit photon propagators and the QED$_L$ prescription, applied at
two different values of the lattice spacing to quantify discretisation
effects. The Mainz/CLS calculation \cite{Chao:2021tvp, Chao:2022xzg}
was the first to use the position-space QED kernel in infinite volume
to produce an estimate for $\ahlbl$ with a total error of about 15\%
in the continuum limit. RBC/UKQCD reported an update of their
calculation, now using their variant of the position-space kernel at a
value of the single lattice spacing in 2023\,\cite{Blum:2023vlm}. The
BMW collaboration published their result \cite{Fodor:2024jyn} obtained
using staggered quarks at three lattice spacing and employing the
position-space kernel of the Mainz group \cite{Asmussen:2022oql}.

The results from the three latest calculations \cite{Chao:2021tvp,
  Blum:2023vlm, Fodor:2024jyn} are all compatible within the quoted
uncertainties, as is evidenced by the right panel of
Fig.\,\ref{fig:PSpoles_HLbLsummary}. One observes a strong
cancellation between the fully connected and ($2+2$) disconnected
contributions (see Fig\,\ref{fig:HLbLDiags}), with the connected one
being nearly twice as large in magnitude. All remaining
quark-disconnected contributions are strongly suppressed and play no
role for the final result. A weighted average of the results published
in \cite{Chao:2021tvp, Chao:2022xzg, Blum:2023vlm, Fodor:2024jyn}
yields
\begin{equation}\label{eq:HLbLlat}
  \left.\ahlbl\right|_{\textrm{lat}}=
  \big(122.5\pm9.0\big)\cdot10^{-11}\,.
\end{equation}
The strange and charm quarks contribute only $2.92(98)\cdot10^{-11}$
to this result.

A comparison of results from Lattice QCD calculations and
phenomenological determinations based on the data-driven dispersive
formalism is shown in the right panel of
Fig.\,\ref{fig:PSpoles_HLbLsummary} along with the 2025 White Paper
estimate based on combinations. Despite the mild tension of
$\approx1.5\sigma$ between the two approaches, one can perform a
weighted average, which yields the result quoted as the 2025 White
Paper estimate, i.e.
\begin{equation}\label{eq:HLbL_WP25}
    \left.\ahlbl\right|_{\textrm{WP25}}=\big(112.6\pm9.6\big)\cdot10^{-11}\quad [8.5\%]\,,
\end{equation}
which is also listed in Table\,\ref{tab:HVPsumm}. In conclusion,
despite substantial technical difficulties, both the data-driven
dispersive formalism as well as lattice QCD calculations have
succeeded in producing model-independent determinations of the HLbL
contribution with a total error just below 10\%. This is an
improvement by a factor two relative to the 2020 White Paper.

%% file: EW.tex
\section{Electroweak contributions}

Electroweak corrections arise from diagrams that contain at least one
of the weak gauge bosons $W^\pm, Z^0$ or a Higgs boson. The fact that
these bosons are so massive compared to the mass of any of the charged
leptons is the main reason for the relative smallness of the
electroweak contribution, $a_\mu^{\rm EW}$. This point was already
made in section~\ref{sec:overview} (see the discussion after
\eq{eq:BSMsensitivity}).
Most efforts to determine electroweak contributions are focussed on
the electron and muon because of the availability of high-precision
experimental measurements. In the following, we will therefore use the
muon as the generic case.
In this section we merely summarise the main relevant features of
computing electroweak contributions. To this end, we largely follow
the presentation in the 2020 and 2025 White Papers
\cite{Aoyama:2020ynm, Aliberti:2025beg}, as well as standard reviews
such as\,\cite{Jegerlehner:2009ry, Jegerlehner:2017gek}.

\begin{figure}
    \centering
    \includegraphics[width=0.97\linewidth]{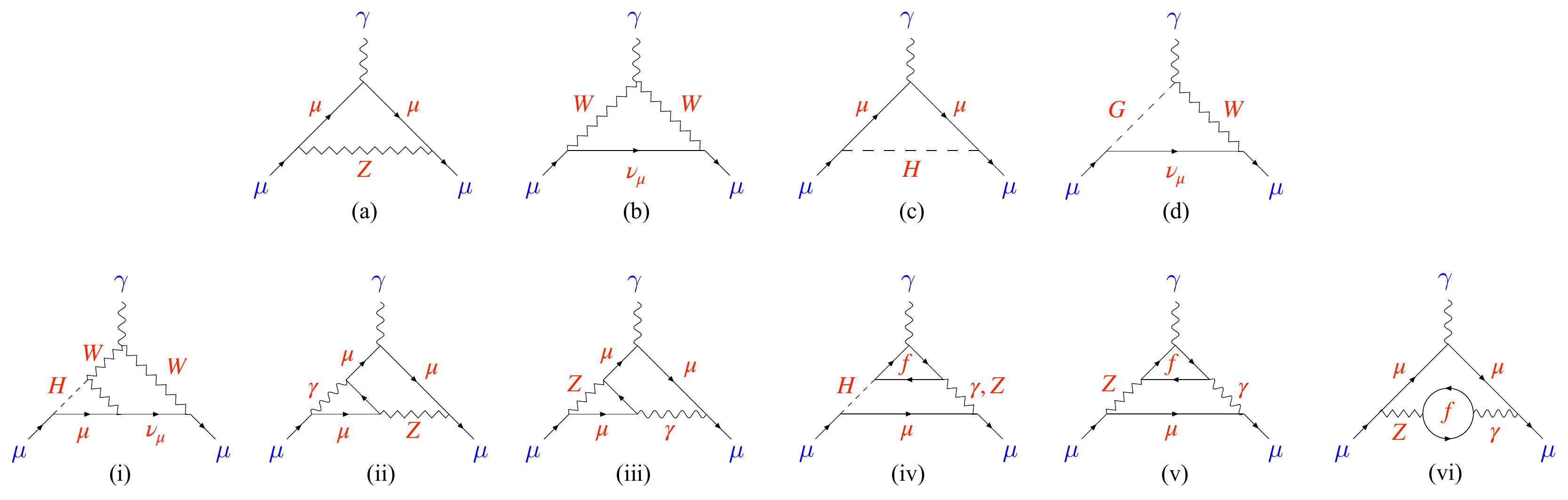}
    \caption{Top: One-loop diagrams that contribute to the electroweak
      contribution to the muon anomalous magnetic moment,
      $a_\mu^{\rm{EW}}$. Diagram~(d) represents the contribution from
      unphysical bosons that arise in certain choices of
      gauge. Bottom: Sample of the most relevant two-loop diagrams
      that contribute to $a_\mu^{\rm{EW}}$. Diagrams~(i)--(iii)
      represent bosonic two-loop contributions, whereas
      diagrams~(iv)--(vi) contain fermion loops.}
    \label{fig:EW2loopDiags}
\end{figure}

The strong suppression induced by the large mass of the $W$ and $Z$
bosons, as well as the relative smallness of the electroweak coupling
$g_2=e\sin\theta_W$ (where $\sin\theta_W$ denotes the electroweak
mixing angle) implies that electroweak contributions can be
principally computed in perturbation theory. The uncertainties that
are quoted for electroweak contributions \cite{Aoyama:2020ynm,
  Aliberti:2025beg} arise from parametric uncertainties, i.e.\ the
errors assigned to input values such as the masses of the $W$, $Z$ and
Higgs bosons or the Fermi constant $\GF$, as well as missing
contributions from higher orders. Diagrams that arise at leading
(one-loop) order are shown in the top row of
Fig.\,\ref{fig:EW2loopDiags}, while the bottom row contains samples of
two-loop diagrams. The evaluation of the latter involves several
interesting features, such as the appearance of large logarithms and
non-perturbative effects. Moreover, anomalies and their cancellation
play an important role.

The proof of renormalisablity of Yang-Mills-type theories by 't Hooft
and Veltman \cite{tHooft:1971akt, tHooft:1971qjg, tHooft:1972tcz} in
1971 inspired several efforts to explicitly compute electroweak
corrections for observables such as lepton anomalous magnetic
moments. Indeed, the very point of these early calculations was to
show that such corrections were finite. The contributions from the
$W$, $Z$ and Higgs bosons, represented by diagrams (a), (b) and (c) in
Fig.\,\ref{fig:EW2loopDiags} have been worked out in
\cite{Jackiw:1972jz, Bars:1972pe, Altarelli:1972nc, Bardeen:1972vi,
  Fujikawa:1972fe} and are as follows:
\be\label{eq:EWoneloopWZH}
   \aew{(2),\,W} =\frac{\GF\,m_\mu^2}{\sqrt{2}\,8\pi^2}\,\frac{10}{3},
   \quad
   \aew{(2),\,Z} =\frac{\GF\,m_\mu^2}{\sqrt{2}\,8\pi^2}\,\left(
   \frac{(1-4\sin^2\theta_W)^2}{3} -\frac{5}{3}\right),\quad
   \aew{(2),\,H} =\frac{\GF\,m_\mu^2}{\sqrt{2}\,8\pi^2}\,
   4\frac{m_\mu^2}{m_H^2}\ln\frac{m_H^2}{m_\mu^2}\,.
\ee
Here, $\GF$ denotes the Fermi constant, while the electroweak mixing
angle is considered in the on-shell scheme, where it is given in terms
of the pole masses of the $W$ and $Z$ bosons,
i.e. $\sin^2\theta_W=1-M_W^2/M_Z^2$.
Besides computing the one-loop contributions, these early references
also sought to verify that the results does not depend on the chosen
gauge, for instance, by studying electroweak corrections in a
generalised renormalisable $R_\xi$ gauge\,\cite{Fujikawa:1972fe}.
The typical size of the leading electroweak contribution is given by
the common prefactor in \eq{eq:EWoneloopWZH} which, using the
tree-level relation
\be
  \frac{\GF}{\sqrt{2}}=\frac{e^2}{8M_W^2}\,\frac{M_Z^2}{M_Z^2-M_W^2}
                      =\frac{\pi\,\alpha}{2\,M_W^2\,\sin^2\theta_W}
\ee
can be rewritten as
\be
  \frac{\GF\,m_\mu^2}{\sqrt{2}\,8\pi^2}=
  \frac{\alpha}{16\pi\,\sin^2\theta_W}\,\frac{m_\mu^2}{M_W^2}\,.
\ee
This illustrates once more the strong suppression by the ratio
$(m_\mu/M_W)^2\simeq 1.7\cdot10^{-6}$. Adding the contributions from
the $W$ and $Z$ bosons yields the one-loop electroweak contribution to
the muon anomalous magnetic moment\footnote{As input for the
evaluation we have used $M_W=80.3692(133)$\,GeV,
$M_Z=91.1880(20)$\,GeV, $M_H=125.20(11)$\,GeV,
$\GF=1.1663788(6)\cdot10^{-5}$\,GeV$^{-2}$
\cite{ParticleDataGroup:2024cfk}.}
\be\label{eq:EWoneloopTOT}
  \aew{(2)} \equiv \aew{(2),\,W} + \aew{(2),\,Z}=
  \frac{\GF\,m_\mu^2}{\sqrt{2}\,8\pi^2}\,\left(
  \frac{5}{3}+\frac{(1-4\sin^2\theta_W)^2}{3}\right)
  = 194.79(1)\cdot10^{-11}\,.
\ee
In the above estimate, the contribution from the Higgs boson,
$\aew{(2),\,H}$, has been neglected
% since it only amounts to $5\cdot10^{-14}$,
due to the suppression relative to the contribution from the $W$ and
$Z$ bosons by a factor proportional to
$({m_\mu^2}/{m_H^2})\ln({m_H^2}/{m_\mu^2})\simeq10^{-5}$ (see
\eq{eq:EWoneloopWZH}).

One prominent feature that arises beyond one-loop level is the
logarithmic enhancement of diagrams that contain heavy particles and a
photon, e.g. diagrams (ii), (iii) or (v) in
Fig.\,\ref{fig:EW2loopDiags}. The resulting large logarithms $\ln
M_{W,Z}^2 /m_f^2$, where $m_f$ is the mass of a light SM fermion,
partially compensate the otherwise expected suppression at two-loop
level. As a consequence, the two-loop contribution lowers the one-loop
estimate of \eq{eq:EWoneloopTOT} by about 20\%. This is in stark
contrast to the situation encountered in the case of electromagnetic
corrections where the entire two-loop contribution to $a_\mu^{\rm
  QED}$ amounts to less than 1\% of the leading-order result.

The logarithmic enhancement of two-loop electroweak contributions was
first discussed in 1992 by Kukhto et al.\,\cite{Kukhto:1992qv} yet
without computing the complete set of diagrams. This was accomplished
in 1995 by Czarnecki, Krause and Marciano\,\cite{Czarnecki:1995wq,
  Czarnecki:1995sz}, which was incidentally the first two-loop
calculation of a SM observable. This work, which was initially
performed in the limit $(1-4\sin^2\theta_W)\to0$, was subsequently
extended in\,\cite{Degrassi:1998es, Czarnecki:2002nt} using an
effective field theory (EFT) approach to determine all two-loop
logarithmic terms without any approximation, as well as the leading
three-loop diagrams.
Furthermore, the dependence of electroweak corrections on the Higgs
mass, which presented a major source of uncertainty before the
discovery of the Higgs boson in 2012, was mapped out in
\cite{Heinemeyer:2004yq} and finally evaluated precisely. The impact
of the Higgs mass measurement was eventually studied in
\cite{Gnendiger:2013pva}.
The full two-loop result for the logarithmic terms reads
\begin{align}\label{eq:EW2loopLogs}
  \aew{(2),\,{\rm{logs}}} =&
  -4\frac{\alpha}{\pi}\ln\frac{M_Z}{m_\mu}\,\aew{(2)}
  +\frac{\GF m_\mu^2}{\sqrt{2}\,8\pi^2} \frac{\alpha}{\pi}\ln\frac{M_Z}{m_\mu}
   \left[ -\frac{47}{9}-\frac{11}{9}(1-4\sin^2\theta_W)^2\right] \nonumber\\
  &+\frac{\GF m_\mu^2}{\sqrt{2}\,8\pi^2} \frac{\alpha}{\pi}
   \sum_{f}\ln\frac{M_Z}{{\rm{max}}(m_f,m_\mu)}\left[ -6\,g_{\rm{A}}^\mu
     g_{\rm{A}}^f\, N_f\, Q_f^2
     +\frac{4}{9}g_{\rm{V}}^\mu g_{\rm{A}}^f\,N_f\, Q_f\right]\,.  
\end{align}
This expression requires some further explanation. The first term on
the right-hand side corresponds to the insertion of a one-loop
diagram, which yields diagram~(iii) in Fig.\,\ref{fig:EW2loopDiags}.
In the EFT treatment, the remaining terms are obtained from an
effective four-fermion vertex which is generated by integrating out
the $Z$ boson. Diagram~(ii) is represented by the second term, while
the third term corresponds to diagrams~(v) and~(vi). Here, the sum
over~$f$ is performed over all SM quarks and leptons, and the factor
of ${\rm max}(m_f,m_\mu)$ arises since the muon is the relevant mass
scale when $f=e$. Furthermore, $g_{\rm{A}}^f\equiv 2I_3^f$ is the
(projection of) weak isospin of fermion~$f$, while $g_{\rm{V}}^f\equiv
2I_3^f-4\sin^2\theta_W Q_f$ denotes the fermion's weak charge (with
$Q_f$ being the electric charge). Finally, the factor $N_f$ counts the
number of degrees of freedom for quarks ($N_f=3$) and leptons
($N_f=1$), respectively.
As was observed in \cite{Czarnecki:1995wq}, the cancellation of the
Adler-Bell-Jackiw anomaly \cite{Adler:1969gk, Bell:1969ts,
  Bardeen:1969md} affects the behaviour of the fermion loop
contributions of diagrams~(v) and~(vi). In the SM, the condition for
anomaly cancellation reads
\begin{equation}
    \sum_f N_f I_3^f Q_f^2 =0
\end{equation}
for each fermion generation. This factor can be identified in the term
proportional to $N_f Q_f^2$ in \eq{eq:EW2loopLogs}. As a consequence,
the term proportional to $\ln M_Z$ cancels, so that the actual
logarithmic enhancement originates from fermion mass ratios instead of
$\ln(M_Z/m_f)$. This can be verified explicitly in the case of the
first two generations. However, the fact that the top quark mass,
$m_t$, is larger than $M_Z$ spoils the cancellation of $M_Z$-dependent
logarithms \cite{DHoker:1992kbg,Czarnecki:1995wq}, leading to a
qualitatively different fermion mass behaviour in the case of the
third generation.

Diagrams~(v) and~(vi) in Fig.\,\ref{fig:EW2loopDiags} contain loops of
light quark flavours~$u,d$ and~$s$. Following the elaborate discussion
of hadronic corrections in section~\ref{sec:LOHVPlat}, it is clear
that a perturbative treatment is not justified. Without going into
much detail, we note that a non-perturbative evaluation has first been
performed in \cite{Peris:1995bb} and improved in \cite{Knecht:2002hr,
  Czarnecki:2002nt, Gnendiger:2013pva, Ludtke:2024ase,
  Hoferichter:2025yih}.
Furry's theorem implies that only the axial component of the $Z$ boson
contributes to diagram~(v), and hence anomaly cancellation plays a
crucial role for the evaluation of the resulting vector-vector-axial
(VVA) interaction. The contribution from the third generation can be
computed in perturbation theory. For the first two generations, one
proceeds by describing the $Z^\ast$-$\gamma$-$\gamma^\ast$ sub-diagram
by a general VVA Green function and express it in terms of two scalar
functions $w_{L,T}(Q^2)$ that depend on the virtuality $Q^2$ of the
virtual $Z$ boson. The resulting contribution to $a_\mu^{\rm EW}$ is
obtained from an integration over $w_{L,T}(Q^2)$, taking additional
constraints from the operator product expansion and
non-renormalisation theorems into account. Specifically for the second
generation, one treats charm and muon loops in perturbation
theory\,\cite{Ludtke:2024ase, Hoferichter:2025yih}.

Diagram~(vi) represents the hadronic contribution from $\gamma$-$Z$
mixing to $a_\mu^{\rm EW}$, which is given by the
expression\,\cite{Hoferichter:2025yih}
\begin{equation}\label{eq:amugammaZ}
    a_\mu^{\gamma Z} = -\frac{\GF\,m_\mu^2}{\sqrt{2}\,8\pi^2}\frac{\alpha}{\pi}
    \cdot\frac{4}{3}(1-4\sin^2\theta_W)\cdot 8\pi^2\bar\Pi^{\gamma Z}(-M_Z^2)\,.
\end{equation}
Here $\bar\Pi^{\gamma Z}(q^2)\equiv\Pi^{\gamma Z}(q^2)-\Pi^{\gamma
  Z}(0)$ is defined in analogy with the HVP amplitude arising from the
current-current correlator of \eq{eq:PmunuMink}, except that one of
the electromagnetic currents has been replaced by the vector component
of the neutral weak current. Hadronic contributions to
$\bar\Pi^{\gamma Z}$ can be expressed in terms of the hadronic shift
in the value of the electroweak mixing angle, $\Delta_{\rm
  had}\sin^2\theta_W(q^2)$, i.e.
\begin{equation}\label{eq:PigammaZ}
    \bar\Pi^{\gamma Z}(q^2) = -\frac{\sin^2\theta_W^2}{4\pi\alpha}\,
    \Delta_{\rm had}\sin^2\theta_W(q^2)\,,\quad
    \Delta_{\rm had}\sin^2\theta_W(q^2)=
    \Delta\alpha_{\rm had}(q^2)-\Delta\alpha_{2,\,\rm had}(q^2)\,.
\end{equation}
Unlike the convention chosen in \eq{eq:PmunuMink}, the definition of
$\Pi^{\gamma Z}$ does not include the electric charge, which accounts
for the factor of $4\pi\alpha$ in \eq{eq:PigammaZ}. The hadronic
contribution to the running of $\alpha$ has been introduced in
\eq{eq:alpha_at_qsq}, and $\Delta\alpha_{2,\,\rm had}$ denotes the
analogue for the electroweak coupling.
After performing the separation in terms of quark flavours, the
expression for $\bar\Pi^{\gamma Z}(-M_Z^2)$ assumes the
form\,\cite{Hoferichter:2025yih}
\begin{equation}\label{eq:gammaZhad}
    \bar\Pi^{\gamma Z}(-M_Z^2) = \frac{1-2\sin^2\theta_W}{8\pi\alpha}\,
    \Delta\alpha^{(5)}_{\rm had}(-M_Z^2)
    -\frac{1}{6\,\sqrt{3}}\bar\Pi^{08}(-M_Z^2)
    -\frac{1}{12}\sum_{f=c,\,b}Q_f\bar\Pi^{ff}(-M_Z^2)\,,
\end{equation}
where the superscript ``$(5)$'' refers to QCD with five active quarks,
and $\bar\Pi^{08}$ parameterises SU(3)-flavour breaking
corrections. The advantage of this expression lies in the fact that
only the contributions from the heavy quark flavours are evaluated
perturbatively via the last term in \eq{eq:EW2loopLogs}. By contrast,
the contributions from light quark flavours are encoded in terms of
$\Delta\alpha^{(5)}_{\rm had}(-M_Z^2)$ and $\bar\Pi^{08}(-M_Z^2)$,
both of which are known beyond perturbation theory. Indeed, the above
expression can be evaluated using recent results for
$\Delta\alpha_{\rm had}(q^2)$ from dispersion theory
\cite{Davier:2019can, Keshavarzi:2019abf, Jegerlehner:2019lxt},
lattice QCD \cite{Ce:2022eix} and perturbative QCD
\cite{Erler:2023hyi, Erler:2024lds}.

To arrive at the full SM estimate it is useful to perform the
following decomposition \cite{Gnendiger:2013pva}:
\be\label{eq:EWfull}
a_\mu^{\rm EW} = a_{\mu,\,\rm bos}^{\rm EW\,(2)}
                +a_{\mu,\,\rm ferm}^{\rm EW\,(4)}
                +a_{\mu}^{\rm EW\,(\geq6)}
\ee
where the superscripts indicate the order in the expansion in powers
of the weak coupling. $a_{\mu,\,\rm bos}^{\rm EW\,(4)}$ collects all
two-loop diagrams without closed fermion loops (i.e. diagrams
(i)--(iii) in Fig.\,\ref{fig:EW2loopDiags}). It contains the
contributions from large logarithms specified in \eq{eq:EW2loopLogs},
as well as non-logarithmic terms and was first calculated in
\cite{Czarnecki:1995sz}.
The fermionic loop contribution $a_{\mu,\,\rm ferm}^{\rm EW\,(4)}$ is
further subdivided into
\begin{equation}\label{eq:EW2loopDecomp}
    a_{\mu,\,\rm ferm}^{\rm EW\,(4)}=
     a_{\mu,\,\rm ferm}^{\rm EW\,(4)}(e,\mu,u,c,d,s)
    +a_{\mu,\,\rm ferm}^{\rm EW\,(4)}(\tau,t,b)
    +a_{\mu,\,{\rm f-rest},\,H}^{\rm EW\,(4)}
    +a_{\mu,\,{\rm f-rest},\,{\rm no}\,H}^{\rm EW\,(4)}\,,
\end{equation}
where $a_{\mu,\,\rm ferm}^{\rm EW\,(4)}(e,\mu,u,c,d,s)$ and
$a_{\mu,\,\rm ferm}^{\rm EW\,(4)}(\tau,t,b)$ arise from diagram~(v) in
Fig.\,\ref{fig:EW2loopDiags}, $a_{\mu,\,{\rm f-rest},\,H}^{\rm
  EW\,(4)}$ denotes contributions from Higgs-dependent fermion loops
(diagram~(iv)), while $a_{\mu,\,{\rm f-rest},\,{\rm no}\,H}^{\rm
  EW\,(4)}$ collects all remaining fermionic two-loop contributions,
for instance, from $\gamma$-$Z$ mixing (diagram (vi)).

Using the estimates for $m_W, m_Z, M_H$ and $m_t$ from the latest
edition of the PDG\,\cite{ParticleDataGroup:2024cfk} the evaluation of
the bosonic and Higgs-dependent fermionic two-loop contributions
evaluate to\,\cite{Czarnecki:2002nt, Heinemeyer:2004yq,
  Gribouk:2005ee, Gnendiger:2013pva, Ishikawa:2018rlv,
  Hoferichter:2025yih}
\begin{equation}
    a_{\mu,\,\rm bos}^{\rm EW\,(4)}=-19.962(3)\cdot10^{-11}\,,
    \quad
    a_{\mu,\,{\rm f-rest},\,H}^{\rm EW\,(4)} =-1.500(2)\cdot10^{-11}\,.
\end{equation}
A remarkable feature of these estimates are the tiny errors, thanks to
the precise knowledge of electroweak input parameters, specifically
the Higgs mass.

The remaining contributions in \eq{eq:EW2loopDecomp} are evaluated in
terms of the standard two-loop diagrams, combined with estimates of
higher-order and non-perturbative QCD corrections. For the fermion
loop diagrams of type~(v) all fermions within one generation must be
included because of anomaly cancellation. The corresponding estimates
for each generation are obtained as\,\cite{Czarnecki:2002nt,
  Peris:1995bb, Knecht:2002hr, Ludtke:2024ase, Hoferichter:2025yih}
\begin{equation}
    a_{\mu,\,\rm ferm}^{\rm EW\,(4)}(e,u,d)=-2.08(3)\cdot10^{-11}\,,\quad
    a_{\mu,\,\rm ferm}^{\rm EW\,(4)}(\mu,c,s)=-4.14(28)\cdot10^{-11}\,,\quad
    a_{\mu,\,\rm ferm}^{\rm EW\,(4)}(\tau,t,b)=-8.12(1)\cdot10^{-11}\,.
\end{equation}
The quoted errors include uncertainties due to neglecting three-loop
contributions considered in \cite{Melnikov:2006qb}.

The non-Higgs part, $a_{\mu,\,{\rm f-rest},\,{\rm no}\,H}^{\rm
  EW\,(4)}$, contains the hadronic part of the $\gamma Z$ interaction
discussed above (see \eq{eq:gammaZhad}). The latter has been evaluated
in \cite{Hoferichter:2025yih} via a combination of dispersion theory
and lattice QCD which, when combined with the earlier results,
yields\,\cite{Czarnecki:2002nt, Gnendiger:2013pva,
  Jegerlehner:2017gek, Ce:2022eix, Hoferichter:2025yih}
\begin{equation}
    a_{\mu,\,{\rm f-rest},\,{\rm no}\,H}^{\rm EW\,(4)}=-4.58(10)\cdot10^{-11}\,.
\end{equation}
Finally, one has to consider higher loop contributions that are not
already included via the QCD corrections. These have been worked out
in \cite{Czarnecki:2002nt} and \cite{Degrassi:1998es} to
leading-logarithmic order, which yields terms of the form $\GF
\alpha^2\ln(M_Z/m_f)\ln(M_Z/m_{f'})$, where $f$ and $f'$ denote light
fermion flavours. Owing to a remarkable numerical cancellation among
three-loop correction observed in \cite{Czarnecki:2002nt} when the
two-loop contributions are expressed in terms of $\GF \alpha$, one
arrives at the estimate
\begin{equation}
    a_\mu^{\rm EW\,(\geq6)}=0.00(20)\cdot10^{-11}\,,
\end{equation}
where the quoted uncertainty is due to higher-order contributions
\cite{Czarnecki:2002nt}. We can now quote the total estimate for the
electroweak contribution to the muon anomalous magnetic moment, by
evaluating the sum in \eq{eq:EWfull}, which yields
\begin{equation}\label{eq:EWmutotal}
    a_\mu^{\rm EW}=154.4(4)\cdot10^{-11}\,.
\end{equation}
This confirms that the electroweak interaction contributes
only a tiny, but significant fraction to the muon anomalous magnetic
moment.

Electroweak corrections for the electron and $\tau$ have also been
worked out. The one-loop contributions can be obtained from
\eq{eq:EWoneloopTOT} by replacing the muon mass with $m_e$ and
$m_\tau$, respectively, which yields
\be
   a_e^{{\rm{EW}}\,{(2)}}=0.045564(2)\cdot10^{-12}\,,\qquad
   a_\tau^{{\rm{EW}}\,{(2)}}=55.096(6)\cdot10^{-8}\,.
\ee
In the case of the electron, the one-loop contribution from the Higgs
boson is tiny. However, for the $\tau$ they can be quite
significant. Corrections of order $(m_\ell/M_{W,Z,H})^2$, worked out
by Studenikin \cite{Studenikin:1998cs}, have been included in the
discussion by Eidelman and Passera \cite{Eidelman:2007sb}.
The two-loop corrections for the electron and $\tau$ cannot be
obtained straightforwardly by making replacements in the corresponding
expressions for the muon. The relevant two-loop formulae for the case
of the electron are listed in \cite{Jegerlehner:2009ry}. A concise
discussion for the $\tau$ can be found in \cite{Eidelman:2007sb}. The
latest estimates, taken from Refs.\,\cite{Jegerlehner:2017zsb}
and\,\cite{Eidelman:2007sb}, respectively, yield
\be\label{eq:EWelectautotal}
   a_e^{\rm EW}=0.030\,53(23)\cdot10{-12}\,,
   \quad
   a_\tau^{\rm EW}=47.4(5)\cdot10^{-8}\,.
\ee
It is important to realise that the estimate for $a_\tau^{\rm EW}$ has
not been updated using the latest results for the input
parameters. This will become an issue when more accurate experimental
measurements of $a_\tau$ are available.

A comparison of eqs.~(\ref{eq:EWmutotal})
and~(\ref{eq:EWelectautotal}) with the hadronic contributions listed
in Table~\ref{tab:HVPsumm} shows that electroweak corrections are much
smaller than the leading contribution from the strong
interaction. However, electroweak contributions gain in prominence for
increasing lepton mass.

%% file: conclusion.tex
\section{Conclusion \label{sec:concl}}

\begin{figure}
    \centering
    \includegraphics[width=0.7\linewidth]{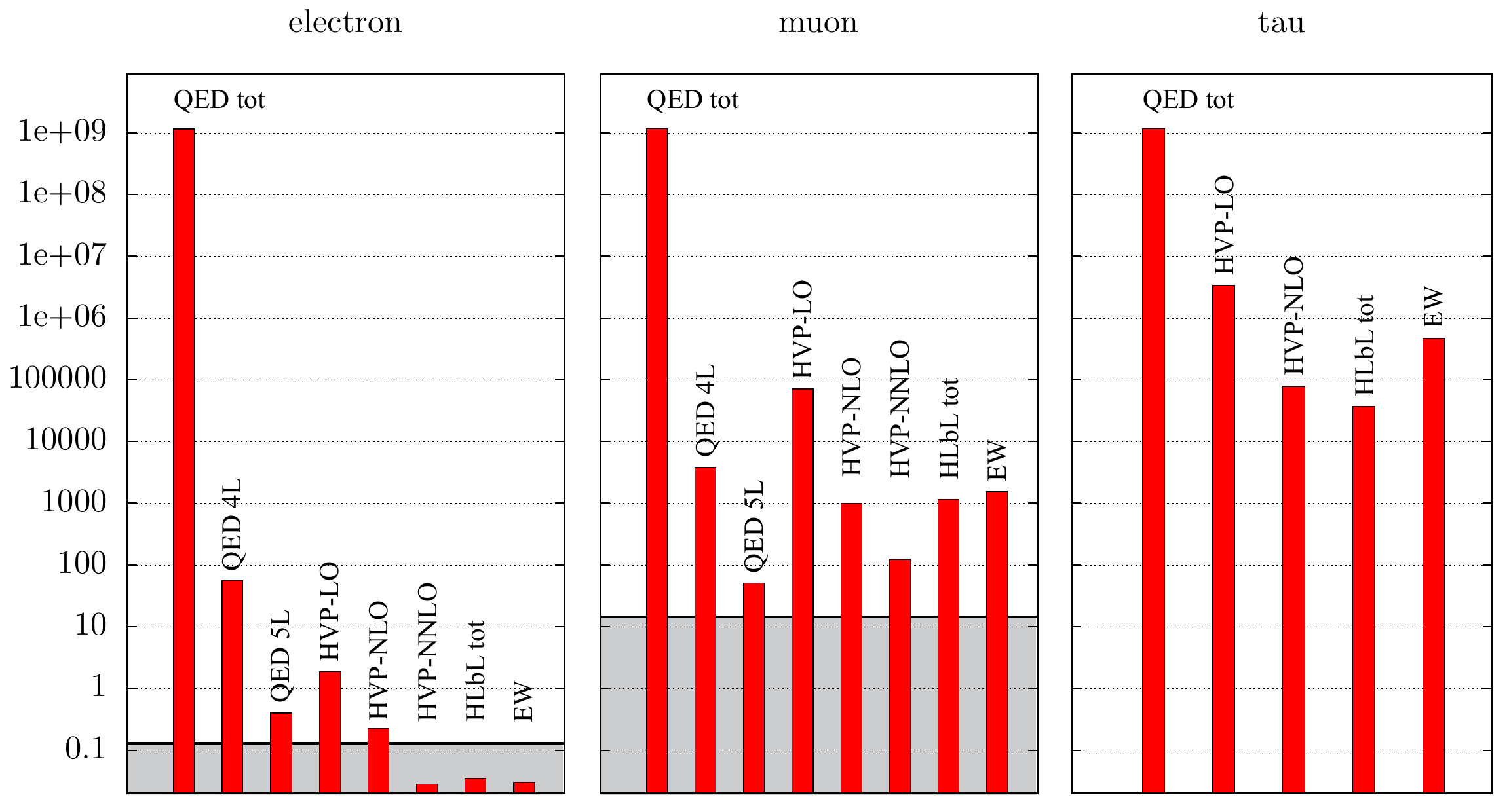}
    \caption{Compilation of the individual contributions to the
      anomalous magnetic moments of the electron, muon and $\tau$, in
      units of $10^{-12}$. The height of the grey bands shown for the
      electron and muon indicate the current experimental sensitivity
      of the direct measurement. Plot inspired by Fig.~3.8 in
      \cite{Jegerlehner:2017gek}.}
    \label{fig:AMMall}
\end{figure}

In this chapter we have discussed lepton anomalous magnetic moments
within the SM, starting with Schwinger's original calculation and
ending with the latest efforts to reliably pin down hadronic
contributions and to benefit from the increased precision in the
knowledge of input quantities.

We are now in a position to take stock, and to this end we present in
Fig.\,\ref{fig:AMMall} a breakdown of all individual SM contributions
to $a_e, a_\mu$ and $a_\tau$, plotted on a common scale in order to
facilitate the comparison of their relative size. Although one should
keep in mind that the actual numerical values may change over time,
they should ideally do so within errors. Figure\,\ref{fig:AMMall}
clearly shows that QED makes by far the biggest contribution,
regardless of lepton flavour. However, with increasing lepton mass the
fraction of the contributions from strong and electroweak interactions
increases substantially. For the muon, the sensitivity of the direct
measurement, shown as the thick horizontal line in
Fig.\,\ref{fig:AMMall}, is high enough to resolve QED effects at
five-loop order (labelled ``QED\,5L''), as well as hadronic effects up
to O$(\alpha^4)$ (i.e. NNLO) and electroweak corrections at two-loop
level. For the electron, higher-order hadronic and electroweak
corrections are more strongly suppressed and currently not resolved by
experiment.
%However, a further increase in experimental sensitivity
%can be expected, which would make knowledge of the six-loop QED
%contribution a necessity.

\begin{table}[t]
    \centering
    \begin{tabular}{c| r@{.}l l | r@{.}l l | r@{.}l l}
    \hline\hline
    & \multicolumn{3}{c|}{$a_e\cdot10^{12}$}
    & \multicolumn{3}{c|}{$a_\mu\cdot10^{10}$}
    & \multicolumn{3}{c}{$a_\tau\cdot10^{8}$} \\
    \hline
    & 1\,159\,652\,181&59(23)$_{\rm Cs}$ & [0.20\,{\rm ppb}] &
    \multicolumn{3}{l|}{$\empty$}
    & \multicolumn{3}{l}{$\empty$}  \\
    \rb{SM} & 1\,159\,652\,180&24(9)$_{\rm Rb}$  & [0.075\,{\rm ppb}] 
    & \multicolumn{2}{l}{\rb{1\,165\,920\,3.3(6.2)}} & \rb{[532\,{\rm
          ppb}]}
    & \multicolumn{2}{l}{\rb{117\,717.1(4.0)}} & \rb{[34\,{\rm ppm}]}
    \\
    \hline
    Exp & 1\,159\,652\,180&59(13) & [0.11\,{\rm ppb}]
    & 1\,165\,920\,7&15(1.45) & [124\,{\rm ppb}] &
    \multicolumn{2}{l}{$-0.057$ to $0.024$} &
    95\%\,C.L. \\
    \hline
    & $-1$&00(26)$_{\rm Cs}$ & $-3.8\sigma$ &
    \multicolumn{3}{l|}{$\empty$}
    & \multicolumn{3}{l}{$\empty$} \\
    \rb{$\rm Exp-SM$} & 0&35(16)$_{\rm Rb}$ & $\phantom{-}2.3\sigma$
    & \multicolumn{2}{l}{\rb{$\phantom{1\,165\,920\,}3.8(6.3)$}} &
    \rb{$\phantom{12}0.6\sigma$}
    & \multicolumn{3}{l}{$\empty$} \\
    \hline\hline
    \end{tabular}
    \caption{Standard Model predictions of lepton anomalous magnetic
      moments versus experimental measurements. For the electron, the
      current SM estimate and uncertainty both depend on the input
      value for the fine-structure constant as measured in atom
      interferometry on Cesium or Rubidium.}
    \label{tab:AMMall}
\end{table}

After adding up all contributions from QED, the strong and the weak
interactions to $a_\ell^{\rm SM}$ one obtains the entries listed in
Table\,\ref{tab:AMMall}. Given that among all leptons, the electron's
anomalous magnetic moment is most strongly dominated by QED
contributions, it is not surprising that its SM estimate is affected
by the choice of input value for the fine-structure constant,
$\alpha$, determined from atom interferometry on either Cesium or
Rubidium. The tension between the SM prediction and experimental
measurement varies between $-3.8$ and $2.3$ standard deviations, which
should be regarded as a manifestation of discrepant measurements of
$\alpha$ rather than a hint of a possible deviation from the SM. The
group that published the latest measurement of $a_e$
\cite{Fan:2022eto} is planning for another reduction of the total
error by an order of magnitude. This would clearly necessitate the
calculation of the six-loop QED contribution. Moreover, the
uncertainty assigned to the HVP contribution $a_e^{\rm hvp}$ would
also have to be reduced, in order to maintain the balance in precision
between measurement and SM prediction.

After the latest update of the SM for the muon anomalous magnetic
moment \cite{Aliberti:2025beg}, no significant tension with the
experimental average \cite{Muong-2:2025xyk} is observed. This is a
major reversal of the situation that persisted for many years and
demonstrates the enormous progress achieved in determining hadronic
contributions. However, the fact that the SM prediction for $a_\mu$
agrees with experiment should not be taken as evidence that there are
no open questions left to study: On the contrary, the tension between
data-driven and lattice QCD determinations of $\alohvp$ must be
understood. New measurements of hadronic cross sections or new
analyses of already collected data are currently in progress at all
major experiments. A significant reduction of the error in $\alohvp$
could be achieved by combining lattice calculations and data-driven
evaluations, provided that the tension among the latter can be
resolved.
An important role in this regards is played by the MUoNE experiment
\cite{MUonE:2016hru} which is designed to measure the leading-order
HVP contribution in elastic muon-electron scattering. Efforts are also
under way to improve the direct measurement of $a_\mu$ beyond the
current precision of 124\,ppb. In addition, the E34 experiment at
J-PARC plans to provide an independent measurement of $a_\mu$
employing the novel technique of cooled muon beams
\cite{Otani:2022wlj}.

There is currently no realistic prospect for a measurement of $a_\tau$
whose precision would be competitive with that of the SM
prediction. Hence, there is currently no incentive to push the
precision of the SM estimate to a similar level as for the electron
and muon.

In this chapter we have not touched at all on contributions that may
arise from BSM physics. A comprehensive and up-to-date overview of
possible BSM scenarios and their implications for $a_\mu$ in
particular can be found in \cite{Athron:2025ets}. While the case for
discussing lepton anomalous magnetic moments within extensions of the
SM may have weakened for now, it could become a lot stronger once the
experimental sensitivity and the precision of the SM prediction
increase further.

%We have therefore refrained from providing a summary table of all
%known numerical values of the various sub-contributions.